\documentclass[12pt]{article}

\usepackage{amssymb}

\newif\ifpdf
\ifx\pdfoutput\undefined
   \pdffalse
   \usepackage{cite}
   \newcommand{\href}[2]{\texttt{#1}}
  \usepackage{epsf}
 \else
   \pdfoutput=1
   \pdftrue
     \usepackage[pdftex]{hyperref,graphicx}
\fi

\newcommand{\Red}[1]{#1}
\newcommand{\OliveGreen}[1]{#1}
\newcommand{\Green}[1]{#1}

\newcommand{\Blue}[1]{#1}
\newcommand{\Maroon}[1]{#1}
\newcommand{\Magenta}[1]{#1}
\newcommand{\Purple}[1]{#1}
\newcommand{\ft}[2]{{\textstyle\frac{#1}{#2}}}

\def\trace{\mathop{\rm Tr}\nolimits}

\def\rmi{{\rm i}}
\def\rmd{{\rm d}}
\newsavebox{\uuunit}
\sbox{\uuunit}
    {\setlength{\unitlength}{0.825em}
     \begin{picture}(0.6,0.7)
        \thinlines
        \put(0,0){\line(1,0){0.5}}
        \put(0.15,0){\line(0,1){0.7}}
        \put(0.35,0){\line(0,1){0.8}}
       \multiput(0.3,0.8)(-0.04,-0.02){12}{\rule{0.5pt}{0.5pt}}
     \end {picture}}
\newcommand {\unity}{\mathord{\!\usebox{\uuunit}}}
\csname @addtoreset\endcsname{equation}{section}
\newcommand{\SU}{\mathop{\rm SU}}
\newcommand{\SO}{\mathop{\rm SO}}

\newcommand{\USp}{\mathop{\rm {}USp}}

\newcommand{\Sl}{\mathop{\rm {}S}\ell }
\newcommand{\Gl}{\mathop{\rm {}G}\ell }
\newcommand{\covder}{\mathfrak{D}}
\def\Ric{R}
\newcommand{\vm}[1]{{\vec m}_{#1}}
\begin{document}

 %%%%%%%%%%%%%%%%%%%%%%%%%%%%%%%%%%%%%%%%%%%%%%%%%%%%%%%%%%%
\begin{titlepage}
\begin{flushright}
KUL-TF-04/33\\
UG-04/03\\
SPIN-04/21\\
ITF-UU-04/39\\
DFTT 27/04\\
 hep-th/0411209
\end{flushright}
\vspace{.5cm}
\begin{center}
\baselineskip=16pt {\LARGE
   The Map Between  \\ \vskip 0.2cm
    Conformal Hypercomplex/Hyper-K{\"a}hler \\ \vskip 0.2cm
     and Quaternionic(-K{\"a}hler) Geometry
}\\
\vskip 3mm % 15mm%27.mm
{\large Eric Bergshoeff$^1$, Sorin Cucu$^2$, Tim de Wit$^1$, \\[2mm]
Jos Gheerardyn$^{2,3}$, Stefan Vandoren$^4$ \\[2mm]
and Antoine Van Proeyen$^2$
  } \\
\vskip 5mm
{\small  $^1$ Center for Theoretical Physics, University of Groningen,\\
       Nijenborgh 4, 9747 AG Groningen, The Netherlands. \\ [2mm]
   %    e.bergshoeff@phys.rug.nl, t.c.dewit@azu.nl \\
  $^2$ Instituut voor Theoretische Fysica, Katholieke Universiteit Leuven,\\
       Celestijnenlaan 200D B-3001 Leuven, Belgium.
      \\[2mm]
  $^3$ Dipartimento di Fisica Teorica, Universit{\`a} di Torino\\
  and I.N.F.N., Sezione di Torino,\\
  via P. Giuria 1, I-10125 Torino, Italy\\[2mm]
  $^4$ Institute for Theoretical Physics, Utrecht University, \\
 Leuvenlaan 4, 3508 TA Utrecht, The Netherlands.
     }
\end{center}
\vfill
\begin{center}
{\bf Abstract}
\end{center}
{\small We review the general properties of target spaces of
hypermultiplets, which are quaternionic-like manifolds, and discuss the
relations between these manifolds and their symmetry generators. We
explicitly construct a one-to-one map between conformal hypercomplex
manifolds (i.e. those that have a closed homothetic Killing vector) and
quaternionic manifolds of one quaternionic dimension less. An important
role is played by `$\xi$-transformations', relating complex structures on
conformal hypercomplex manifolds and connections on quaternionic
manifolds. In this map, the subclass of conformal hyper-K{\"a}hler manifolds
is mapped to quaternionic-K{\"a}hler manifolds. We relate the curvatures of
the corresponding manifolds and furthermore map the symmetries of these
manifolds to each other.
 }
\vspace{2mm} \vfill \hrule width 3.cm {\footnotesize \noindent e-mails:
 e.bergshoeff@phys.rug.nl, sorin.cucu@ua.ac.be, t.c.dewit@azu.nl,
gheerard@to.infn.it, \\ \phantom{e-mails: }
s.vandoren@phys.uu.nl, antoine.vanproeyen@fys.kuleuven.ac.be }
\end{titlepage}
\addtocounter{page}{1}
 \tableofcontents{}
\newpage
%%%%%%%%%%%%%%%%
\section{Introduction}

Ever since Einstein's theory of gravity, differential geometry has
entered research areas in theoretical physics in various places and
contexts. The idea of using geometry to describe and unify the forces of
nature continues to play an important role in present day high energy
physics. After the invention of supersymmetry, superstrings and their
compactifications, connections were found with the holonomy groups and
Killing spinors of Riemannian manifolds. It was shown that supersymmetric
sigma models were deeply related to geometries with complex structures.
K{\"a}hler manifolds appear already in $N=1$ theories in 4
dimensions~\cite{Zumino:1979et}. Some theories with 8 supersymmetries,
i.e. $N=4$ in 2 dimensions and $N=2$ in 3, 4, 5 or 6 dimensions, exhibit
a hypercomplex structure. It was shown that Lagrangians for rigid
supersymmetry lead to hyper-K{\"a}hler~\cite{Alvarez-Gaume:1981hm} manifolds
and Lagrangians for supergravity theories lead to quaternionic-K{\"a}hler
manifolds~\cite{Bagger:1983tt}. Many aspects of such geometries are by
now well-known and studied by the supersymmetry, supergravity and
superstring community, and this has led to new insights and useful
applications in these fields. Recently, it has been shown that a
generalization of hyper-K{\"a}hler manifolds is possible for rigid $N=2$
supersymmetric theories if one does not demand that the field equations
are derivable from an action~\cite{Bergshoeff:2002qk,Gheerardyn:th2004}.
Such a situation arises when in the field equations for the scalar
fields, a connection is chosen different from a Levi-Civita connection
derivable from a metric on the sigma model target manifold. The
corresponding geometry is called hypercomplex~\cite{Salamon:1986}, which
differs from a hyper-K{\"a}hler geometry by the fact that it does not
necessarily possess a covariantly constant metric preserved by the
connection induced by the hypercomplex structure, as we explain in detail
below. We will further refer to a Hermitian metric that is covariantly
constant with respect to the latter connection as a `good metric'. The
manifolds mentioned above are schematically represented in
Table~\ref{tbl:quatlikeMan}.
\begin{table}[ht]
  \caption{\it Quaternionic-like manifolds of real dimension $4r$.
These are the manifolds that have a quaternionic structure
satisfying~(\ref{JJ}) and~(\ref{covconstJq}). The holonomy group is
indicated. Generalizations for different signatures are obvious. The
table is essentially taken over from~\cite{AM1996}. The dot notation
means e.g. $\SU(2)\cdot \USp(2r)=\frac{\SU(2)\times
\USp(2r)}{\mathbb{Z}_2}$.} \label{tbl:quatlikeMan}
\begin{center}
  \begin{tabular}{|c||c|c||l|}\hline
      & no good  metric & with a good metric& \\ \hline\hline
    no $\SU(2)$ & \textit{\textbf{hypercomplex}} & \textit{\textbf{hyper-K{\"a}hler}}& rigid  \\
    curvature & $\Gl(r,\mathbb{H})$ & $\USp(2r)$&supersymmetry \\ \hline
    non-zero $\SU(2)$ & \textit{\textbf{quaternionic}} & \textit{\textbf{quaternionic-K{\"a}hler}}& \\
    curvature & $\SU(2)\cdot \Gl(r,\mathbb{H})$  & $\SU(2)\cdot\USp(2r)$ & supergravity \\
\hline\hline
 & field equations & action& \\ \hline
  \end{tabular}
\end{center}
\end{table}
There it is indicated that hypercomplex and hyper-K{\"a}hler manifolds occur
in rigid supersymmetric theories, while quaternionic and
quaternionic-K{\"a}hler manifolds occur in supergravity. Manifolds without a
good metric occur as long as the field equations are not derived from a
conventional action. The conventional actions in supersymmetry and
supergravity involve a metric on the manifold of scalars, and the field
equations naturally involve the Levi-Civita connection. Hypercomplex and
quaternionic manifolds are endowed with a connection that is not
necessarily the Levi-Civita connection. For this reason we say they are
not derived from an action.

The supersymmetry algebra on hypermultiplets requires that we use
torsionless connections. However, hyper-K{\"a}hler torsion (HKT) manifolds,
which appear as the moduli space of supersymmetric multi-black holes
\cite{Gibbons:1997iy,Michelson:1999zf}, belong also to the class of
hypercomplex manifolds. Indeed, they carry a triplet of complex
structures with vanishing Nijenhuis tensor, implying that there exists a
torsionless connection. This is similar to the way in which we used
two-dimensional non-linear sigma models on group manifolds
\cite{Spindel:1988sr} for the example
 in~\cite[Appendix C]{Bergshoeff:2002qk}. Also there, the connection with torsion
corresponding to the three-form field strength is different from the
torsionless connection used in the definition of hypercomplex manifolds.

Supergravity theories can be constructed by the `superconformal tensor
calculus'~\cite{Ferrara:1977ij,Kaku:1978nz,Kaku:1978ea}, which gives
insight in the geometry of the relevant sigma models. These geometries
are obtained as projective spaces, related to the dilatation symmetry in
the superconformal group, and with possible further projections related
to the R-symmetry group, see~\cite{VanProeyen:2001wr} for a review of
these principles. In particular, this has been used
in~\cite{Galicki:1992tm,deWit:1998zg,deWit:1999fp,deWit:2001dj} in the
context of $N=2$ theories with hypermultiplets, constructing the link
between hyper-K{\"a}hler manifolds and quaternionic-K{\"a}hler manifolds. This
construction starts from a conformal hyper-K{\"a}hler manifold. We will use
the name \emph{conformal manifold} for a manifold which has a closed
homothetic Killing vector. For a manifold with coordinates $q^X$ and
affine connection $\Gamma _{XY}{}^Z$, this is a vector $k^X$,
satisfying\footnote{The factor $3/2$ is a choice of normalization which
is convenient for applications of 5-dimensional supergravity theories.
The translation between formulations appropriate to supergravities in
other dimensions has been considered in~\cite{Rosseel:2004fa}.}
\begin{equation}
  { {\mathfrak{D}}}_{ Y} k^{ X} \equiv \partial_{ Y}
k^{ X} + {{ \Gamma}_{{ Y}{ Z}}}{}^{ X} k^{ Z} = \ft32 \delta_{ Y}{}^{ X}.
 \label{homothetic}
\end{equation}
The presence of such a vector allows the definition of conformal
symmetries as a basic step for the superconformal tensor
calculus~\cite{Sezgin:1995th}.

In this paper, we will formulate and prove the 1-to-1
correspondence\footnote{As will be explained below, the correspondence is
actually 1-to-1 between families (or `equivalence classes') of
manifolds.} (locally) between conformal hypercomplex manifolds of
quaternionic dimension $n_H+1$ and quaternionic manifolds of dimension
$n_H$. Furthermore, we show that this 1-to-1 correspondence is also
applicable between the subset of hypercomplex manifolds that are
hyper-K{\"a}hler and the subset of quaternionic manifolds that are
quaternionic-K{\"a}hler. In the mathematics literature the map between
quaternionic-K{\"a}hler and hyper-K{\"a}hler manifolds is constructed by Swann
\cite{Swann}, and its generalization to quaternionic manifolds is treated
in \cite{PedersenPS1998}. Here, we give explicit expressions for the
complex structures and connections that are needed to apply these results
in the context of the conformal tensor calculus in supergravity.

To explain the 1-to-1 mapping let us first repeat the basic definitions.
A \emph{hypercomplex manifold} is a manifold with a hypercomplex
structure. This means that in any local patch\footnote{We use the integer
$r$ for the quaternionic dimension of any quaternionic-like manifold (the
number of hypermultiplets). In the application to the map, this $r$ can
be $n_H$ or $n_H+1$ depending on whether we are considering the
quaternionic space or the hypercomplex space, respectively.} there are
coordinates $q^X$ with $X=1,\dots ,4r$ and a triplet of complex
structures $\vec{J}_X{}^Y$ that satisfy the algebra of the imaginary
quaternions,
%\begin{equation}
%J^{\bar \alpha} J^{\bar \beta} =-\delta^{\bar \alpha\bar \beta}\unity
%_{4r}+\varepsilon^{\bar \alpha\bar \beta\bar \gamma} J^{\bar \gamma}.
%\label{JJ}
%\end{equation}
which is that for any vectors $\vec A$ and $\vec B$,
\begin{equation}
\vec A\cdot \vec J\, \vec B\cdot \vec J = -\unity _{4r}\vec A\cdot \vec B
+ (\vec A\times \vec B)\cdot \vec J, \qquad \trace \vec{J}=0.
  \label{JJ}
\end{equation}
This defines an almost hypercomplex structure.\footnote{Note that the
tracelessness is implied by the first relation.} The closure of the
(rigid) supersymmetry algebra on hypermultiplets requires that the
complex structures are covariantly constant using a \textit{torsionless}
affine connection $\Gamma _{XY}{}^Z=\Gamma _{YX}{}^Z$:
\begin{equation}
  \covder_X\vec J_Y{}^Z\equiv \partial_X\vec J_Y{}^Z
  -\Purple{\Gamma _{XY}{}^W}\vec J_W{}^Z
  +\Purple{\Gamma _{XW}{}^Z}\vec J_Y{}^W=0
  %+2\,\OliveGreen{\vec \omega _X}\times \vec J_Y{}^Z=0
. \label{covconstJhc}
\end{equation}
Stated otherwise, the hypercomplex structures should be
integrable.\footnote{In the context of G-structures, condition
(\ref{covconstJhc}) would rather be called the one-integrability of the
hypercomplex structure \cite{AM1996}.}

A \emph{quaternionic manifold} is defined by a local span of these three
complex structures. This means that the three complex structures can at
any point $q$ be rotated as $\vec J'= R(q) \vec J$, where $R$ is a
$3\times 3$ matrix of $\SO(3)$. The covariant constancy condition (i.e.
the integrability) should then also be covariant with respect to these
rotations. This implies that one needs a connection $\vec{\omega }_X$ and
the condition (\ref{covconstJhc}) is replaced by
\begin{equation}
  \covder_X\vec J_Y{}^Z\equiv \partial_X\vec J_Y{}^Z
  -\Purple{\Gamma _{XY}{}^W}\vec J_W{}^Z
  +\Purple{\Gamma _{XW}{}^Z}\vec J_Y{}^W
  +2\,\OliveGreen{\vec \omega _X}\times \vec J_Y{}^Z=0
. \label{covconstJq}
\end{equation}
Infinitesimal $\SO(3)$ rotations are parametrized by 3 angles
$\vec{\ell}(q)$,
\begin{equation}
  \delta_{\rm SU(2)} \vec{J}_X{}^Y=\vec{\ell }\times\vec{J}_X{}^Y,\qquad
  \delta_{\rm SU(2)} \vec{\omega }_X=-\ft12\partial _X\vec{\ell }
    +\vec{\ell }\times\vec{\omega}_X.
 \label{SO3quat}
\end{equation}

The connections in (\ref{covconstJq}) are not unique. Indeed, they can be
changed simultaneously depending on an arbitrary one-form $\xi =\xi
_X\rmd q^X$ as \cite{AM1996,Fujimura:1976,Oproiu1984}
\begin{equation}
\Purple{\Gamma _{XY}{}^Z}\rightarrow \Purple{\Gamma _{XY}{}^Z}+2\delta
^Z_{(X}\Blue{\xi_{Y)}}-2\vec J_{(X}{}^Z\cdot \vec J_{Y)}{}^W\Blue{\xi
_W},\qquad  \OliveGreen{\vec \omega_X} \rightarrow \OliveGreen{\vec
\omega_X } +\vec J_X{}^W\Blue{\xi _W}.
 \label{changeGammaomega}
\end{equation}
These transformations relate different connections on the same manifold
and therefore define equivalence classes. Furthermore,  it will turn out
that for the conformal hypercomplex manifolds there are related
transformations between complex structures and connections. These define
equivalence classes between conformal hypercomplex structures.

There is however one important difference between the transformations on
the quaternionic and on the hypercomplex space. On the quaternionic
manifold, the $\xi$-transformation yields different torsionless
connections for a given (integrable) quaternionic structure (i.e. for a
given span of three complex structures). Otherwise stated, an integrable
quaternionic structure does not determine the connection uniquely. On the
conformal hypercomplex space, the transformations relate different
(integrable) hypercomplex structures in a continuous fashion. To the best
of our knowledge, these transformations between hypercomplex structures
on conformal manifolds have not yet been discussed in the literature. The
map from conformal hypercomplex to quaternionic manifolds is subject to
these transformations, and is 1-to-1 for the equivalence classes.

We give an explicit construction of this correspondence by relating the
geometric quantities in the corresponding manifolds, i.e. the complex
structures, affine connections, curvatures, Killing vectors and moment
maps. We prove that the existence of a `good metric' for a hypercomplex
manifold is equivalent to the existence of a `good metric' in the
corresponding quaternionic space. Also the symmetries of related
manifolds are mapped 1-to-1.

Our formulation here focuses purely on the geometrical aspects. Our
results have been applied already to $N=2$ matter coupled $D=5$
supergravity~\cite{Bergshoeff:2004kh}. The results of this paper are
applicable to the geometry of hypermultiplets independent of whether they
are defined in 3, 4, 5 or 6 dimensions.

\smallskip

In Sect.~\ref{ss:QuatLikeMan} we give a summary of the properties of the
geometries with a triplet of complex structures (`quaternionic-like
manifolds'). In particular we discuss the curvatures determining the
holonomy groups. We will devote special attention to the
$\xi$-transformations and to the properties of the Ricci tensor. At the
end of that section we will look at those manifolds that have a Hermitian
Ricci tensor, which include all hypercomplex, hyper-K{\"a}hler and
quaternionic-K{\"a}hler manifolds. Most results in this section are due to
\cite{AM1996}.

The new work starts in Sect.~\ref{ss:map}, where we construct the map
discussed above. We start by considering conformal hypercomplex manifolds
and identify new transformations between connections and the hypercomplex
structure that respect the conditions for hypercomplex manifolds. These
spaces are then reduced to a submanifold that turns out to be
quaternionic. Coordinates are chosen in view of the gauge fixing of
dilatation, $\SU(2)$ symmetry and special ($S$) supersymmetry in the
superconformal context. We construct the geometric building blocks in
this suitable basis. These are the complex structures and affine
connections. The freedom of $\xi $-transformations comes naturally out of
this map. Inversely, we associate such a conformal hypercomplex manifold
to any quaternionic manifold. This finishes the proof that the mapping
between these manifolds is 1-to-1. In a further subsection, we focus on
the manifolds with a good metric, i.e. hyper-K{\"a}hler and
quaternionic-K{\"a}hler manifolds. We show that these are also 1-to-1 related
and give explicit expressions for the connections. Finally, in this
section we also construct the vielbeins and the spin connections (i.e.
the connections that transform under local general linear quaternionic
transformations).

The relation between curvatures of the conformal hypercomplex and the
quaternionic manifolds is discussed in Sect.~\ref{ss:curvatures}. As
recalled in Sect.~\ref{ss:QuatLikeMan}, the curvatures of
quaternionic-like manifolds are characterized by a symmetric `Weyl
tensor' ${\cal W}_{ABC}{}^D$, where $A=1,\ldots ,2r$ are indices in the
tangent space. We will therefore connect the Weyl tensors of the
hypercomplex and quaternionic spaces.

Symmetries of the manifolds that preserve also the hypercomplex structure
are called triholomorphic. Such symmetries of conformal hypercomplex
manifolds descend to quaternionic symmetries of quaternionic manifolds
(which is a statement of preservation of the span of complex structures).
We define the moment maps of quaternionic manifolds, and we show in
Sect.~\ref{ss:symmetries} that these are determined in terms of
components of the symmetries of the hypercomplex manifolds in directions
that are projected out by the gauge fixing.

We summarize and overview all results of this paper in
Sect.~\ref{ss:discussion}. We have written this in such a way that this
section can be read by itself by a reader familiar with quaternionic
geometry (that is recapitulated in Sect.~\ref{ss:QuatLikeMan}). We
illustrate the results with a schematic picture,
Fig.~\ref{fig:mapoverview}. After this summary, we discuss some remaining
issues and give some remarks.   In the main part of the paper, we choose
the signatures in the way that is most relevant for supergravity, i.e.
such that the scalars and the graviton have positive kinetic energies.
This selects quaternionic-K{\"a}hler manifolds with a positive definite
metric and a negative definite scalar curvature. These are obtained in
the map by starting with hyper-K{\"a}hler manifolds of quaternionic signature
$(-+\cdots +)$. In the discussion section, we indicate how our formulae
can be used for other signatures.

An appendix gives some useful formulae for calculus with complex
structures and Hermitian tensors.

%%%%%%%%%%%%%%%%%%%%%%%%%%%%%%%%%%%%%%%%%%%%%%%%%%%
\section{Quaternionic-like Manifolds}
\label{ss:QuatLikeMan}
%%%%%%%%%%%%%%%%%%%%%%%%%%%%%%%%%%%%%%%%%%%%%%%%%%%

In this section, we review the four different geometries that are used in
this paper. They correspond to hypercomplex, hyper-K{\"a}hler, quaternionic
and quaternionic-K{\"a}hler manifolds, and are distinguished by the
properties of their holonomy groups and the presence of a preserved
metric, as summarized in Table~\ref{tbl:quatlikeMan}. This section is an
extension of Appendix~B of~\cite{Bergshoeff:2002qk}, a paper where these
geometries are discussed in the context of five dimensional conformal
hypermultiplets, and which itself makes heavy use of the pioneering
paper~\cite{AM1996}.

We use here the name `quaternionic-like manifolds' for all the manifolds
in Table~\ref{tbl:quatlikeMan}. In fact, the formulae for quaternionic
manifolds are the most general ones, and in this respect one can argue to
just use `quaternionic manifolds' as general terminology. The subtlety is
that in principle for hypercomplex manifolds one admissible basis of
quaternionic structures is selected, while in quaternionic manifolds only
the local span of complex structures is used. For all practical purposes,
the formulae of quaternionic manifolds are the most general, and can be
applied with $\vec{\omega }_X=0$ for hypercomplex manifolds.

Subsection \ref{ss:genpropquat} gives these general formulae for an
arbitrary quaternionic manifold. The properties of the curvatures are
presented. The $\xi $-transformations mentioned in the introduction are
treated in more detail in Subsect.~\ref{ss:xi}. Special features for the
case of hypercomplex manifolds are given in Subsect.~\ref{ss:hc}, for
hyper-K{\"a}hler manifolds in Subsect.~\ref{ss:hk} and for
quaternionic-K{\"a}hler manifolds in Subsect.~\ref{ss:qk}. In these 3 cases,
the Ricci tensor is Hermitian. We give general properties of
quaternionic-like manifolds with Hermitian Ricci tensor in
Subsect.~\ref{ss:hermRicci}.

\subsection{General properties of quaternionic-like manifolds}
 \label{ss:genpropquat}
The common property of all quaternionic-like manifolds, say of dimension
$4r$, is the existence of a quaternionic structure, i.e.\ a triplet of
endomorphisms $\vec{J}$, realizing the algebra of the imaginary
quaternions (\ref{JJ}).

In general relativity, it is often convenient to introduce (locally) a
set of one-forms that define an orthonormal frame: $e^a= e_\mu {}^a\rmd
x^\mu $. Their components $e_\mu{}^a$ are the so-called `vielbeins'. As
such, the so-called \emph{flat} index $a$ takes values in the orthogonal
group, and $\mu$ is the one-form index, also known as the \emph{curved}
index. In the present context, the orthogonal group is substituted by the
groups mentioned in Table~\ref{tbl:quatlikeMan}. Objects with flat
indices will be said to take values in the tangent bundle.

Therefore, we locally introduce the coordinates $q^X$ with $X=1,\dots
,4r$, and we assume the existence of an invertible vielbein $f_X^{iA}$,
two matrices $\rho_A{}^B$ and $E_i{}^j$ (with $i=1,2$ and $A=1,\ldots
,2r$) that satisfy
\begin{equation}
\rho\, \rho^{\ast} = - \unity_{2r}, \qquad E\, E^{\ast} = -\unity_2,
\end{equation}
together with the reality condition
\begin{equation}
(f_X^{iA})^{\ast} = f_X^{jB} E_j {}^i \rho_B {}^A.
 \label{realfunctions}
\end{equation}
One can choose a basis such that $E_i{}^j=\varepsilon _{ij}$, see e.g.
\cite{deWit:1985px}. We will further always use this basis.

The transformations on variables with an $A$ index are restricted by the
reality condition  to $\Gl (r,\mathbb{H})=\SU^*(2r)\times $U$(1)$. The
inverse vielbein, denoted by $f^X_{iA}$, satisfies
\begin{equation}\label{inv-vielbein}
f^{iA}_Y f^X_{iA} = \delta_Y^X ,\qquad f^{iA}_Xf^X_{jB}=\delta^i_j
\delta^A_B,
\end{equation}
and can be used to define the quaternionic structure as
\begin{equation}
\vec J_X{}^Y \equiv -\rmi f_X^{iA}\vec \sigma _i{}^j
f_{jA}^Y,\label{defJf}
\end{equation}
where $\vec \sigma$ are the three Pauli matrices. These matrices
$\vec{J}$ satisfy (\ref{JJ}). We use here a~slight change of notation
with respect to~\cite{Bergshoeff:2002qk} in that we will indicate the
triplets by a vector symbol, rather than an $\alpha $ index
in~\cite{Bergshoeff:2002qk} (which will be needed below to indicate a
coordinate set).

All quaternionic-like manifolds admit connections with respect to which
the vielbeins are covariantly constant by definition. These are a
torsionless affine connection $\Gamma_{XY}{}^Z$, a $\Gl(r,\mathbb{H})$
connection ${\omega}_{XA}{}^B$ on the tangent bundle and possibly an
$\SU(2)$ connection $\omega_{Xi}{}^j$, such that
\begin{equation}
 \covder_Xf_Y^{iA}\equiv \partial _X f_Y^{iA}-\Purple{\Gamma _{XY}{}^Z}
 f_Z^{iA}+f_Y^{jA}
  \OliveGreen{\omega _{Xj}{}^i} +f_Y^{iB}\Red{\omega_{XB}{}^A}=0,
 \label{covconstf}
\end{equation}
and similarly for the inverse vielbein. %Note that in the
%hypercomplex/hyper-K{\"a}hler case, the $\SU(2)$ connection is trivial. We
%will mainly work in a gauge in which the $\SU(2)$ connection vanishes.

This implies that the quaternionic structure is covariantly constant with
respect to the affine connection and the $\SU(2)$
connection~\footnote{\label{ijtoalpha}One can make the transition from
doublet to vector notation by using the sigma matrices, ${\omega}_{X
i}{}^j = \rmi \vec \sigma _i{}^j\cdot \vec \omega_X$, and similarly $\vec
\omega_X = -\frac 12\rmi\vec \sigma _i{}^j {\omega}_{Xj}{}^i$. This
transition between doublet and triplet notation is valid for any triplet
object as e.g.\ the complex structures.},
\begin{equation}
  \covder_X\vec J_Y{}^Z\equiv \partial_X\vec J_Y{}^Z
  -\Purple{\Gamma _{XY}{}^W}\vec J_W{}^Z
  +\Purple{\Gamma _{XW}{}^Z}\vec J_Y{}^W+2\,\OliveGreen{\vec \omega _X}\times \vec J_Y{}^Z=0
. \label{covconstJ}
\end{equation}
The $\Gl(r,\mathbb{H})$-connection ${\omega}_{XA}{}^B$ on the tangent
bundle can then be determined by requiring~(\ref{covconstf}):
\begin{equation}
  \omega _{XA}{}^B=\ft 12 f_Y^{iB}\left( \partial _X f_{iA}^Y+\Gamma
  _{XZ}{}^Yf_{iA}^Z - \OliveGreen{\omega _{Xi}{}^j} f^Y_{jA}
  \right) .
 \label{determineOm}
\end{equation}

A useful concept in describing the integrability of almost
quaternionic-like structures is the Nijenhuis tensor. %$N^{\bar \alpha
%\bar \beta{}_{XY} {}^Z}$, defined for any combination of two complex
%structures.
We will use only the `diagonal' Nijenhuis tensor
(normalization for later convenience)
\begin{equation}
N_{XY}{}^Z\equiv \ft16 \vec J_{[X}{}^W\cdot \left( \partial _{|W|}\vec
J_{Y]}{}^Z-
 \partial _{Y]}\vec J_{W}{}^Z\right).
 \label{Nijenhuisdiag}
\end{equation}
A quaternionic manifold is defined by requiring that the Nijenhuis tensor
satisfies
\begin{equation}
  (1-2\,r)\,N_{XY}{}^Z=-\vec J_{[X}{}^ZN_{Y]V}{}^W\cdot \vec J_W{}^V,
 \label{conditQuat}
\end{equation}
which is equivalent to requiring that
\begin{equation}
  N_{XY}{}^Z=-\vec J_{[X}{}^Z\cdot  \OliveGreen{{\vec \omega}^{\rm Op} _{Y]}},
 \label{N=omega}
\end{equation}
for some ${ \vec \omega}^{\rm Op} _{X}$. Since the trace of the Nijenhuis
tensor vanishes, $\vec \omega^{\rm Op}_X$ obeys $\vec J_X{}^Y \cdot \vec
\omega_Y^{\rm Op}=0$. Hence, (\ref{conditQuat}) is automatically
satisfied. The condition (\ref{N=omega}) can be solved for ${ \vec
\omega}^{\rm Op} _{X}$, and used to define an $\SU(2)$ connection
\begin{equation}
  (1-2\,r)\,{\vec \omega}^{\rm Op} _X =N_{XY}{}^Z\vec J_Z{}^Y.
 \label{valueTilOmQ}
\end{equation}
The condition (\ref{conditQuat}) ensures that there exists an appropriate
affine connection such that (\ref{covconstJ}) is satisfied. This
connection is called the \emph{Oproiu connection} \cite{Oproiu1977},
\begin{eqnarray}
\Purple{\Gamma ^{\rm Op}{}_{XY}{}^Z}&\equiv&  \Purple{\Gamma^{\rm
Ob}{}_{XY}{}^Z}-\vec J_ {(X}{}^Z\cdot \Green{{\vec\omega}^{\rm Op} _{Y)}}, \label{Oproiu}\\
\Purple{\Gamma^{\rm Ob}{}_{XY}{}^Z}&\equiv &-\ft16 \left( 2
\partial_{(X}\vec J_ {Y)}{}^W+\vec J_{(X}{}^U\times
\partial_{|U|} \vec J_{Y)}{}^W\right) \cdot \vec J_W{}^Z. \label{Obata}
\end{eqnarray}
$\Purple{\Gamma^{\rm Ob}{}_{XY}{}^Z}$ is the \emph{Obata connection},
which is the solution if ${\vec \omega}_X=0$.

Conversely, any two connections $\Gamma_{XY}{}^Z$ and $\vec \omega _X$
that satisfy~(\ref{covconstJ}), necessarily imply the
condition~(\ref{conditQuat}) on the Nijenhuis tensor. Moreover, as shown
in~\cite{Bergshoeff:2002qk}, they must be related to the connections
defined by~(\ref{valueTilOmQ}) and~(\ref{Oproiu}) by means of the
$\xi$-transformations mentioned in (\ref{changeGammaomega}).

%In conclusion, \emph{a quaternionic manifold is a $4r$-dimensional
%manifold $\mathcal{M}$, with a hypercomplex structure with Nijenhuis
%tensor satisfying~(\ref{conditQuat})}.

We proceed by discussing the curvature tensor of quaternionic manifolds.
We first give our conventions:
\begin{eqnarray}
 R_{XYZ}{}^W &\equiv & 2
\partial_{[X}\Gamma_{Y]Z}{}^W + 2 \Gamma_{V[X}{}^W \Gamma_{Y]Z}{}^V
,\nonumber\\
{\cal R}_{XYB}{}^A &\equiv & 2\partial_{[X} \omega_{Y]B}{}^A
+ 2\omega_{[X|C|}{}^A \omega_{Y]B}{}^C, \nonumber\\
 \vec {\cal R}_{XY}  &\equiv & 2\partial_{[X}\vec \omega _{Y]}
+2\vec \omega _X\times \vec \omega _Y.
 \label{defCurv}
\end{eqnarray}
The integrability condition of~(\ref{covconstf}) implies that the total
curvature on the manifold is the sum of the $\SU(2)$ curvature and the
$\Gl (r,\mathbb{H})$ curvature. This shows that the (restricted) holonomy
splits in these two factors:\footnote{This follows also from the
Ambrose-Singer theorem~\cite{Ambrose:1953}, which says that the Lie
algebra of the restricted holonomy group of the frame bundle coincides
with the algebra generated by the curvature. The direct product structure
of the holonomy group is then reflected in these
relations.\label{fn:AmbrSinger}}
\begin{eqnarray}\label{curvdecomp}
  \Purple{R_{XYW}{}^Z}&=\OliveGreen{R^{\SU(2)}{}_{XYW}{}^Z}&+\,
  \Red{R^{\Gl(r,\mathbb{H})}{}_{XYW}{}^Z} \nonumber\\
&=-\vec J_W{}^Z\cdot\OliveGreen{\vec {\cal R} _{XY}}& +\,
L_W{}^Z{}_A{}^B\,\Red{{\cal R}_{XYB}{}^A },\label{RdecompJ}
\end{eqnarray}
where matrices $L_A{}^B$ appear. They are defined in Appendix
\ref{app:projmatrices}, where also various useful properties are
exhibited. This implies
\begin{equation}
4r\,\vec {\cal R}_{XY}= R_{XYZ}{}^W\vec J_W{}^Z,\qquad
 \mathcal{R}_{XYA}{}^B=\ft12 L_W{}^Z{}_A{}^BR_{XYZ}{}^W.
 \label{nearpropSU2J}
\end{equation}
Furthermore, we define the Ricci tensor as
\begin{equation}
R_{XY}=R_{ZXY}{}^Z.
\end{equation}
For an arbitrary affine connection, it has both a symmetric and
antisymmetric part. In general, the antisymmetric part can be traced back
to the curvature of the $\mathbb{R}$ part in
$\Gl(r,\mathbb{H})=\Sl(r,\mathbb{H})\times \mathbb{R}$. Indeed, using the
cyclicity condition, we find
\begin{equation}
  \Ric_{[XY]}=R_{Z[XY]}{}^Z=-\ft12R_{XYZ}{}^Z=-{\cal R}_{XY}^{\mathbb{R}},
  \qquad {\cal R}_{XY}^{\mathbb{R}}\equiv {\cal R}_{XYA}{}^A.
 \label{Ricas}
\end{equation}
Therefore, the antisymmetric part of the Ricci tensor follows completely
from the $\mathbb{R}$ part.

The separate curvature terms in (\ref{RdecompJ}) do not satisfy the
cyclicity condition (unless the $\SU(2)$ curvature vanishes), and thus
are not bona-fide curvatures. Another splitting of the full curvature can
be made where both terms separately satisfy the cyclicity condition,
\begin{equation}
  R_{XYW}{}^Z=R^{\rm Ric}{}_{XYW}{}^Z+R^{(\rm W)}{}_{XYW}{}^Z.
 \label{RRBRW}
\end{equation}
The first part only depends on the Ricci tensor of the full curvature,
and is called the `\emph{Ricci part}'. It is defined by
%using (\ref{defS}):
\begin{equation}
  R^{\rm Ric}{}_{XYZ}{}^W\equiv 2\delta _{[X}{}^WB_{Y]Z}
  -2\delta _Z{}^W B_{[XY]}
  -4\vec J_Z{}^{(W}\cdot \vec J_{[X}{}^{V)}B_{Y]V},
 \label{defRB}
\end{equation}
where~\cite{Musso:1992}
\begin{equation}
 B_{XY}\equiv  \frac{1}{4r}\left( \delta _X{}^Z\delta _Y{}^W-\Pi _{XY}{}^{ZW}\right) \Ric_{(ZW)}+\frac{1}{4(r+2)}\Pi
  _{XY}{}^{ZW}\Ric_{(ZW)}+\frac{1}{4(r+1)}\Ric_{[XY]}.
 \label{defB}
\end{equation}
Here, we have introduced a projection operator
\begin{equation}
 \Pi _{XY}{}^{ZW}\equiv \ft14\left( \delta _X{}^Z\delta _Y{}^W+\vec J_X{}^Z
  \cdot \vec J_Y{}^W   \right),
\label{defPi}
\end{equation}
whose properties are discussed in the Appendix. The symmetric part of
$B_{XY}$ can be considered as the candidate for a `good metric' for
quaternionic manifolds. Indeed, if there is a good metric, then it is
proportional to the symmetric part of this tensor as we will see below in
(\ref{BinqK}).

The Ricci part does satisfy the cyclicity property, and its Ricci tensor
is just $R_{XY}$. We can further split it as
\begin{equation}
  R^{\rm Ric}{}_{XYZ}{}^W=
  \left( R^{\rm Ric}_{\rm symm}+R^{\rm Ric}_{\rm antis}\right)
  _{XYZ}{}^W,
 \label{RRicsplit}
\end{equation}
where the first term is the construction~(\ref{defRB}) using only the
symmetric part of $B$ (the symmetric part of the Ricci tensor), and the
second term uses only the antisymmetric part of $B$ (i.e. of the Ricci
tensor).

The second term in (\ref{RRBRW}) is defined as the remainder, and its
Ricci tensor is zero. For this reason, it is called the `\emph{Weyl
part}'~\cite{AM1996}. Following again the discussion
in~\cite{Bergshoeff:2002qk}, one can rewrite the Weyl part in terms of a
symmetric and traceless tensor $\mathcal{W}_{ABC}{}^D$, such
that~\footnote{The case of four-dimensional quaternionic manifolds must
be treated separately, but they are defined such that (\ref{RdecompRBWA})
is still satisfied. See~\cite{Bergshoeff:2002qk} for more details.}
\begin{equation}
  R_{XYZ}{}^W=\Blue{R^{\rm Ric}{}_{XYZ}{}^W} - \ft 12 f^{iA}_X
\varepsilon_{ij}f^{jB}_Yf_Z^{kC}f^W_{kD}
  \Magenta{\mathcal{W}_{ABC}{}^D},
 \label{RdecompRBWA}
\end{equation}
with
\begin{equation}
\mathcal{W}_{CDB}{}^A \equiv \ft12 \varepsilon ^{ij} f^X_{jC} f^Y_{iD}
f_{kB}^Z f_W^{kA} {R^{(\rm W)}}_{XYZ}{}^W.\label{WfromRW}
\end{equation}

The $\Gl(r,\mathbb{H})$ curvature can also be decomposed in its Ricci and
Weyl part:
\begin{eqnarray}
 \mathcal{R}_{XYA}{}^B & = & \mathcal{R}^{\rm Ric}{}_{XYA}{}^B+ \mathcal{R}^{({\rm W})}{}_{XYA}{}^B,\nonumber\\
&&  \mathcal{R}^{\rm Ric}{}_{XYA}{}^B\equiv
 \ft12 L_W{}^Z{}_A{}^BR^{\rm Ric}{}_{XYZ}{}^W=
  2\delta_A{}^B B_{[YX]}+4L_{[X}{}^V{}_A{}^B B_{Y]V},
  \nonumber\\
% \label{RRicHPr}
& & \mathcal{R}^{({\rm W})}{}_{XYA}{}^B\equiv\ft12
  L_W{}^Z{}_A{}^BR^{({\rm W})}{}_{XYZ}{}^W=-f_X^{iC}\varepsilon
  _{ij}f_Y^{jD}\mathcal{W}_{CDA}{}^B.
% \label{RWHPr}
\end{eqnarray}
On the other hand, the $\SU(2)$ curvature is determined only by the Ricci
tensor, or, equivalently, by the tensor $B_{XY}$:
\begin{equation}
  \vec {\mathcal{R}}_{XY} = 2\vec J_{[X}{}^Z B_{Y]Z}.
 \label{RSU2B}
\end{equation}

We can summarize the different curvature decompositions in the following
scheme:
\begin{equation}
  \begin{array}{rccccc}
  R_{XYZ}{}^W= & \big( R^{\rm Ric}_{\rm symm}  & + & R^{\rm Ric}_{\rm antis}  & + &
  R^{({\rm W})}\big)_{XYZ}{}^W
  \\
  &\multicolumn{3}{c}{\phantom{.}\hspace{5mm}\leavevmode
\ifpdf
  \includegraphics{pijlen.pdf}
 \else
  \epsfxsize=85mm
 \epsfbox{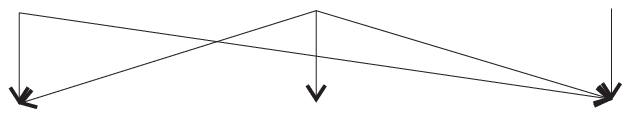}
 \fi
 \hspace{-18mm}} \\
  = & \big( R^{\SU(2)} & + & R^{\mathbb{R}} & + &
R^{\Sl(r,\mathbb{H})}\big)_{XYZ}{}^W.
\end{array}
 \label{RelatDecompR}
\end{equation}
The terms in the second line depend only on specific terms of the first
line as indicated by the arrows. This is the general scheme and thus
applicable for quaternionic manifolds, which is the general case, but for
specific other quaternionic-like manifolds some parts are absent as can
be seen in Table~\ref{tbl:CurvQuatlikeMan}.
\begin{table}[t]
  \caption{\it The curvatures in quaternionic-like manifolds. The first line gives
  the decomposition according to Ricci and Weyl curvatures, while the second line
  gives the decomposition in accordance with the holonomy groups.}
\label{tbl:CurvQuatlikeMan}
\begin{center}
  \begin{tabular}{||c|c||}\hline\hline
   \textit{\textbf{hypercomplex}} & \textit{\textbf{hyper-K{\"a}hler}}  \\
    $R^{\rm Ric}_{\rm antis}+ R^{({\rm W})}$
      & $R^{({\rm W})}$ \\
    $R^{\mathbb{R}} + R^{\Sl(r,\mathbb{H})}$
      & $ R^{\Sl(r,\mathbb{H})}$ \\
 \hline
    \textit{\textbf{quaternionic}} & \textit{\textbf{quaternionic-K{\"a}hler}} \\
    $R^{\rm Ric}_{\rm symm}+R^{\rm Ric}_{\rm antis}+ R^{({\rm W})}$
      & $R^{\rm Ric}_{\rm symm}+ R^{({\rm W})}$ \\
    $R^{\SU(2)} + R^{\mathbb{R}} + R^{\Sl(r,\mathbb{H})}$
      & $R^{\SU(2)} + R^{\Sl(r,\mathbb{H})}$ \\
\hline\hline
  \end{tabular}
\end{center}
\end{table}
%As explained further below, in both decompositions, the first terms are
%absent for hypercomplex and hyper-K{\"a}hler. The second terms are absent for
%hyper-K{\"a}hler and quaternionic-K{\"a}hler. The $\SU(2)$ curvature is determined
%by both parts of the Ricci curvature, but gets no contribution from the
%Weyl part (for $r=1$ this is part of the definition of quaternionic
%manifolds). The $\mathbb{R}$ curvature originates only from the antisymmetric
%part of the Ricci tensor, see~(\ref{Ricas}). The $\Sl(r,\mathbb{H})$
%curvature gets contributions from all 3 terms, see~(\ref{RRicHPr})
%and~(\ref{RWHPr}).

%%%%%%%%%%%%%%%%%%%%%%%%%%%%%%%%%%%%%%%%%%%%%%%%%%%%
\subsection{The $\xi $-transformations}
 \label{ss:xi}
%%%%%%%%%%%%%%%%%%%%%%%%%%%%%%%%%%%%%%%%%%%%%%%%%%%%
The requirement (\ref{covconstJq}) for a fixed complex structure does not
determine the connections uniquely. Indeed, the affine and $\SU(2)$
connections can be changed simultaneously depending on an arbitrary
one-form $\xi=\xi _X \rmd q^X$ as
\begin{equation}
  \tilde \Gamma_{XY}{}^Z= \Gamma _{XY}{}^Z+S_{XY}{}^{WZ}\xi _W,\qquad
\vec {\tilde \omega}_X =\vec \omega_X +\vec J_X{}^W\xi _W.
 \label{tildeGammaomega}
\end{equation}
Here we have introduced the $S$-symbols, which can be read of from
(\ref{changeGammaomega}). In terms of the projection operator in
(\ref{defPi}), they are
\begin{equation}
   S_{XY}{}^{ZW}\equiv 2\delta ^Z_{(X}\delta ^W_{Y)}-2 \vec J_X{}^{(Z}\cdot
   \vec J_Y{}^{W)}= 4\delta ^Z_{(X}\delta ^W_{Y)}-8 \Pi _{(XY)}{}^{ZW}.
\label{propS}
\end{equation}
Further properties are given in (\ref{Scontr})--(\ref{SS}).

Obviously, since the complex structures do not transform, the Nijenhuis
tensor is invariant, and so the geometry remains quaternionic. When
transforming the affine and the $\SU(2)$ connection as
in~(\ref{tildeGammaomega}), also the connection $\omega_{XA}{}^B$
transforms, according to its value in (\ref{determineOm}):
\begin{equation}
\tilde \omega_{XA}{}^B=\omega_{XA}{}^B+\ft12 L_Y{}^Z{}_A{}^B
S_{XZ}{}^{YW}\xi_W. \label{tildeomegaAB}
\end{equation}
Note in particular that the $\mathbb{R}$ connection transforms as
\begin{equation}
\tilde  \omega_{XA}{}^A= \omega_{XA}{}^A+2(r+1) \xi_X.
 \label{xitransformomU1}
\end{equation}

Clearly, all curvatures will transform under these deformations of the
connections, with terms at most quadratic in $\xi_X$. A direct
computation shows that the full Riemann curvature transforms as
\begin{equation}
 R_{XYZ}{}^W(\tilde \Gamma ) =  R_{XYZ}{}^W(\Gamma )+2S_{Z[Y}{}^{WU}
  D_{X]}\xi _U+2S_{T[X}{}^{WU}S_{Y]Z}{}^{TV} \xi _U\xi _V.
  \label{xicurvature}
\end{equation}
Defining furthermore\footnote{The quadratic terms in these equations can
be understood from the consistency of applying these formulae with
$\Gamma $ and $\tilde \Gamma $ interchanged.}
\begin{equation}
  \eta _{XY}  \equiv   -D_X\xi _Y
 +\ft12S _{(XY)}{}^{UV}\xi _U\xi _V= -\tilde D_X\xi _Y
-\ft12S _{XY}{}^{UV}\xi _U\xi _V,
 \label{defetaxi}
\end{equation}
we find
\begin{eqnarray}
\Ric_{XY}(\tilde \Gamma ) & = &  \Ric_{XY}(\Gamma )+4r\eta _{(XY)}
 +8\Pi _{(XY)}{}^{UV}\eta _{UV} - 4(r+1)\partial _{[X}\xi _{Y]},\nonumber\\
\vec{\cal R}_{XY}(\tilde{\vec \omega} )
& = & \vec{\cal R}_{XY}( {\vec\omega}) -2\vec J_{[Y}{}^Z \eta _{X]Z}, \nonumber\\
 B_{XY}(\tilde\Gamma )&=&B_{XY}(\Gamma )+\eta _{XY}.
 \label{transfoxieta}
\end{eqnarray}
Using the last expression, one finds that the Ricci part, (\ref{defRB}),
transforms as the full curvature (\ref{xicurvature}). Therefore, the Weyl
part, $R^{(\rm W)}$, does not transform, and the $\mathcal{W}$-tensor is
invariant.

The $\xi$-transformations can be used to fix the form of the $\mathbb{R}$
connection. Indeed, looking at (\ref{xitransformomU1}), we see we can
transform away the $\mathbb{R}$ connection completely. The $\mathbb{R}$
curvature then vanishes. Furthermore, there are residual
$\xi$-transformations depending on a scalar function $f(q)$, i.e. $\xi
_X=\partial _Xf $ that leave the $\mathbb{R}$ curvature invariant.

An alternative $\xi$-choice yields the Oproiu connection that satisfies
\begin{equation}
  \vec J_X{}^Y\cdot \vec \omega^{\rm Op} _Y=0,
 \label{gaugechoiceOproiu}
\end{equation}
which leads to the connection
(\ref{Oproiu})~\cite{AM1996,Bergshoeff:2002qk}.

%%%%%%%%%%%%%%%%%%%%%%%%%%%%%%%%%%%%%%%%%%%%%%%%%%%%
\subsection{Hypercomplex manifolds}
 \label{ss:hc}
%%%%%%%%%%%%%%%%%%%%%%%%%%%%%%%%%%%%%%%%%%%%%%%%%%%%

Hypercomplex manifolds were introduced in~\cite{Salamon:1986}. A very
thorough paper on the subject is~\cite{AM1996}. Examples of homogeneous
hypercomplex manifolds that are not hyper-K{\"a}hler, can be found
in~\cite{Spindel:1988sr,Joyce:1992}, and are further discussed
in~\cite{Bergshoeff:2002qk}. Non-compact homogeneous manifolds are dealt
with in~\cite{Barberis:1996}. Various aspects have been treated in two
workshops with mathematicians and
physicists~\cite{QuatWorksh1,QuatWorksh2}.

A hypercomplex manifold has no $\SU(2)$ connection:
\begin{equation}
{\vec \omega}_X=0. \label{vecomega0}
\end{equation}
This implies that we do not allow for local $\SU(2)$ redefinitions of the
hypercomplex structure.

The quaternionic structure should thus be covariantly constant with
respect to the affine connection only. The unique solution
of~(\ref{covconstJ}) with~(\ref{vecomega0}) is
\begin{equation}
  \Purple{\Gamma _{XY}{}^Z}=\Purple{\Gamma ^{\rm
  Ob}{}_{XY}{}^Z}+N_{XY}{}^Z,
 \label{solGammaDJ0}
\end{equation}
where the first term is symmetric and the second one (the Nijenhuis
tensor) is antisymmetric in $XY$. The first term is called the
\emph{Obata connection}~\cite{Obata}, and is given in terms of the
complex structures and their derivatives in (\ref{Obata}). As mentioned
before, and motivated by the supersymmetry algebra, we consider
torsionless (symmetric) connections, which requires
\begin{equation}
N_{XY}{}^Z=0. \label{N=0}
\end{equation}

By definition, \emph{a hypercomplex manifold is a $4r$-dimensional
manifold $\mathcal{M}$, equipped with a hypercomplex structure with
vanishing Nijenhuis tensor}.

Clearly, hypercomplex manifolds are those quaternionic manifolds with
vanishing $\SU(2)$ connection. If the quaternionic manifold has a
non-vanishing $\SU(2)$ connection, it might still be possible to define
from it a hypercomplex manifold. For instance, consider the class of
quaternionic manifolds with vanishing $\SU(2)$ curvature, i.e.,
\begin{equation}
\vec{\cal R}_{XY}=0.
\end{equation}
This only requires the $\SU(2)$ connection to be pure gauge. The complex
structures are then covariantly constant with respect to this $\SU(2)$
connection, and one can still act with $\xi$-transformations. However, a
small calculation shows that one can redefine the complex structures with
a local SO(3) matrix $R$ according to ${\vec J}'=R J$, such that no
$\SU(2)$ connection is needed anymore. In such a basis, also the freedom
of doing $\xi$-transformations is fixed. The resulting manifold is then
hypercomplex.

One can in fact further relax the vanishing condition on the $\SU(2)$
curvature by requiring only
\begin{equation}
  \vec{{\cal R}}_{XY}=2\vec{J}_{[X}{}^Z\eta _{Y]Z},
 \label{genhypercomplex}
\end{equation}
for some $\eta_{XY}$ that can be written as (\ref{defetaxi}). In
particular, the $\SU(2)$ connection is non-vanishing. But now, looking at
(\ref{transfoxieta}), this implies that we can do a $\xi$-transformation
such that the $\SU(2)$ curvature vanishes, and so we are back in the
situation discussed above.

In summary, for those quaternionic manifolds that satisfy
(\ref{genhypercomplex}) we can associate and define a hypercomplex
manifold by making use of the quaternionic, local span of the complex
structures, together with the $\xi$-transformations.

Further properties of hypercomplex manifolds can be derived. Since they
have vanishing $\SU(2)$ curvature, the Riemann tensor can be decomposed
as
\begin{equation}
  R_{XYW}{}^Z = - \ft 12 f^{iA}_X \varepsilon_{ij}f^{jB}_Yf_W^{kC}f^Z_{kD}
  W_{ABC}{}^D.
 \label{RinWHC}
\end{equation}
The tensor $W$ is defined as
\begin{equation}
W_{CDB}{}^A \equiv \varepsilon ^{ij} f^X_{jC} f^Y_{iD} {\cal R}_{XYB}
{}^A = \ft12 \varepsilon ^{ij} f^X_{jC} f^Y_{iD} f_{kB}^Z f_W^{kA}
R_{XYZ}{}^W,\label{WfromR}
\end{equation}
and is symmetric in its lower indices. It is however not traceless, and
its trace determines the Ricci tensor,
\begin{equation}
  \Ric_{XY}=\Ric_{[XY]}= \ft 12  \varepsilon_{ij}f^{iB}_Xf_Y^{jC}
  W_{ABC}{}^A=- {\cal R}_{XYA}{}^A.
 \label{RicasHC}
\end{equation}
The tensor $B$ is then antisymmetric, and is just the last term of
(\ref{defB}). This form of the Ricci tensor implies that it is Hermitian
\begin{equation}
  \Pi _{XY}{}^{UV}\Ric_{UV}= \Ric_{XY}.
 \label{PiRU1hc}
\end{equation}
This follows from
\begin{equation}
  \vec J_X{}^Z f_Z{}^{iA}= -\rmi \vec\sigma _j{}^if^{jA}_X,\qquad
\vec\sigma _k{}^i\cdot \vec\sigma _\ell
{}^j\varepsilon_{ij}=-3\varepsilon _{k\ell }.
 \label{auxeqnsPiRU1hc}
\end{equation}

The tensor $W$ in (\ref{WfromR}) should not be confused with the
traceless tensor ${\cal W}$ defined in (\ref{WfromRW}). The precise
relation is given by
\begin{equation}
  \mathcal{W}_{ABC}{}^D=W_{ABC}{}^D-\frac{3}{2(r+1)}\delta
  ^D_{(A}W_{BC)E}{}^E.
 \label{calWW}
\end{equation}

Thus, for hypercomplex manifolds, the Ricci tensor is antisymmetric and
Hermitian. Conversely, a quaternionic manifold with antisymmetric and
Hermitian Ricci tensor is necessarily hypercomplex. Indeed, using the
general result (\ref{FJTis-JF}) for any Hermitian bilinear form,
(\ref{RSU2B}) then implies that the $\SU(2)$ curvature vanishes, and so a
basis can be chosen such that it is hypercomplex. We come back to the
hermiticity properties of quaternionic-like manifolds at the end of this
section.

Nowhere in this section have we assumed the existence of a (covariantly
constant) metric. When such a tensor exists, hypercomplex manifolds are
promoted to hyper-K{\"a}hler manifolds.

%%%%%%%%%%%%%%%%%%%%%%%%%%%%%%%%%%%%%%%%%%%%%%%%%%%%%
\subsection{Hyper-K{\"a}hler manifolds}
 \label{ss:hk}
%%%%%%%%%%%%%%%%%%%%%%%%%%%%%%%%%%%%%%%%%%%%%%%%%%%%%

The crucial difference between hyper-K{\"a}hler and hypercomplex geometries
is that hyper-K{\"a}hler manifolds admit a Hermitian metric $g$. This
involves 3 conditions:
\begin{enumerate}
  \item $g$ should be Hermitian. This can be expressed as
 \begin{equation}
  \vec J_X{}^Zg_{ZY}= - \vec J _Y{}^Zg_{ZX}.
 \label{hermMetric}
\end{equation}
 \item $g$ should be invertible.
 \item $g$ should be covariantly constant using the Obata connection.
\end{enumerate}

If this metric is preserved using the Obata connection, then the
hypercomplex manifold is promoted to a hyper-K{\"a}hler manifold.
Equivalently, when the Levi-Civita connection preserves the quaternionic
structure, then the manifold is hyper-K{\"a}hler. It is clear from the
discussion of the previous section, that the Levi-Civita and Obata
connection on a hyper-K{\"a}hler manifold must coincide.

The Ricci tensor of the Levi-Civita connection is always symmetric.
Combined with the fact that the Obata connection has an antisymmetric
Ricci tensor, it follows that hyper-K{\"a}hler manifolds are Ricci flat,
\begin{equation}
R_{XY}=0.
\end{equation}
Using (\ref{Ricas}), this is equivalent to saying that the trace of
$W_{ABC}{}^D$ vanishes, and so,
\begin{equation}
{\mathcal W}_{ABC}{}^D=W_{ABC}{}^D.
\end{equation}
As a consequence, the curvature takes values in $\USp(2r)$.

The existence of a metric allows us to define the covariantly constant
antisymmetric tensors
\begin{equation}
  C_{AB}=\ft12 f_{iA}^X g_{XY} \varepsilon ^{ij} f_{jB}^Y
  , \qquad C^{AB}=\ft12 f^{iA}_X g^{XY} \varepsilon _{ij} f^{jB}_Y,
 \label{defC}
\end{equation}
which satisfy
\begin{equation}
 C_{AC} C^{BC} =   \delta_A{}^B ,
 \label{epsCup}
\end{equation}
and which can be used to raise and lower indices according to
the NW--SE convention similar to $\varepsilon_{ij}$:
\begin{equation}
A_A = A^BC_{BA} ,\qquad A^A = C^{AB} A_B.
 \label{NWSE}
\end{equation}

The integrability condition on $C_{AB}$ implies
\begin{equation}
{\cal R}_{XY[A}{}^CC_{B]C}=0,
\end{equation}
and it follows that
\begin{equation}
W_{ABCD}\equiv W_{ABC}{}^EC_{ED},
\end{equation}
is fully symmetric in its four lower indices.

%%%%%%%%%%%%%%%%%%%%%%%%%%%%%%%%%%%%%%%%%%%%%%%%%%%%%%%%
\subsection{Quaternionic-K{\"a}hler manifolds}
 \label{ss:qk}
%%%%%%%%%%%%%%%%%%%%%%%%%%%%%%%%%%%%%%%%%%%%%%%%%%%%%%%%
For some basic references on quaternionic-K{\"a}hler manifolds, we refer to
\cite{Ishihara:1974,Salamon:1982}, or the earlier references
\cite{Obata,Wolf:1963,Bonan:1964a,Bonan:1964b,Alekseevsky:1968}.

Similar to hyper-K{\"a}hler spaces, quaternionic-K{\"a}hler manifolds admit a
Hermitian and invertible metric satisfying (\ref{hermMetric}). The
connection that preserves this metric, i.e. the Levi-Civita connection,
must be related to the Oproiu connection (\ref{Oproiu}) by a
$\xi$-transformation. It is a well known fact that
\emph{quaternionic-K{\"a}hler spaces are Einstein}:
\begin{equation}
  R_{XY}=\frac{1}{4r} g_{XY} R .
 \label{Einstein}
\end{equation}

{}From the Bianchi identity, one easily shows that the Ricci scalar is
constant. The Ricci tensor is obviously symmetric, such that the
$\mathbb{R}$ part of the $\Gl(r,\mathbb{H})$ curvature is zero. The
$B$-tensor in (\ref{defB}) can easily be computed to be
\begin{equation}
  B_{XY}=\ft14\nu g_{XY},\qquad \nu
  \equiv \frac{1}{4r(r+2)}R.
   \label{BinqK}
\end{equation}
Using (\ref{RSU2B}), one then finds that the $\SU(2)$ curvature is
proportional
to the quaternionic 2-form: %[B.70]
\begin{equation}
\OliveGreen{\vec{\cal R}_{XY}{}} =\ft12\nu \vec J_{XY}.
 \label{RlambdaJ}
\end{equation}
The Ricci part of the curvature is determined by the curvature of a
quaternionic projective space of the same dimension:
\begin{eqnarray}
\left(R^{{\mathbb H}P^n}\right)_{XYWZ} &\equiv &\ft 12 g_{Z[X}g_{Y]W}+\ft
12 \vec J_{XY}\cdot \vec J_{ZW}-\ft 12 \vec J_{Z[X}\cdot \vec J_{Y]W} \nonumber\\
&=&\ft 12 \vec J_{XY}\cdot \vec J_{ZW}+L_{[ZW]}{}^{AB}L_{[XY]AB}.
 \label{RHPn}
\end{eqnarray}
The full curvature decomposition is then
\begin{equation}
 R_{XYWZ}  = \nu (R^{{\mathbb H}P^n})_{XYWZ}+ \ft 12 L_{ZW}{}^{AB}\mathcal{W}
 _{ABCD}L_{XY}{}^{CD},
 \label{RdecompHPnW}
\end{equation}
with $\mathcal{W}_{ABCD}$ completely symmetric. In supergravity, the
supersymmetry connects the value of $\nu $ to the normalization of the
Einstein term in the action. This fixes the value of $\nu $ to $-\kappa
^2$, where $\kappa$ is the gravitational coupling constant. The
quaternionic-K{\"a}hler manifolds appearing in supergravity thus have
negative scalar curvature, and this implies that all such manifolds that
have at least one isometry are non-compact.

Properties of the connections of all the quaternionic-like manifolds are
summarized in Table~\ref{tbl:ConnQuatlikeMan}.
\begin{table}[t]
  \caption{\it The affine connections in quaternionic-like manifolds}
\label{tbl:ConnQuatlikeMan}
\begin{center}
  \begin{tabular}{||c|c||}\hline\hline
   \textit{\textbf{hypercomplex}} & \textit{\textbf{hyper-K{\"a}hler}}  \\
    Obata connection & Obata connection \\
                     & = Levi-Civita connection  \\ \hline
    \textit{\textbf{quaternionic}} & \textit{\textbf{quaternionic-K{\"a}hler}} \\
    Oproiu connection or  & Levi-Civita connection = \\
    other related by $\xi _X$ transformation & connection related to
    Oproiu \\
    & by a particular choice of $\xi _X$ \\
\hline\hline
  \end{tabular}
\end{center}
\end{table}

%%%%%%%%%%%%%%%%%%%%%%%%%%%%%%%%%%%%%%%%%%%%%%%%%%%%%
\subsection{Hermitian Ricci tensor}
 \label{ss:hermRicci}
%%%%%%%%%%%%%%%%%%%%%%%%%%%%%%%%%%%%%%%%%%%%%%%%%%%%%

In this subsection, we study the properties of quaternionic-like
manifolds with a Hermitian Ricci tensor. First of all, it is immediate
from the relation between $B$ and the Ricci tensor, see~(\ref{defB}),
that $B$ is Hermitian if and only if $\Ric$ is Hermitian.

The important relation we now look at is (\ref{RSU2B}). Using
(\ref{PiJ}), one can see that, for Hermitian $B$, the antisymmetric part
of $B$ does not contribute. We can therefore conclude that an
antisymmetric and Hermitian Ricci tensor is equivalent to the requirement
of hypercomplex. Indeed, the $\SU(2)$ curvature is then zero, and we use
the argument in Sect.~\ref{ss:hc}. The other direction was also shown in
that section.

Furthermore, for the symmetric part of $B$, the antisymmetrization
in~(\ref{RSU2B}) is automatic in the right-hand side due
to~(\ref{FJTis-JF}). For a Hermitian Ricci tensor, we thus have,
\begin{equation}
  \vec{\cal R}_{XY}= 2\vec J_{X}{}^ZB_{(YZ)}, \qquad
  \Ric_{(XY)}=4(r+2)B_{(XY)}.
 \label{HermitRB}
\end{equation}
It is appropriate to define a `candidate metric',
\begin{equation}
  g_{XY}= \frac{4}{\nu }B_{(XY)}=-\frac{1}{\nu }h_{XY},
 \label{candidateMetric}
\end{equation}
where $\nu $ is an undetermined number. We define $h_{XY}$ such that this
number can be avoided in most of our formulae. With the usual
normalization in supergravity where $\kappa =1$, this is the metric
anyway. We now have
\begin{equation}
 \vec{\cal R}_{XY}= -\ft12\vec J_{X}{}^Zh_{YZ}.
 \label{SU2isJ}
\end{equation}

The identification of the symmetric part of the Ricci tensor as a metric
becomes even more appropriate due to the property that it is covariantly
constant. To prove this, we start with acting with a $D_U$ derivative
on~(\ref{SU2isJ}) and antisymmetrizing in $[XYU]$ using the Bianchi
identity for the $\SU(2)$ curvature. This leads to
\begin{equation}
  3\vec J_{[X}{}^ZD_Uh_{Y]Z}=\vec J_{X}{}^ZD_Uh_{YZ}+\vec J_{U}{}^ZD_Yh_{XZ}+\vec J_{Y}{}^ZD_Xh_{UZ}=0.
 \label{JDhcyclic}
\end{equation}
Multiplying this with $(\vec J_V{}^U\times \vec J_W{}^Y)$ and taking the
symmetric part in $(XW)$, after using several times the antisymmetry of
$\vec J_X{}^Zh_{YZ}$, it leads to
\begin{equation}
  D_V h_{XW}=0.
 \label{hHermcovconst}
\end{equation}

We can summarize this section as follows:
\begin{enumerate}
  \item If the Ricci tensor is Hermitian and the symmetric part is invertible,
then it defines a good metric. Therefore the antisymmetric part of the
Ricci tensor is zero in this case (with respect to the Levi-Civita
connection of $h_{XY}$). On the other hand, if the symmetric part is
zero, then a Hermitian Ricci tensor implies zero $\SU(2)$ curvature.
  \item A quaternionic manifold is quaternionic-K{\"a}hler if and only if the
  Ricci tensor is Hermitian and its symmetric part is non-degenerate.
\end{enumerate}
Thus, there are 3 cases of Hermitian Ricci tensors on quaternionic-like
manifolds:
\begin{enumerate}
  \item symmetric part is invertible (quaternionic-K{\"a}hler manifold):
  there is no antisymmetric part,
  \item symmetric part is zero, i.e. antisymmetric Ricci (hypercomplex
  manifold),
  \item symmetric part is non-zero but non-invertible. Then an
  antisymmetric part is still possible.
\end{enumerate}

%%%%%%%%%%%%%%%%%%%%%%%%%%%%%%%%%%%%%%%%%%%%%%%%%%%%%
\section{The Map}
\label{ss:map}
%%%%%%%%%%%%%%%%%%%%%%%%%%%%%%%%%%%%%%%%%%%%%%%%%%%%%
We now start to discuss the map between the hypercomplex/hyper-K{\"a}hler and
qua\-ter\-ni\-o\-nic(-K{\"a}hler) manifolds. The first will be called the
\emph{large space} and will be taken to be $4(n_H+1)$ real dimensional,
while the latter will be of real dimension $4n_H$ and be called the
\emph{small space}. Objects on the large space will be denoted by hats,
either on their indices or on the objects themselves or on both.

The content of this section is as follows. In Subsect.~\ref{CHC} we
discuss some special properties of conformal hypercomplex manifolds that
are important for our discussion. We show that the holonomy of such
manifolds is further restricted and we discuss a continuous deformation
of the hypercomplex structure. After choosing coordinates that are
adapted to our setting and rewriting the hypercomplex structure, we prove
in Subsects.~\ref{ss:maphcquat} and \ref{propquat} that the large space
contains a $4n_H$ dimensional quaternionic subspace. This is the central
part of our discussion. Moreover, in Subsect.~\ref{ss:SU2} we clarify the
origin of the quaternionic local $\SU(2)$ symmetry. In the following
Subsect.~\ref{ss:mapqh} we show how a quaternionic manifold can be used
to construct a hypercomplex one. Subsect.~\ref{ss:maphkqk} considers the
map for hypercomplex manifolds that possess a `good metric', and are
therefore hyper-K{\"a}hler manifolds. The conditions for this metric to be a
`good metric' are equivalent to the condition that the quaternionic
manifold has a `good metric' and is thus promoted to a
quaternionic-K{\"a}hler manifold. Therefore, we prove that the image of the
map is a quaternionic-K{\"a}hler manifold if and only if the original
manifold is a conformal hyper-K{\"a}hler manifold. We show that in order for
the affine connection to agree with the Levi-Civita connection, as it
should be for quaternionic-K{\"a}hler manifolds, one has to choose a
particular $\xi $-transformation between the allowed connections for
quaternionic manifolds. This choice is different from the one that leads
to the Oproiu connection. We also prove the inverse map: for all
quaternionic manifolds we will define a candidate metric, and if this is
a good metric, which is the condition that the Ricci tensor is Hermitian
and invertible, then we can construct a good metric for the conformal
hypercomplex manifold. After having completed our discussion of the map,
we conclude in Subsect.~\ref{mapviel} by listing the vielbeins and all
connection coefficients on the large and small space in the adapted
coordinates.

%%%%%%%%%%%%%%%%%%%%%%%%%%%%%%%%%%%%%%%%%%%%%%%%%%%%%
\subsection{Hypercomplex manifolds with conformal symmetry}
\label{CHC}
%%%%%%%%%%%%%%%%%%%%%%%%%%%%%%%%%%%%%%%%%%%%%%%%%%%%%
The starting point of our map is given by hypercomplex manifolds that
admit a conformal symmetry. By definition, this means there exists a
closed homothetic Killing\footnote{Although there is not necessarily a
metric defined, we use the same terminology.} vector $k^{\widehat X}$,
satisfying (\ref{homothetic}).

Given this homothetic Killing vector $k^{\widehat X}$, three more vectors
can be constructed naturally:
\begin{equation}
  \vec k^{\widehat X} \equiv \ft13 {\widehat {\vec J}}_{\widehat Y}{}^{\widehat X}
k^{\widehat Y},
 \label{defveck}
\end{equation}
which generate an $\SU(2)$ algebra and satisfy %[2.87]:
\begin{equation}
 \widehat \covder_{ \widehat Y} \vec k^{ \widehat X}=\ft12{\widehat {\vec J}}_{ \widehat Y}
{}^{ \widehat X} .
 \label{DkSU2}
\end{equation}
It then follows that, under dilatations and $\SU(2)$ transformations, the
hypercomplex
structure is scale invariant and rotated into itself, %[2.90]
\begin{eqnarray}\label{transf-J}
\left(\mathcal L_{\Lambda_D k} \widehat{\vec J}\right){}_{\widehat
X}{}^{\widehat Y}\equiv\Lambda_D k^{\widehat Z} \partial_{ \widehat Z}\,
{\widehat {\vec J}}_{\widehat X} {}^{\widehat Y} -\Lambda_D
\partial_{\widehat Z} k^{\widehat Y}\, {\widehat {\vec J}}_{\widehat
X}{}^ {\widehat Z}+\Lambda_D \partial_{\widehat X} k^{\widehat Z}\,
{\widehat {\vec J}}_{\widehat Z}{}^{\widehat
Y} &=&0, \\
\left(\mathcal L_{\vec \Lambda\cdot \vec k}\widehat{ \vec
J}\right){}_{\widehat X}{}^{\widehat Y}\equiv ({\vec \Lambda}\cdot {\vec
k}^{\widehat Z})  \partial_{\widehat Z}\, {\widehat {\vec J}}_{\widehat
X}{}^{\widehat Y}- ({\vec \Lambda}\cdot
\partial_{\widehat Z} {\vec k}^{\widehat Y})\,{\widehat {\vec J}}_{\widehat X}{}^{\widehat Z}+
({\vec \Lambda}\cdot \partial_{\widehat X}{\vec k}^{\widehat Z})\,
{\widehat {\vec J}}_{ \widehat Z}{}^{\widehat Y} &=&-{\vec \Lambda}\times
{\widehat {\vec J}}_{\widehat X}{}^{\widehat Y}.\nonumber
\end{eqnarray}
Here, we introduced parameters $\Lambda_D$ and ${\vec \Lambda}$ (local in
spacetime but not dependent on the coordinates of the quaternionic space)
to generate the infinitesimal dilatations and $\SU(2)$ transformations on
the coordinates,
\begin{equation}
\delta _Dq^{ \widehat X}= \Lambda _D\Red{k^{\widehat X}},\qquad \delta
_{\SU(2)}q^{\widehat X}= \ft 13 \vec \Lambda\cdot(\Red{k^{\widehat
Y}}\widehat{\vec J}_{\widehat Y}{}^{\widehat X})= \vec
\Lambda\cdot\Red{\vec k^{\widehat X}}.
\end{equation}

Notice that all this follows from (\ref{homothetic}) and the covariant
constancy of the quaternionic structure.

%The integrability conditions for (\ref{homothetic}) and (\ref{DkSU2})
%then read
We demand (here and everywhere below) that the vectors $k$ and $\vec{k}$
are `symmetries' (see (\ref{RkI}) below, which is mathematically the
statement that they define affine transformations). This leads to
\begin{equation}
k^{\widehat X}\widehat R_{\widehat X\widehat Y\widehat Z}{}^{\widehat
W}=0,\qquad  \vec k^{\widehat X}\,\widehat R_{\widehat X\widehat
Y\widehat Z}{}^{\widehat W}=0. \label{konRis0}
\end{equation}
When the connection is metric, then these equations are integrability
conditions for (\ref{homothetic}) and (\ref{DkSU2}) using the symmetries
of the Riemann tensor.

Hence, the four vector fields now introduced are zero eigenvectors of the
curvature. This implies that the holonomy of a $4(n_H+1)$ dimensional
conformal hypercomplex manifold is contained in $\SU(2)\cdot
\Gl(n_H,\mathbb H)$, which can easily be understood as follows. On a
hypercomplex manifold, we can group all vectors of a given fibre into
quaternions. Let us call the fibre $F$ at a certain point $p$. Given then
a vector $X\in F$, we may construct a quaternion as $\{X,\widehat
J^1(p)X,\widehat J^2(p)X,\widehat J^3(p)X\}\equiv \{X,\widehat{\vec
J}(p)X\}$. In general, the holonomy group yields a $\Gl(n_H+1,\mathbb H)$
action on $F$ that is generated by the curvature. Since $\{k(p),3\vec
k(p)\}$ form such a quaternion, the relation (\ref{konRis0}) implies that
the holonomy group should leave that quaternion fixed and thus should be
included in $\SU(2)\cdot \Gl(n_H,\mathbb H)$.

Moreover, if one looks at the components of the hypercomplex structures
that lie along the other quaternions, it is easy to see that the holonomy
group induces an $\SU(2)\cdot \Gl(n_H,\mathbb H)$ action on these
components. This observation is the heart of our construction, since it
strongly suggests that the submanifold along the other $4n_H$ directions
is quaternionic.\footnote{More exactly, the four vector fields $k$ and
$\vec k$ generate a four dimensional foliation of the hypercomplex space,
and we will show that the space of leaves carries a quaternionic
structure.}

As will be motivated in the following sections, the existence of the
vector fields $k$ and $\vec k$ implies that we can define a continuous
family of hypercomplex structures. Suppose that we start with a conformal
hypercomplex manifold with closed homothetic Killing vector $k$ and
complex structure $\widehat{\vec J}$. We can define
\begin{equation}
  (\widehat{\vec J}_{\widehat \xi})_{ \widehat X}{}^{\widehat Y}=\widehat{\vec{J}}_{\widehat X}{}^{\widehat Y}
   +\ft23\left[ \widehat {\vec{J}}_{\widehat X}{}^{\widehat Z}(\widehat \xi
  _{\widehat Z}k^{\widehat Y})-
  (\widehat \xi _{\widehat X}k^{\widehat Z})\widehat {\vec{J}}_{\widehat Z}{}^{\widehat
  Y}\right],
 \label{transfoJxi}
\end{equation}
for a one-form with components $\widehat \xi _{\widehat X}$ that
satisfies $k^{\widehat X}\widehat \xi _{\widehat X}=\vec{k}^{\widehat
X}\widehat \xi _{\widehat X}=0$.

The new complex structures still satisfy the quaternionic algebra and
have vanishing Nijenhuis tensor if
\begin{equation}
 \partial _{[\widehat{X}}\widehat{\xi }_{\widehat{Y}]} \mbox{ is Hermitian, and } \qquad
{\cal L}_k\widehat{\xi }_{\widehat{X}}={\cal L}_{\vec k}\widehat{\xi
}_{\widehat{X}}=0.
%  k^{\widehat{X}}\partial _{[\widehat{X}}\widehat{\xi }_{\widehat{Y}]}=0,\qquad
%\vec{k}^{[\widehat{X}}\partial _{\widehat{X}}\widehat{\xi
%}_{\widehat{Y}]}=0.
 \label{hatxiconditions}
\end{equation}
The last requirement is automatic if the first two are satisfied.
%
%
%where $\widehat u$ is (one half of) the external derivative of
%$\widehat\xi$, i.e. $\widehat u_{\widehat X\widehat
%Y}=\partial_{[\widehat X}\widehat \xi_{\widehat Y]}$. The new almost
%hypercomplex structure $\widehat{\vec J}_{\xi}$ has vanishing Nijenhuis
%tensor if
%\begin{eqnarray}
% \mbox{Linear order in }\widehat \xi  & : & \widehat u _{[\widehat X\widehat Y]}\mbox{ is Hermitian}, \nonumber\\
% \mbox{Quadratic order in }\widehat \xi  & : & k^{\widehat X}\widehat u_{\widehat X\widehat Y}=0
% \qquad \mbox{i.e.}\qquad  \mathcal L_k \widehat \xi=0.
% \label{tildeJN}
%\end{eqnarray}
%The combination of the two also implies $\vec{k}^{\widehat X}\widehat
%u_{\widehat X\widehat Y}=0$ or $\mathcal L_{\vec k}\widehat  \xi=0$.
In conclusion, we can construct a new hypercomplex structure using
(\ref{transfoJxi}) with a one-form that is constant along the flows of
$k$ and $\vec k$ and whose external derivative is Hermitian. One can
moreover show that the new hypercomplex manifold is again conformal with
the same vector field $k$. As far as we know this
$\widehat{\xi}$-transformation has not been given before in the
mathematical literature.

%Same way: coordinate-independent way recognition of $\SU(2)$
%transformations:
%\begin{equation}
%  \delta \widehat{\vec{J}}_{\widehat X}{}^{\widehat Y}= -\ft13\vec{k}^{\widehat Y}\partial _{\widehat X}\vec{\ell} +
%  \widehat{\vec{J}}_{\widehat X}{}^{\widehat Z}\left( \vec{k}^{\widehat Y}\cdot \partial _{\widehat Z}\vec{\ell }\right)
%  -\vec{k}^{\widehat Y}\times \partial _{\widehat X}\vec{\ell }+\vec{\ell
%  }\times\widehat{\vec{J}}_{\widehat X}{}^{\widehat Y},
% \label{SU2coordindep}
%\end{equation}
%with the conditions $k^{\widehat X}\partial _{\widehat X}\vec{\ell }=0$
%and for any $\vec{A}$, also $\vec{A}\cdot \vec{k}^{\widehat X}\partial
%_{\widehat X}\vec{\ell }=0$.
%%%%%%%%%%%%%%%%%%%%%%%%%%%%%%%%%%%%%%%%%%%%%%%%%%%%%%%%%%%%%%%%%%%%%%%%
\subsection{The map from hypercomplex to quaternionic}
 \label{ss:maphcquat}
%%%%%%%%%%%%%%%%%%%%%%%%%%%%%%%%%%%%%%%%%%%%%%%%%%%%%%%%%%%%%%%%%%%%%%%%
We now start to construct a map between $4(n_H+1)$-dimensional
hypercomplex/hyper-K{\"a}hler manifolds, admitting a conformal symmetry, and
$4n_H$-dimensional quaternionic(-K{\"a}hler) manifolds.

%%%%%%%%%%%%%%%%%%%%%%%%%%%%%%%%%%%%%%%%%%%%%%%%%%%%%%%%
\subsubsection{Suitable coordinates and almost complex structures}\label{ss:coords}
%%%%%%%%%%%%%%%%%%%%%%%%%%%%%%%%%%%%%%%%%%%%%%%%%%%%%%%%

We first construct a set of coordinates adapted to our setting. These
coordinates should allow us to solve explicitly the constraints imposed
by conformal symmetry. The primary object is the homothetic Killing
vector~(\ref{homothetic}). Therefore, first of all, we will choose one
coordinate such that the vector $k$ has a convenient form. One can always
find local coordinates $q^{\widehat X}=\{z^0,y^p\}$, where
$p=1,...,4n_H+3$, such that the components of the homothetic Killing
vector are
\begin{equation}\label{dilat-coord}
k^{\widehat{X}}=3z^0 \delta _0^{\widehat{X}}.
\end{equation}
This is obtained by choosing at any point the first coordinate in the
direction of the vector $k^{\widehat{X}}$. The factor $3$ in
(\ref{dilat-coord}) is a convenient choice for later purposes. Notice
that arbitrary coordinate transformations $y'{}^p(y^q)$ trivially
preserve (\ref{dilat-coord}). We will make use of this freedom below.

Having singled out the `dilatation direction', we now proceed similarly
for the $\SU(2)$ vector fields~(\ref{defveck}). Frobenius' theorem tells
that the three-dimensional hypersurface spanned by the direction of the
three $\SU(2)$ vector fields can be parametrized by coordinates
$(z^\alpha)_{\alpha= 1,2,3}$, such that $\vec k^{\widehat X}$ is only
non-zero for $\widehat X$ being one of the indices $\alpha $. Note that
these vectors point in different directions than $k^{\widehat X}$, due
to~(\ref{defveck}), and the fact that the complex structures square to
$-\unity $. Therefore, they do not coincide with the direction `0' chosen
above. The other $4n_H$ coordinates are indicated by $q^X$. Thus we have
at this point
\begin{eqnarray}
&&q^{\widehat X}=\left\{z^0 ,\,y^p\right\} = \left\{ \Red{z^0,\,
z^\alpha},\, \Blue{q^X}\right\}, \qquad \alpha =1,2,3,\quad
X=1,\ldots ,4n_H,\nonumber\\
&&  k^{\widehat X}=3z^0 \delta _0^{\widehat{X}}, \qquad \vec k^0=\vec
k^X=0.
 \label{splitcoord}
\end{eqnarray}

Now, as $\vec k^{\widehat{X}}=\ft13k^{\widehat{Y}}\widehat{\vec
J}_{\widehat{Y}}{}^{\widehat{X}}$ and due to our particular choice of
coordinates, we find that
\begin{equation}
\widehat{\vec J}_0{}^\alpha = \frac1{z^0}\vec k^\alpha,\qquad
\widehat{\vec J}_0{}^0=0,\qquad \widehat{\vec J}_0{}^X=0.
\label{firstcompJ}
\end{equation}

Generically, we allow ${\vec k}^\alpha $ to depend on $q^X$ (as it is the
case also in group manifold reductions~\cite{Pons:2003ka}). We assume it
to be invertible as a three by three matrix, and define $\vm\alpha $ as
the inverse, in the sense that we have for any vectors $\vec{A}$ and
$\vec{B}$,
\begin{equation}\label{kinv}
\vec{A}\cdot \vm\alpha \, \vec{k}^\alpha\cdot \vec{B}=\vec{A}\cdot
\vec{B}\qquad \mbox{or}\qquad \vec k^\alpha \cdot \vm\beta =\delta
^\alpha _\beta .
\end{equation}
It is convenient to introduce
\begin{equation}
  \vec A_X \equiv\frac{1}{z^0} \widehat {\vec J}_X{}^0.
 \label{defAXalpha}
\end{equation}
Using ~(\ref{firstcompJ}), we can complete the table of complex
structures by requiring the quaternionic algebra (\ref{JJ}). In terms of
$\vec A_X$, we find
\begin{equation}
   \begin{array}{lll}
    \widehat{\vec J}_0{}^0=0 ,\qquad \qquad & \widehat{\vec J}_\alpha{}^0=
    -z^0\vm\alpha  ,
     & \widehat{\vec J}_X{}^0=z^0 \vec A_X,\\
    \widehat{\vec J}_0{}^\beta=\frac{1}{z^0}\vec k^\beta   ,& \widehat{\vec J}_\alpha{}^\beta=
     \vec k^\beta \times\vm\alpha ,\qquad&
    \widehat{\vec J}_X{}^\beta=\vec A_X \times \vec k^\beta
                               +\vec J_X{}^Z(\vec A_Z\cdot k^\beta)  ,    \\
    \widehat{\vec J}_0{}^Y=0 ,& \widehat{\vec J}_\alpha{}^Y=0 ,& \widehat{\vec J}_X{}^Y=\vec
    J_X{}^Y.
  \end{array}
 \label{allhatJ}
\end{equation}
All dependence on $z^0$ of these complex structures is explicitly shown
in this formula. The last equation says that the components of the
hypercomplex structures on the large space that lie along the small space
satisfy the algebra of the imaginary quaternions.

Thus we have decomposed any almost hypercomplex structure on the large
space and we find that it is expressed in 3 different quantities: the
vectors $\vec{k}^\alpha $ [and their inverses, see (\ref{kinv})], vectors
$\vec{A}_X$, which are arbitrary so far, and an almost quaternionic
structure on the small space, $\vec{J}_X{}^Y$.

%%%%%%%%%%%%%%%%%%%%%%%%%%%%%%%%%%%%%%%%%%%%%%%%%%%%%%%%
\subsubsection{The complex structures and Obata connection}\label{ss:cstrObata}
%%%%%%%%%%%%%%%%%%%%%%%%%%%%%%%%%%%%%%%%%%%%%%%%%%%%%%%%

As was explained in Sect.~\ref{ss:genpropquat}, the almost hypercomplex
structure $\widehat{\vec{J}}$ is hypercomplex if the Nijenhuis tensor
vanishes, or, equivalently, if there exists a torsionless connection
$\widehat{\Gamma }$ such that the complex structures are covariantly
constant. The latter is then the Obata connection, see (\ref{Obata}).

In practice it is easier to first compute that connection. First of all,
the relation~(\ref{homothetic}) in our coordinate
ansatz~(\ref{splitcoord}) gives rise to
\begin{equation}
\widehat\Gamma_{\widehat{X}0}{}^{\widehat{Y}}=\frac 1{z^0}\left(\frac12
\delta_{\widehat{X}}^{\widehat{Y}}-\delta_{\widehat{X}}^0\delta_0^{\widehat{Y}}\right)
% ,
%\end{equation}
%or
%\begin{equation}
%\widehat\Gamma_{00}{}^0=-\frac1{2z^0}  ,\qquad\widehat\Gamma_{00}{}^p=0,
%\qquad\widehat\Gamma_{p0}{}^{\widehat{Y}}=\frac1{2z^0}\delta_p^{\widehat{Y}}
 .  \label{Gammanometric}
\end{equation}
Further, we can immediately find some more information on the coordinate
dependence of the basic quantities. Multiplying the first relation
of~(\ref{transf-J}) by $k^{\widehat X}$ yields that the $\SU(2)$ vectors
commute with the homothetic Killing vector field.
Using~(\ref{splitcoord}), this yields
\begin{equation}
\partial_0 {\vec k}^\alpha =0,\qquad
%\label{partial0ku}
%\end{equation}
%The equation for the inverses, (\ref{kinv}), then implies
%\begin{equation}
\partial_0\vm\alpha=0.\label{partial0}
\end{equation}
One may further use the second line of~(\ref{transf-J}) with $\vec
\Lambda $ replaced by $\vm\alpha$ and obtain the $z^\alpha $ dependence
of the $\SU(2)$ vector fields that reflect the $\SU(2)$ algebra. One can
write the corresponding equation in various forms:
\begin{equation}
  \vec k^\gamma \times
\partial _\gamma \vec k^\alpha= \vec k^\alpha,\qquad
  \partial _{[\alpha } \vm{\beta ]}=
 - \ft12\vm\alpha \times \vm\beta,\qquad
  \vm{[\alpha } \cdot \partial _{\beta ]}\vec k^\gamma =-\ft12(\vm\alpha \times \vm\beta)\cdot \vec k^\gamma .
 \label{SU2algebra}
\end{equation}

%The covariant derivative on the $\SU(2)$ Killing vectors (\ref{DkSU2})
%yields some other components of the connection.
The connection coefficients can then be written as
\begin{equation}
\begin{array}{ll}
\widehat\Gamma_{00}{}^0 = -\ft{1}{2z^0},&
\widehat\Gamma_{00}{}^p = 0,\\
\widehat\Gamma_{0p}{}^0 = 0,&\widehat\Gamma_{0q}{}^p = \frac{1}{2z^0}
\delta^p_q , \\ [2mm]
 \widehat\Gamma_{\alpha\beta}{}^0 =
\ft12\widehat{g}_{\alpha\beta},& \widehat\Gamma_{\alpha\beta}{}^\gamma =
- \vm{(\alpha}\cdot
\partial_{\beta)} \vec{k}^\gamma,\\
\widehat\Gamma_{\alpha\beta}{}^X = 0 ,&\\
\widehat\Gamma_{X\alpha}{}^0 = \ft12z^0 \vec A_X \cdot \vm\alpha=
\ft{1}{2} \widehat{\vec J}_X{}^0 \cdot \vm\alpha ,\quad&
\widehat\Gamma_{X\alpha}{}^\beta  = \ft{1}{2} \widehat{\vec J}_X{}^\beta
\cdot \vm\alpha -\vm\alpha \cdot \partial _X\vec{k}^\beta
,\\
\widehat\Gamma_{X\alpha}{}^Y = \ft{1}{2} \widehat{\vec J}_X{}^Y\cdot
\vm\alpha,%-\vm\alpha \partial_X \vec{k}^p
\\
 \multicolumn{2}{l}{\widehat{\Gamma}_{XY}{}^0=\ft12\widehat{g}_{XY},\qquad
 \widehat\Gamma_{XY}{}^\alpha
 =-(\partial_{(X} \vec A_{Y)})\cdot \vec k^\alpha + \widehat\Gamma_{XY}{}^W \vec A_W\cdot \vec k^\alpha-\ft12
 h_{Z(X}\vec J_{Y)}{}^Z\cdot \vec k^\alpha
 %A_{(Y}^\beta\vec{k}^\alpha \partial_{X)}\vm\beta
,}\\
\multicolumn{2}{l}{ \widehat{\Gamma}_{XY}{}^Z    = \Gamma^{\rm
Ob}{}_{XY}{}^Z+\ft{1}{3} \widehat{\vec J}
   {}_{(X}{}^\delta\cdot\left(  \vm\delta \delta _{Y)}^Z
    +\ft12 \vm\delta \times \vec J{}_{Y)}{}^Z\right).}
\end{array}
\label{summGammabl}
\end{equation}
Here $\Gamma^{\rm Ob}$ is the Obata connection defined by~(\ref{Obata})
using the $\vec J$ complex structures, while $\widehat{\Gamma }$ is the
Obata connection using $\widehat{\vec J}$.

We have also introduced the following convenient notation:
\begin{equation}\label{defhatg}
\widehat{g}_{\alpha\beta}\equiv 2\widehat\Gamma_{\alpha\beta}{}^0, \qquad
\widehat{g}_{\widehat{X}\widehat{Y}}\equiv
2\widehat{\Gamma}_{\widehat{X}\widehat{Y}}{}^0,\qquad h_{XY}\equiv
\frac{1}{z^0} \widehat{g}_{XY}+ \vec A_X\cdot\vec A_Y.
\end{equation}
Note that although we have not introduced a metric, we use here
suggestive notation since a `good metric' coincides with these
definitions, as we will show in Sect.~\ref{ss:proposalmetric}. Hence, we
have used $\widehat g$ as a shorthand for a complicated function of
$z^0$, $\vec{k}^\alpha$, $\vec{A}_X$ and $\vec{J}_X{}^Y$. Considering the
$\widehat{X}=0$ and $\widehat{Y}=\alpha $ components of~(\ref{DkSU2}) in
the basis~(\ref{splitcoord}) and using~(\ref{allhatJ}) leads to
\begin{equation}\label{inversekg}
\vec{k}_\alpha\equiv\widehat{g}_{\alpha \beta}\vec{k}^\beta=
-z^0\vm\alpha.
\end{equation}
This implies
\begin{equation}
\widehat{g}_{\alpha \beta}=-\ft1{z^0}\vec{k}_{\alpha}\cdot
\vec{k}_\beta=-z^0\vm\alpha\cdot  \vm\beta .
 \label{galphabeta}
\end{equation}

The requirements of covariantly constant $\widehat{\vec{J}}$ lead to
requirements on the coordinate dependence of the quantities
$\vec{k}^\alpha$, $\vec{A}_X$ and $\vec{J}_X{}^Y$. From the requirements
that $\widehat {\covder}_0 \widehat{\vec J}=0 $ and
$\widehat{\covder}_\alpha \widehat{\vec J}=0$, we find that
\begin{eqnarray}
\partial _0\vec{A} _X&=&0, \qquad
\left( \partial _\alpha +\vm\alpha \times \right) \vec A_X+\partial _X\vm\alpha =0,\label{dXveck} \\
\partial_0\vec J_X{}^Y&=&0,\qquad \left( \partial _\alpha +\vm\alpha \times \right) \vec J_X{}^Y=0.
\label{dXJ}
\end{eqnarray}
%The final equation implies that the quaternionic structure on the small
%space should rotate in a specific way, when taken along the flow of the
%$\SU(2)$ vector fields.
Note that the integrability condition for the second relation in
(\ref{dXveck}) yields
\begin{equation}
  \left( \partial _\alpha +\vm\alpha \times \right) \big[\vec
  R(-\ft12\vec A)\big]_{XY}=0, \qquad \mbox{with}\qquad  \big[\vec R(\vec V)\big]_{XY}\equiv 2\partial _{[X}\vec
  V_{Y]}+ 2\vec V_X\times \vec V_Y.
 \label{integrdXk}
\end{equation}

A main non-trivial result comes from
$\widehat{\mathfrak{D}}_X\widehat{\vec J}_Y{}^0=0$. We find
\begin{equation}
  \big[\vec R(-\ft12\vec A )\big]_{XY}=\ft12 h_{Z[X} \vec J_{Y]}{}^Z,
  \label{RhJA}
\end{equation}
with $h$ as in (\ref{defhatg}). We can calculate this expression using
the definition of the Obata connection $\widehat{\Gamma }_{\alpha \beta
}{}^0$ from (\ref{Obata}) and our particular decomposition
(\ref{allhatJ}). This leads to
\begin{equation}
  h_{XY}=-\ft13\left( 4\vec J_{(X}{}^Z\cdot \big[\vec R(-\ft12\vec A )\big]_{Y)Z}
  +(\vec J_X{}^U\times\vec
  J_Y{}^Z)\cdot \big[\vec R(-\ft12\vec A )\big] _{UZ}\right) .
   \label{hinvecR}
\end{equation}
However, this equation can also be obtained from solving $h$ from
(\ref{RhJA}). Thus, the equation implies by itself that the matrix $h$
that appears in (\ref{hinvecR}) is necessarily the quantity defined in
(\ref{defhatg}). Therefore, the integrability condition is equivalent to
the requirement that there should be a symmetric matrix $h$ such that
(\ref{RhJA}) is satisfied.

Finally, the vanishing of the components along the small space of the
Nijenhuis tensor of the hypercomplex structure $\widehat{J}$ implies that
the Nijenhuis tensor in the small space should be
\begin{equation}
  6 N_{XY}{}^Z= -\widehat{\vec J}{}_{[X}{}^\alpha  \cdot
  \partial _\alpha \widehat{\vec J}{}_{Y]}{}^Z=\widehat{\vec J}{}_{[X}{}^\alpha  \cdot\vm\alpha
    \times {\vec J}_{Y]}{}^Z=-\left(2\vec A_{[X}+\vec A_W\times\vec J_{[X}{}^W\right) \cdot {\vec J}_{Y]}{}^Z .
 \label{Nij1}
\end{equation}
This equation is the basic equation that determines that the small space
is quaternionic. We will further elaborate on this in
Sect.~\ref{ss:proofQuat}.

The conclusions of Sect.~\ref{ss:coords} can now be completed. There are
integrable complex structures on the conformal hypercomplex space for
functions
\begin{equation}
  \vec k^\alpha(q^X,z^\alpha),\qquad \vec{A}_X(q^X,z^\alpha),\qquad
  \vec{J}_X{}^Y(q^X,z^\alpha).
 \label{depzaqX}
\end{equation}
The $\vec{k}^\alpha $ should satisfy the $\SU(2)$ algebra, i.e.
(\ref{SU2algebra}), while the $z^\alpha $ dependence of the other
quantities is determined by these vector fields using (\ref{dXveck}) and
(\ref{dXJ}). Moreover there are the conditions that there should be a
symmetric tensor $h_{XY}$, which is given by (\ref{hinvecR}), such that
(\ref{RhJA}) is satisfied. The Nijenhuis tensor of the complex structures
in the small space should satisfy (\ref{Nij1}), which, as we will show in
the next section, has the meaning that it is a quaternionic structure
with $\SU(2)$ connection determined by $\vec{A}_X$.

We remark that, combining the previous results, the $z^\alpha
$-dependence of all quantities can be calculated. E.g the following are
useful results:
\begin{equation}
\partial _\alpha h_{XY}=0,\qquad
   \left( \partial _\alpha + \vm\alpha \times \right)
   \vec J_X{}^\beta -\vec J_X{}^\gamma \left( \vm\alpha \cdot \partial
   _\gamma \vec k^\beta \right) =0.
 \label{dalphaJXbeta}
\end{equation}

%%%%%%%%%%%%%%%%%%%%%%%%%%%%%%%%%%%%%%%%%%%%%%%%%%%%%%%%%%%%
\subsection{The embedded quaternionic space}\label{propquat}
%%%%%%%%%%%%%%%%%%%%%%%%%%%%%%%%%%%%%%%%%%%%%%%%%%%%%%%%%%%%

%%%%%%%%%%%%%%%%%%%%%%%%%%%%%%%%%%%%%%%%%%%%%%%%%%%%%%%%%%%%%%%%%%%%%%%%
\subsubsection{Proof that the small space is quaternionic}
\label{ss:proofQuat}
%%%%%%%%%%%%%%%%%%%%%%%%%%%%%%%%%%%%%%%%%%%%%%%%%%%%%%%%%%%%%%%%%%%%%%%%
As already explained in Sect.~\ref{ss:QuatLikeMan}, contrary to the
hypercomplex case where the $\SU(2)$ connection is trivial, in the
quaternionic case there is a non-trivial $\SU(2)$ connection. This means
that parallel transport with respect to the affine connection rotates the
three complex structures into each other. As a consequence, the
integrability condition for the complex structures differs from the
hypercomplex case, as the (diagonal) Nijenhuis tensor should now be
proportional to the $\SU(2)$ connection~(\ref{N=omega}). This is exactly
what we obtained with (\ref{Nij1}), leading to the $\SU(2)$ Oproiu
connection satisfying (\ref{gaugechoiceOproiu}),
\begin{equation}
  \vec { \omega}^{\rm Op} _X = -\ft16\left( 2\vec A_X + \vec A_Y\times \vec
  J_X{}^Y\right).
 \label{vecomegaOp}
\end{equation}
Hence, this shows that the small space is quaternionic.

The corresponding affine connection is,  according to~(\ref{Oproiu}),
\begin{eqnarray}
\Gamma^{\rm Op}{}_{XY}{}^Z&=&\Gamma^{\rm Ob}{}_{XY}{}^Z
-\vec{J}_{(X}{}^Z\cdot\vec{\omega}^{\rm Op}_{Y)}\label{exprOp}\\
&=&\widehat{\Gamma}{}_{XY}{}^Z-\ft1{3} \vec A_V\cdot \vec J
{}_{(X}{}^V\delta_{Y)}^Z +\ft2{3} \vec A_{(X} \cdot \vec
J{}_{Y)}{}^Z+\ft1{3} \vec A_V\cdot \vec J{}_{(X}{}^V\times \vec
J{}_{Y)}{}^Z,\nonumber
\end{eqnarray}
where we used the last equation of~(\ref{summGammabl}).

As we have discussed in Sect.~\ref{ss:xi}, there is a family of
torsionless connections that are compatible with that given structure.
Related to that freedom, all $\SU(2)$ connections can be written as
\begin{equation}
  \vec \omega _X = -\ft16\left( 2\vec A_X + \vec A_Y\times \vec
  J_X{}^Y\right)+\vec J_X{}^Y \xi _Y.
 \label{anySU2conn}
\end{equation}
A particular choice that will be useful below is
\begin{equation}
  \xi _X= \ft16 \vec J_X{}^Y\cdot \vec A_Y,
 \label{specialxi}
\end{equation}
such that we find the connections
\begin{equation}
  \vec \omega _X=-\ft12\vec A_X, \qquad \Gamma_{XY}{}^Z=\widehat{\Gamma}_{XY}{}^Z + \vec A_{(X} \cdot \vec
J{}_{Y)}{}^Z.
 \label{specialchoiceOmega}
\end{equation}
This choice of a quaternionic connection will turn out to be special for
two different reasons. First of all, if there is a good metric on the
small space (i.e. if the space is quaternionic-K{\"a}hler) this connection
will correspond to the Levi-Civita connection. Secondly, we will show
that for this choice the $\mathbb{R}$ curvatures for both the large and
the small space will be equal, $\widehat {\cal R}(\mathbb{R})={\cal
R}(\mathbb{R})$ . In the hypercomplex space this curvature is
proportional to the Ricci tensor\footnote{Observe that (\ref{konRis0})
implies that the Ricci tensor has only components in the directions of
the small space.}, which is Hermitian. This implies that the $\mathbb{R}$
curvature on the small space is Hermitian.

%%%%%%%%%%%%%%%%%%%%%%%%%%%%%%%%%%%%%%%%%%%%%%%%%%%%%%%%%%%%
\subsubsection{The local $\SU(2)$}\label{ss:SU2}
%%%%%%%%%%%%%%%%%%%%%%%%%%%%%%%%%%%%%%%%%%%%%%%%%%%%%%%%%%%%

One may wonder what the origin is of the local $\SU(2)$ invariance on the
embedded quaternionic manifold. We will show that this local invariance
is already present in \textit{conformal} hypercomplex manifolds, but the
transformations on the complex structures are more complicated than
simple vector rotations. Then, we will show how this induces the expected
local $\SU(2)$ on the quaternionic manifold, with $\vec{\omega }_X$ as
gauge field.

Considering the equations of Sect.~\ref{ss:maphcquat}, they are nearly
all invariant under a usual vector rotation $\delta \vec{V}=\vec{\ell
}\times \vec{V}$ for all 3-vectors $\vec{V}$, especially if $\vec{\ell }$
only depends on the coordinates of the small space $q^X$. One troublesome
equation is (\ref{dXveck}). $\vec{A}_X$ is an arbitrary quantity in the
construction, a `black box'. It turns out that the local invariance can
be obtained by adding a gauge-type transformation for $\vec{A}_X$. Thus,
we consider for the elementary quantities the following $\SU(2)$
transformations:
\begin{eqnarray}
    \delta_{\rm SU(2)} \vec A_X &=& \partial_X \vec{\ell}   + \vec{\ell} \times \vec A_X, \qquad \delta_{\rm SU(2)}\vec J_X{}^Y=\vec{\ell} \times \vec J_X{}^Y,\nonumber\\
   \delta_{\rm SU(2)}\vec k^\alpha&=&\vec{\ell} \times \vec k^\alpha,
   \qquad\qquad\qquad \delta_{\rm SU(2)} \vm\alpha =\vec{\ell} \times \vm\alpha, \label{delSU2A}
\end{eqnarray}
where the parameter $\vec{\ell}(q^X)$ cannot depend on $z^0$ or
$z^\alpha$. The main requirement for $\vec{A}_X$ was (\ref{RhJA}), in
which the curvature of $\vec{A}$ appears. This equation is thus
consistent if $h_{XY}$ is invariant under the $\SU(2)$ transformations. A
long calculation using the definition of the Obata connection shows that
$\widehat{\Gamma}_{\widehat{X}\widehat{Y}}{}^0$ is not invariant, but
precisely transforms such that $h_{XY}$ defined as in (\ref{defhatg}) is
invariant. This proves the local $\SU(2)$ symmetry of the conformal
hypercomplex manifold.

Note that, due to the transformations of $\vec{A}_X$ some components of
$\widehat{J}$ do not transform as an ordinary vector. These are
\begin{eqnarray}
  \delta_{\rm SU(2)} \widehat{\vec{J}}_X{}^0&=&z^0\partial _X\vec{\ell} + \vec{\ell}\times\widehat{\vec{J}}_X{}^0, \nonumber\\
 \delta_{\rm SU(2)} \widehat{\vec{J}}_X{}^\beta &=&-\vec k^\beta \times \partial _X\vec{\ell}
+\vec J_X{}^Z\left( \vec k^\beta \cdot \partial _Z\vec{\ell}\right)
 + \vec{\ell}\times\widehat{\vec{J}}_X{}^\beta .
 \label{modifiedSU2J}
\end{eqnarray}
It turns out that the full Nijenhuis tensor is invariant under this
$\SU(2)$.

The complex structures in the small space thus transform as ordinary
vectors. We have seen that $\vec{A}_X$ is the gauge field of these
$\SU(2)$ transformations. In the $\xi $-gauge where $\vec{A}_X$ is
proportional to $\vec{\omega }_X$, see (\ref{specialchoiceOmega}), this
is thus the expected $\SU(2)$ gauge field, and we find
\begin{equation}
\delta_{\rm SU(2)} \vec \omega_X=-\ft 12 \partial_X \vec{\ell}+\vec{\ell}
\times \vec \omega_X.
\end{equation}
Hence, $\vec \omega_X$ transforms as an true connection since
(\ref{covconstJq}) now transforms covariantly.

Another $\xi $-choice, as e.g. the Oproiu choice
(\ref{gaugechoiceOproiu}), is not invariant under $\SU(2)$. Hence, if we
take this connection, it implies that the remaining $\SU(2)$ contains a
compensating $\xi $-transformation, which is
\begin{equation}
  \xi _X=\ft16\vec{J}_X{}^Y\cdot\partial _Y\vec{\ell}.
 \label{xiSU2Oproiu}
\end{equation}
This compensating transformation also contributes to the $\SU(2)$
transformation of the affine Oproiu connection $\Gamma^{\rm Op}$ , such
that it cancels other terms that follow from its definition
(\ref{Oproiu}). Thus the final affine connection is $\SU(2)$ invariant,
as one should expect for covariance of the covariant derivative of the
complex structures (\ref{covconstJq}).

Hence, the quaternionic local $\SU(2)$ is naturally included into the
map.

\subsection{The map from quaternionic to hypercomplex}\label{ss:mapqh}
%%%%%%%%%%%%%%%%%%%%%%%%%%%%%%%%%%%%%%%%%%%%%%%%%%%%%%

\subsubsection{Uplifting a quaternionic manifold}
 \label{ss:upliftingq}

By now, we are in a position to discuss the inverse procedure, namely the
construction of a conformal hypercomplex manifold starting from a
quaternionic space. We may at this point choose a value for $\xi$, and we
will see below that we have to choose one such that the $\mathbb{R}$
curvature (i.e. the antisymmetric part of the Ricci tensor) is Hermitian.
This is always possible, as it is clear from (\ref{xitransformomU1}) that
we may even choose a $\xi$ such that the $\mathbb{R}$ connection
vanishes.

We now consider a space of 4 real dimensions bigger than the one of the
quaternionic space. The extra coordinates are labeled $z^0$ and
$z^\alpha$. Let $\vec k^\alpha(z^\alpha ,q^X)$ denote left-invariant
vector fields on the $\SU(2)$ group manifold, i.e. satisfying
(\ref{SU2algebra}). For now, the dependence on $q^X$ is not fixed. One
may take the $\SU(2)$ vectors independent of $q$, but an arbitrary
dependence is allowed. It will be fixed later. Then construct their
inverse $\vm\alpha(z,q)$ using (\ref{kinv}). Furthermore, we define $\vec
A_X(q)=-2\vec \omega_X(q)$ for $\vec \omega_X(q)$ the $\SU(2)$ connection
for the chosen value of $\xi$.

We take the $\vec{k}^\alpha$, $\vec{J}_X{}^Y$ and $\vec{A}_X$ independent
of $z^0$. The $z^\alpha $ dependence of $\vec{k}^\alpha $ is determined
by its $\SU(2)$ property (\ref{SU2algebra}). The complex structures are
taken to be covariant constant in their $z^\alpha $ dependence in the
sense of (\ref{dXJ}). This means in fact that when we change the value of
$z^\alpha $ we go to a different choice of complex structures. These
different complex structures are related by an $\SU(2)$ rotation. This
implies that also the $\SU(2)$ connection should change, and indeed this
is in agreement with the comparison of (\ref{delSU2A}) and
(\ref{dXveck}). The latter equation determines the $q^X$-dependence of
the $\SU(2)$ vector fields $\vec{k}^\alpha $. One also notices that the
curvature of $A_X$ is taken to be a covariant vector as shown in
(\ref{integrdXk}).

With these ingredients, we can  construct an almost hypercomplex
structure as in (\ref{allhatJ}). In order for this structure to be
integrable, the only remaining condition is (\ref{RhJA}). The condition
states that there should exist a symmetric object $h$ such that the
$\SU(2)$ curvature is related to the hypercomplex structure as indicated.
In a quaternionic-K{\"a}hler manifold this is satisfied with $h$ being the
metric. In an arbitrary quaternionic manifold, we have (\ref{RSU2B}). We
thus just need that the antisymmetric part of $B$ does not contribute to
this equation. That is a condition of the form (\ref{FJTis-JF}) for the
antisymmetric part of $B$, which is the antisymmetric part of the Ricci
tensor or $\mathbb{R}$ curvature. Hence it says that this ${\cal
R}^{\mathbb{R}}$ should be Hermitian, which we can obtain by a
$\xi$-transformation as mentioned in the beginning of this section.

With these choices, (\ref{allhatJ}) defines a hypercomplex structure on a
manifold parametrized by the coordinates $\{z^0,z^\alpha,q^X \}$.
Moreover, it is easy to see that the vector field $3z^0\partial_0$
satisfies~(\ref{homothetic}), hence the manifold is actually conformal
hypercomplex.

\subsubsection{A $\widehat{\xi }$-transformation for conformal hypercomplex manifolds}
\label{ss:hatxi}

Suppose we have constructed a hypercomplex manifold from a quaternionic
manifold with a Hermitian $\mathbb{R}$ curvature ${\cal R}^{\mathbb{R}}$.
We perform a $\xi$-transformation with Hermitian $\partial _{[X}\xi
_{Y]}$. Then the new $\mathbb{R}$ curvature ${\cal R}^{\prime\mathbb{R}}$
is also Hermitian. Hence, we can use again the procedure of
Sect.~\ref{ss:upliftingq} to obtain a hypercomplex structure on the large
space. As the $\xi$-gauge modified the $\SU(2)$ connection, the vector
$\vec A'_X$ is different from $\vec A_X$, and hence the new hypercomplex
structure differs from the original one. This defines a new mapping,
dependent on a one-form $\hat \xi$, between hypercomplex structures on
conformal hypercomplex manifolds, see (\ref{schemexihat}):
\begin{equation}
 \setlength{\unitlength}{0.25mm}
 \begin{picture}(120,120)
 \put(20,85){\vector(0,-1){50}}
 \put(35,100){\vector(1,0){50}}
 \put(100,35){\vector(0,1){50}}
 \put(40,20){\vector(1,0){40}}
 \put(20,100){\makebox(0,0){$\widehat{\vec J}$}}
 \put(20,20){\makebox(0,0){$\vec J,\, \vec \omega$}}
 \put(100,20){\makebox(0,0){$\vec J, \,\vec \omega'$}}
 \put(100,100){\makebox(0,0){$\widehat{\vec J}'$}}
 \put(35,60){\makebox(0,0){$\scriptstyle \vec A_X$}}
 \put(85,60){\makebox(0,0){$\scriptstyle \vec A_X'$}}
 \put(60,110){\makebox(0,0){$\scriptstyle \hat\xi$}}
 \put(60,30){\makebox(0,0){$\scriptstyle \xi$}}
 \end{picture}
 \label{schemexihat}
\end{equation}
This is the transformation that was announced in (\ref{transfoJxi}).
Remember that the $\xi $-trans\-for\-ma\-tions are defined by one forms
$\xi _X\rmd q^X$ that depend only on the quaternionic coordinates $q^X$.
This says that $\hat \xi_X=\xi_X$ must be constant along the flows of $k$
and $\vec k$, and that $\hat \xi$ must transform a Hermitian $\hat{\cal
R}^{\mathbb{R}}$ in another Hermitian $\hat {\cal R}^{\mathbb{R}}$,
yielding the conditions in~(\ref{hatxiconditions}). Hence, we have
re-derived the results of Sect.~\ref{CHC}, and shown how its origin is
necessary for the consistency of this picture. The transformation
(\ref{transfoJxi}) is in these coordinates, with complex structures as in
(\ref{allhatJ}), simply given by
\begin{equation}
  \delta(\widehat{\xi }) \vec{A}_X = 2\vec{J}_X{}^Z\widehat{\xi }_Z.
 \label{hatxiA}
\end{equation}

%%%%%%%%%%%%%%%%%%%%%%%%%%%%%%%%%%%%%%%%%%%%%%%%%%%%%%%
\subsection{The map from hyper-K{\"a}hler to quaternionic-K{\"a}hler spaces}
\label{ss:maphkqk}
%%%%%%%%%%%%%%%%%%%%%%%%%%%%%%%%%%%%%%%%%%%%%%%%%%%%%%%
We will now restrict to the case in which there is a compatible metric.
We will show that starting from a hyper-K{\"a}hler space, the small space
will be automatically quaternionic-K{\"a}hler and vice-versa. The existence
of a metric will moreover remove the freedom that was implied by the
$\xi$-transformation, since it unambiguously specifies the torsionless
connection as being the Levi-Civita connection. Hence, the connection on
the small space will be determined uniquely, given the structures that
are defined on that manifold.

We will show how to construct the quaternionic metric from the one in the
hyper-K{\"a}hler space, and show how the affine connection reduces to the
Levi-Civita connection.

%%%%%%%%%%%%%%%%%%%%%%%%%%%%%%%%%%%%%%%%%%%%%%%%%%%%%%%%%%%%%%
\subsubsection{Decomposition of a hyper-K{\"a}hler metric}\label{ss:proposalmetric}
%%%%%%%%%%%%%%%%%%%%%%%%%%%%%%%%%%%%%%%%%%%%%%%%%%%%%%%%%%%%%%

As we already mentioned in Sect.~\ref{ss:QuatLikeMan}, a hypercomplex
manifold is promoted to a hyper-K{\"a}hler space if there is a Hermitian
metric $\widehat{g}_{\widehat{X}\widehat{Y}}$ that preserves the Obata
connection, i.e.
\begin{eqnarray}
  \partial _{\widehat Z} \widehat{g} _{\widehat{X}\widehat{Y}}
  -2\widehat{\Gamma }_{\widehat{Z}(\widehat{X}}{}^{\widehat{W}}\widehat{g}
  _{\widehat{Y})\widehat{W}}&=&0,
 \label{condQuatK}\\
  \widehat{\vec J}_{\widehat{X}}{}^{\widehat{Z}}\widehat{g}_{\widehat{Z}\widehat{Y}}
  +\widehat{\vec J}_{\widehat{Y}}{}^{\widehat{Z}}\widehat{g}_{\widehat{Z}\widehat{X}}&=&0.
 \label{hermitHatg}
\end{eqnarray}
We can now split these equations in various parts in the
basis~(\ref{splitcoord}). The first equation for $\widehat{Z}=0$,
using~(\ref{Gammanometric}), determines the $z^0$ dependence of the
various parts of the metric. Considering furthermore the $\widehat{Z}=p$,
$\widehat{X}=\widehat{Y}=0$ part leads to
\begin{equation}
\rmd \widehat{s}^2\equiv \widehat{g}_{\widehat{X}\widehat{Y}}\rmd
q^{\widehat X}\rmd q^{\widehat Y} =-\frac1{z^0}\widehat{h}_{00}(\rmd
z^0)^2-2\partial_p\widehat{h}_{00}\rmd z^0\rmd
y^p+z^0\widehat{h}_{pq}\rmd y^p\rmd y^q,
\end{equation}
where the $\widehat{h}$ are independent of $z^0$. Furthermore we want to
invoke the $\widehat{X}=\widehat{Y}=0$ part of~(\ref{hermitHatg}), which
using~(\ref{allhatJ}) and the invertibility of $\vm\alpha $ implies that
$\partial _\alpha\widehat{h}_{00}=0$.

One can further simplify the metric by a redefinition
\begin{equation}
z^0{}'=z^0 \widehat{h}_{00}(q) , \qquad z^\alpha {}'=z^\alpha , \qquad
q^X{}'=q^X.
\end{equation}
This redefinition preserves our previous coordinate
choices~(\ref{splitcoord}). In particular, $\vec k$ has still only
components along $z^\alpha$ since $\partial _\alpha\widehat{h}_{00}=0$.
This redefinition accomplishes that (using the new, primed, $z^0$
coordinate, but omitting the primes from now on)
\begin{equation}
  \widehat{g}_{00}=-\frac{1}{z^0},\qquad \widehat{g}_{0p}=0.
 \label{g000p}
\end{equation}
The $\widehat{Z}=p$, $\widehat{X}=q$, $\widehat{Y}=0$ part
of~(\ref{condQuatK}) leads to
\begin{equation}
  \widehat{g}_{\widehat{X}\widehat{Y}}= 2\widehat{\Gamma}_{\widehat{X}\widehat{Y}}{}^0.
 \label{hatg}
\end{equation}
This coincides with~(\ref{defhatg}). At that time we have made a
definition for arbitrary hypercomplex manifolds. Here we prove that any
good metric on these manifolds is of the form~(\ref{hatg}) after choosing
suitable coordinates.

Using the table of affine connections, (\ref{summGammabl}) and
(\ref{defhatg}) we thus obtain
\begin{eqnarray}\label{QK}
\rmd \widehat s^2  & = & -\frac{(\rmd z^0)^2}{z^0}
 +\Big\{ z^0h_{XY} (q)\rmd q^X \rmd q^Y
 \nonumber\\
&&\qquad \qquad + \widehat{g}_{\alpha\beta} [\rmd z^\alpha - \vec
A_X(z,q)\cdot \vec k^\alpha \rmd q^X][\rmd z^\beta - \vec A_Y(z,q)\cdot
\vec k^\beta \rmd q^Y]\Big\} .
\end{eqnarray}
The metric therefore is a cone~\footnote{This follows from first
extracting the $z^0$ dependence from ${\hat g}_{\alpha\beta}$, and then
defining the radial variable $r^2=z^0$.} and since this cone is
hyper-K{\"a}hler, this implies that the base is a tri-Sasakian manifold, see
e.g. \cite{Boyer:1998sf}.

We now check the remaining conditions in (\ref{condQuatK}) and
(\ref{hermitHatg}). For both it turns out that the only non-trivial
components are the ones where the indices are restricted to those in the
small space.  We then find
\begin{eqnarray}
  (\ref{hermitHatg}) %\widehat{\vec J}_{(\widehat{X}}{}^{\widehat{Z}}\widehat{g}_{\widehat{Y})\widehat{Z}}=0
  & \leftrightarrow&
  \vec J_{(X}{}^Z h_{Y)Z}=0,
 \label{hermgIsHermh}
\\
   (\ref{condQuatK})& \leftrightarrow&\partial _Z h_{XY} -2\Gamma _{Z(X}{}^Wh_{Y)W}=0,\label{condhconst}
\end{eqnarray}
where $\Gamma_{XY}{}^Z$ is given in (\ref{specialchoiceOmega}). We can
thus state that the metric $h$ on the quaternionic space is Hermitian if
and only if the metric $\widehat{g}$ on the hypercomplex manifold is
Hermitian. The second result, (\ref{condhconst}), states that $h$ is
covariantly constant using as connection $\Gamma _{XY}{}^Z$. Therefore,
it is the Levi-Civita of $h$. This connection also  preserves the
quaternionic structure, because we have shown in
(\ref{specialchoiceOmega}) that it is equivalent
 to the Oproiu connection up to $\xi $-transformations.

We have thus shown that the only $\xi $-choice that can be taken for
quaternionic-K{\"a}hler manifolds is the one mentioned in
Sect.~\ref{ss:proofQuat}, and in particular that
\begin{equation}
 \vec{\omega} _X =-\ft12\vec{A}_X.
 \label{vecomegaisA}
\end{equation}

It is seen from (\ref{QK}) that the induced metric in the small space is
\begin{equation}
  g_{XY}= z^0 h_{XY}= -\frac{1}{\nu }h_{XY}= \frac{1}{\kappa ^2}h_{XY}.
 \label{gquat}
\end{equation}
The overall factors do not play a role in the equations in this section.
This metric does not depend on $z^\alpha $, see (\ref{dalphaJXbeta}).

%%%%%%%%%%%%%%%%%%%%%%%%%%%%%%%%%%%%%%%%%%%%%%%%%%%%
\subsubsection{The inverse map}\label{ss:invmapqkhk}
%%%%%%%%%%%%%%%%%%%%%%%%%%%%%%%%%%%%%%%%%%%%%%%%%%%%

%In~(\ref{defhatg}) we will also define $h_{XY}$. In
%section~\ref{ss:curvatures}, we will derive for general
%quaternionic-K{\"a}hler manifolds the relation~(\ref{Bsymm}) for $B$.
%Comparing with the general relation~(\ref{BinqK}) shows that if the
%quaternionic manifold has a good metric, it should be equal to this
%$h_{XY}$. This is not a coincidence. As we have shown in
%section~\ref{ss:hermRicci}, the symmetric and Hermitian Ricci tensor (and
%hence $B$) serves as a metric and it will turn out that in our
%coordinates $h_{XY}$~(\ref{defhatg}) is just this Ricci tensor.

The inverse map is of course a special case of the discussion in
Sect.~\ref{ss:mapqh}. Therefore, we will not repeat the complete lifting
process but merely point out the facts specific to this case.

As the small space carries both a quaternionic structure $\vec J$
together with a good metric $h$, there is a unique torsionless connection
compatible with that structure. Hence, we choose as a triplet of vectors
$\vec A_X=-2\vec \omega_X$, where $\vec \omega_X$ is the $\SU(2)$
connection that corresponds to the Levi-Civita connection on the small
space, and use it in~(\ref{allhatJ}). If we introduce $z^0$ and
$z^\alpha$ dependence as in Sect.~\ref{ss:mapqh}, the resulting almost
hypercomplex structure $\widehat {\vec J}$ is automatically integrable.
The reason for this is that Eq.~(\ref{RhJA}) is now always satisfied
since the Ricci tensor, and hence the tensor $B$~(\ref{defB}) cannot have
an antisymmetric part. Therefore, the large space is already
hypercomplex.

We can moreover construct a good metric $\widehat g$ on the large space,
starting from the good metric $h$ on the small space, using the following
table:
\begin{eqnarray}
\widehat g_{00}&=&-\frac1{z^0},\qquad\qquad\:\, \widehat{g}_{0\alpha
}=\widehat{g}_{0X}=0,\qquad  \widehat g_{\alpha
\beta}=-z^0\vm\alpha \cdot \vm\beta ,\nonumber\\
\widehat g_{\alpha X}&=&z^0\vec A_X\cdot \vm\alpha,\qquad \widehat
g_{XY}=z^0\left(h_{XY}-\vec A_X\cdot \vec A_Y\right).
\label{metricinverse}
\end{eqnarray}

Combining this with the remarks of Sect.~\ref{ss:proposalmetric}, we
conclude that a conformal hypercomplex manifold is hyper-K{\"a}hler if and
only if the corresponding quaternionic space is quaternionic-K{\"a}hler.

%%%%%%%%%%%%%%%%%%%%%%%%%%%%%%%%%%%%%%%%%%%%%%%%%%%%%%%%%%
\subsection{The map for the vielbeins and related connections}\label{mapviel}
%%%%%%%%%%%%%%%%%%%%%%%%%%%%%%%%%%%%%%%%%%%%%%%%%%%%%%%%%%
In the previous part, we did not yet consider the objects that are
related to the $\Gl(r,\mathbb{H})$ structure of the manifolds. To discuss
these, one needs the vielbeins. These vielbeins are also the starting
point for the supersymmetry transformations of hypermultiplets. We have
first to choose a suitable basis to express these objects.

%%%%%%%%%%%%%%%%%%%%%%%%%%%%%%%%%%%%%%%%%%%%%%%%%%%%
\subsubsection{Coordinates on the tangent space} % associated vector bundle
%%%%%%%%%%%%%%%%%%%%%%%%%%%%%%%%%%%%%%%%%%%%%%%%%%%%
Having singled out the $z^0$ and $z^\alpha$ coordinates in which the
quaternionic structure is given by (\ref{allhatJ}), the structure group
of the frame bundle reduces from $\Gl(n_H+1,\mathbb{H})$ to
$\SU(2)\cdot\Gl(n_H,\mathbb{H})$, which consists of the set of frame
reparametrizations on the small space. Similarly, the vielbein ${\widehat
f}_{\widehat X}^{i\widehat A}$ and its inverse ${\widehat f}^{\widehat
X}_{i\widehat A}$ can be decomposed. To do so, we split the index
$\widehat A$ into $(i,A)$ with $i=1,2$ and $A=1,\dots,2n_H$.

To be more precise, we split any $\Gl(n_H+1,\mathbb{H})$-vector $\zeta
^{\widehat{A}}(q^{\widehat{X}})$ in $(\zeta ^i,\zeta ^A)$, which is a
vector of $\SU(2)\cdot\Gl(n_H,\mathbb{H})$, and where $\zeta^i$ is
defined by
\begin{equation}
\sqrt{\ft12z^0 } \, \zeta ^i\equiv -\rmi \varepsilon ^{ij}\widehat f^0
_{j\widehat{A}} \zeta ^{\widehat{A}}.
 \label{defzetai}
\end{equation}
This choice of basis is guided by applications in supergravity, in which
$\zeta^{\widehat A}$ are the superpartners of the coordinate scalar
fields $q^{\widehat X}$. We will see that the choice of frame in the
tangent space is useful for the identification of the quaternionic
tangent space within the full hypercomplex tangent space. The
formula~(\ref{defzetai}) moreover implies
\begin{equation}
  \widehat f^0_{ij} = -\rmi\varepsilon_{ij}\sqrt{\ft12z^0},\qquad \widehat f^0_{iA}=0.
 \label{fchi}
\end{equation}
The $\rmi$ factors are needed for the reality
conditions~(\ref{realfunctions}) where the $\rho _i{}^j$ components  of
$\rho _{\widehat A}{}^{\widehat B}$ are $\rho _i{}^j=
-E_i{}^j=-\varepsilon _{ij}$, and $\rho $ has no off-diagonal elements,
i.e. $\rho_i^A=\rho^A_i=0$. This is convenient for the formulation of
quaternionic manifolds that appear in supergravity, see the discussion in
Sect.~\ref{ss:discussion}.

Writing the first entry of~(\ref{allhatJ}) in terms of vielbeins,
(\ref{defJf}) yields
\begin{equation}
{\widehat f}_0^{ij}=\rmi \varepsilon^{ij}\sqrt{\frac{1}{2z^0}}.
\label{f0ij}
\end{equation}
Furthermore, we redefine $\zeta^A$ according to
\begin{equation}
  \zeta^{\prime A} = \zeta^A - \rmi\sqrt{2z^0}\widehat f^{iA}_0 \zeta^j\varepsilon
  _{ji}.
 \label{Sinvzeta}
\end{equation}
Using $\widehat{\vec J}_0{}^Y=0$ and~(\ref{f0ij}), this implies
\begin{equation}
  f^X_{i{\widehat A}}\zeta ^{\widehat A}=f^X_{iA}\zeta ^{\prime A}.
 \label{prooffXij}
\end{equation}
Therefore, we have in the new basis $\zeta^{\widehat{A}} =
(\zeta^i,\zeta^{\prime A})$,
\begin{equation}
  \widehat f^X_{ij}=0.
 \label{fXij0}
\end{equation}
After having established this basis, we can drop the primes.
Using~(\ref{inv-vielbein}) and the form of components of the quaternionic
structure in~(\ref{allhatJ}) combined with~(\ref{defJf}), we find the
following table determining the vielbeins of the large space in terms of
those of the small space, $z^0$, $\vec k^\alpha $ and $\vec{A}_X$:
\begin{equation}
  \begin{array}{lll}
\widehat f^0_{ij} = -\rmi\varepsilon _{ij}\sqrt{\ft12z^0}, \qquad &
  \widehat f^\alpha_{ij} =\sqrt{\frac{1}{2z^0}}\vec k^\alpha \cdot \vec
\sigma_{ij},\qquad &
 \widehat f^X_{ij}=0, \\
\widehat f^0_{iA}=0,\qquad & \widehat{f}^\alpha _{iA}=f^X_{iA}\vec A_X\cdot \vec k^\alpha,\qquad & \widehat f_{iA}^X=f_{iA}^X,\\
 \widehat f_0^{ij}=\rmi\varepsilon^{ij}\sqrt{\frac{1}{2z^0}},\qquad & \widehat f_\alpha^{ij}
  = - \sqrt{\frac{z^0}{2}} \vm\alpha\cdot
\vec\sigma^{ij}
 ,\qquad &\widehat f_X^{ij}= \sqrt{\frac{z^0}{2}} \vec A_X\cdot
\vec\sigma^{ij} ,
 \\
\widehat f_0^{iA}=0,\qquad  & \widehat f_\alpha^{iA}=0,\qquad &\widehat
f_X^{iA}=f_X^{iA},
  \end{array} \label{allf}
\end{equation}
where $\vec \sigma_i{}^j$ denote the Pauli matrices, $\vec
\sigma^{ij}=\varepsilon^{ik}\vec \sigma_k{}^j=\vec \sigma ^{ji}$ and
$\vec \sigma_{ij}=\vec \sigma_i{}^k\varepsilon_{kj}=\vec \sigma _{ji}$.

If there is a good metric on the hyper-K{\"a}hler space, this implies that
the scalar product defined by this metric carries over to the tangent
space, as was shown in~(\ref{defC}). In the
coordinates~(\ref{splitcoord}), this symplectic metric decomposes as
\begin{equation}
 \widehat C_{AB} =  C_{AB}, \qquad \widehat C_{ij} = \varepsilon _{ij}, \qquad \widehat C_{iA} = 0 .
 \label{mapCAB}
\end{equation}
Note that the tangent space metric $C_{AB}$, which can be used to raise
and lower flat indices on the small space, corresponds to the metric
$g_{XY}=z^0 h_{XY}$.

%%%%%%%%%%%%%%%%%%%%%%%%%%%%%%%%%%%%%%%%%%%%%%%%%%%%%%%%%%%%%%%%%%%%%%%
\subsubsection{Connections on the hypercomplex space} \label{ss:connhc}
%%%%%%%%%%%%%%%%%%%%%%%%%%%%%%%%%%%%%%%%%%%%%%%%%%%%%%%%%%%%%%%%%%%%%%%

Since we already know the reduction rules for the
$\widehat{\vec{J}}_{\widehat{X}}{}^{\widehat{Y}}$ in terms of $z^0$,
$\vec k^\alpha $, $\vec A_X$ and ${\vec J}_X{}^Y$, see~(\ref{allhatJ}),
and the form of $\hat \Gamma$, we can calculate the induced connection
$\widehat{\omega}$ on the tangent space from~(\ref{determineOm}) for the
hatted quantities (remember that in the large space there is no $\SU(2)$
connection) using the reduction rules for the vielbeins~(\ref{allf}). The
induced connection has the following component expressions:
\begin{equation}
   \begin{array}{ll}
  \widehat{\omega }_{0 i}{}^j=\widehat{\omega }_{0 i}{}^A=\widehat{\omega }_{0
  A}{}^j=0,\qquad &\widehat{\omega }_{0 A}{}^B=\ft12f_Y^{iB}\partial _0 f_{iA}^Y+ \ft{1}{2z^0}
  \delta_A^B,\\
\widehat{\omega }_{\alpha i}{}^A=\widehat{\omega }_{\alpha
A}{}^j=0,\qquad & \widehat{\omega }_{Xi}{}^j=-\rmi\frac{z^0}{2}\vec
A_X\cdot
\vec \sigma _i{}^j,\\
\widehat{\omega }_{\alpha i}{}^j=\rmi\frac{z^0}{2}\vm\alpha \cdot\vec
\sigma _i{}^j,\qquad
  & \widehat{\omega }_{\alpha A}{}^B=\ft12 f_Y^{iB}\partial _\alpha f_{iA}^Y, \\
\widehat\omega_X{}_i{}^A = \rmi\sqrt{\frac{1}{2z^0}} \varepsilon_{ik}
 f_X^{kA} ,\quad &  \widehat{\omega}_{XA}{}^i= -
\rmi\sqrt{\frac{z^0
}{2}}\varepsilon ^{ij}f_{jA}^Y h_{YX},\\
\multicolumn{2}{l}{
\widehat{\omega}_{XA}{}^B=\ft12f_Y^{iB}\partial_Xf_{iA}^Y+\ft12f_Y^{iB}f_{iA}^Z\left(
\widehat{\Gamma}_{XZ}{}^Y +\ft1{2}\vec A_Z\cdot \vec{J}_X{}^Y \right).}
  \end{array}
 \label{allOmega}
\end{equation}

%%%%%%%%%%%%%%%%%%%%%%%%%%%%%%%%%%%%%%%%%%%%%%%%%%%%%%
\subsubsection{Connections on the quaternionic space}
\label{ss:connectionsQuat}
%%%%%%%%%%%%%%%%%%%%%%%%%%%%%%%%%%%%%%%%%%%%%%%%%%%%%%
By now, we have completely specified our reduction ansatz. This enabled
us to give the component expressions for the Obata connection on the
hypercomplex manifold in terms of the objects appearing in the ansatz. We
now determine an expression for the $\Gl(n_H,\mathbb{H})$ connection
components on the lower dimensional quaternionic space.

Using (\ref{allOmega}), we can dimensionally reduce the
$\Gl(n_H+1,\mathbb{H})$ connection as
\begin{equation}
\widehat{\omega}_{XA}{}^B= \omega^{\rm
Op}{}_{XA}{}^B+\ft1{6}f_{iA}^Z\left[ f_{(X}^{iB}\vec J{}_{Z)}{}^V  \cdot
\vec A_V +f_{Y}^{iB}\vec A_{(Z}\cdot \vec J{}_{X)}{}^Y-f_{Y}^{iB}\vec
A_V\cdot J{}_{(Z}{}^V\times \vec J{}_{X)}{}^Y \right] ,
\end{equation}
where $\omega^{\rm Op}{}_{XA}{}^B$ is the $\Gl(n_H,\mathbb{H})$
connection corresponding to the Oproiu connection (\ref{exprOp}) and the
$\SU(2)$ connection (\ref{vecomegaOp}).

As we have already explained quite extensively, there is a family of
possible connections on a quaternionic manifold, related to each other by
$\xi$-transformations, which for the $\Gl(n_H,\mathbb{H})$ connection is
given by (\ref{tildeomegaAB}). The transformation (\ref{specialxi}) gives
the drastic simplification
\begin{equation}
\omega_{XA}{}^B= \widehat{\omega}_{XA}{}^B. \label{eq:su2-connection}
\end{equation}
While in quaternionic manifolds this is just one possible choice, we have
seen in Sect.~\ref{ss:proposalmetric} that in quaternionic-K{\"a}hler
manifolds this $\xi $-choice is imposed.

%%%%%%%%%%%%%%%%%%%%%%%%%%%%%%%%%%%%%%%%%%%
\section{Curvatures}\label{ss:curvatures}
%%%%%%%%%%%%%%%%%%%%%%%%%%%%%%%%%%%%%%%%%%%

In the `large space' many components of the curvatures are zero due to
the existence of the homothetic Killing vector field and the $\SU(2)$
isometries,~(\ref{homothetic}) and~(\ref{DkSU2}), yielding
(\ref{konRis0}):
\begin{equation}
k^{\widehat{X}}\widehat{R}_{\widehat X \widehat Y \widehat Z}{}^{\widehat
W}= \vec{k}{}^{\widehat{X}}\widehat{R}_{\widehat X \widehat Y \widehat
Z}{}^{\widehat W}=0 .
\end{equation}
This (together with the cyclicity properties of the curvatures) shows
that the only possible non-zero components of the curvature of the
conformal hypercomplex manifolds are $\widehat{R}_{XYZ}{}^{\widehat W}$.
For the Ricci tensor, this implies that the only nonvanishing components
are along the quaternionic directions:
\begin{equation}
  \widehat{R}_{XY}=\widehat{R}_{ZXY}{}^Z,
 \label{RiccionlyXY}
\end{equation}
and this is antisymmetric as in general for hypercomplex manifolds. The
other part of the curvature, as shown in Table~\ref{tbl:CurvQuatlikeMan},
is the Weyl part. The latter is generally determined by a traceless
tensor, see~(\ref{WfromRW}). But for hypercomplex manifolds there is also
a generically non-traceless tensor
$\widehat{W}_{\widehat{A}\widehat{B}\widehat{C}}{}^{\widehat{D}}$, whose
trace part determines the antisymmetric part of the Ricci tensor, and
whose traceless part is $\widehat{\cal
W}_{\widehat{A}\widehat{B}\widehat{C}}{}^{\widehat{D}}$. The vanishing of
all the curvature components mentioned above implies that the
non-vanishing parts of
$\widehat{W}_{\widehat{A}\widehat{B}\widehat{C}}{}^{\widehat{D}}$ are
only $\widehat{W}_{ABC}{}^{\widehat{D}}$, i.e. $\widehat{W}_{ABC}{}^D$
and $\widehat{W}_{ABC}{}^i$. The latter will not be important for the
reduction to the small space. Note, from (\ref{calWW}), that the
traceless part,
$\widehat{\mathcal{W}}_{\widehat{A}\widehat{B}\widehat{C}}{}^{\widehat{D}}$,
has as non-zero components
\begin{eqnarray}
&&\widehat{\mathcal{W}}_{ABC}{}^D   =\widehat{W}_{ABC}{}^D
-\frac{3}{2(n_H+2)}\delta ^D_{(A}\widehat{W}_{BC)E}{}^E, \nonumber\\
&&\widehat{\mathcal{W}}_{iAB}{}^j =
\widehat{\mathcal{W}}_{AiB}{}^j=\widehat{\mathcal{W}}_{ABi}{}^j=
-\frac{1}{2(n_H+2)}\delta ^j_i\widehat{W}_{ABE}{}^E, \qquad
 \widehat{\mathcal{W}}_{ABC}{}^i   =\widehat{W}_{ABC}{}^i .
 \label{hatcalW}
\end{eqnarray}
This implies that the Weyl tensor of the hypercomplex space,
$\widehat{{\cal W}}$, depends also on the trace $\widehat{W}_{ABE}{}^E$.

We now start the reduction to the small space. We use the $\xi$-choice
(\ref{specialchoiceOmega}), which gives the easiest formulae for the map,
as explained above. The formulae for other $\xi $-choices follow from
(\ref{xicurvature})--(\ref{transfoxieta}). An expression for the $\SU(2)$
curvature can be found by considering (\ref{RhJA}). With
(\ref{specialchoiceOmega}), this implies
\begin{equation}\label{RisJ}
\vec{\mathcal{R}}_{XY}=-\ft12 \vec{J}_{[X}{}^Zh_{Y]Z} ,
\end{equation}
which is the equivalence between the $\SU(2)$ curvature and the
quaternionic two-forms in the quaternionic-K{\"a}hler case. One can derive
the curvature components $\widehat{R}_{XYZ}{}^W$ of the hypercomplex
space as a function of the curvature $R_{XYZ}{}^W$ of $\Gamma _{XY}{}^Z$
by a long but straightforward calculation. We obtain
\begin{eqnarray}
  \widehat R_{XYZ}{}^W =   R_{XYZ}{}^W+\ft12\delta _{[X}^Wh_{Y]Z}
  -\ft12\vec J_Z{}^{W}\cdot \vec J_{[X}{}^{V}h_{Y]V}
  -\ft12\vec J_Z{}^{V}\cdot \vec J_{[X}{}^{W}h_{Y]V}.
 \label{hatRRRRicS}
\end{eqnarray}
Taking the trace of this expression gives the Ricci tensor
\begin{equation}
  \widehat R_{XY}=  R_{XY}+\frac{2n_H+1}2h_{XY}
  +\frac12\vec{J}_X{}^Z\cdot\vec{J}_Y{}^Wh_{ZW}.
 \label{decompricci}
\end{equation}
The left hand side is antisymmetric, which determines the antisymmetric
part of the quaternionic Ricci tensor $R_{XY}$,
\begin{equation}
\widehat{R}_{XY}=R_{[XY]}.
\end{equation}
As the $\mathbb{R}$ curvature is, up to a sign, equal to this
antisymmetric part, see~(\ref{Ricas}), we find that the $\mathbb{R}$
curvatures are the same for the large and for the small space. As
explained in Sect.~\ref{ss:hatxi}, they can both be transformed to zero,
using a $\widehat{\xi }$-transformation in the large space and a $\xi
$-transformation with $\xi _X=\widehat{\xi }_X$ in the small space.

The symmetric part of~(\ref{decompricci}) determines the symmetric part
of the Ricci tensor for the quaternionic manifold. Using this
in~(\ref{defB}) gives
\begin{equation}
  B_{XY}=\frac1{4(n_H+1)}\widehat{R}_{XY}-\frac14h_{XY}.
 \label{BisRplush}
\end{equation}
Then we see that the extra terms in (\ref{hatRRRRicS}) constitute the
Ricci part of the curvature using the symmetric part of $B$:
\begin{equation}
  R^{\rm Ric}_{{\rm symm}\,XYZ}{}^W=-\ft12\delta _{[X}^Wh_{Y]Z}
  +\ft12\vec J_Z{}^{W}\cdot \vec J_{[X}{}^{V}h_{Y]V}
  +\ft12\vec J_Z{}^{V}\cdot \vec J_{[X}{}^{W}h_{Y]V}.
 \label{RRicsymm}
\end{equation}
We can then identify
\begin{eqnarray}
  \widehat R_{XYZ}{}^W =   \left( R- R^{\rm Ric}_{\rm symm}\right) _{XYZ}{}^W
  = \left( R^{({\rm
  W})}+R^{\rm Ric}_{\rm antis}\right)_{XYZ}{}^W.
 \label{hatRRRic}
\end{eqnarray}
For the full hypercomplex manifold, we have to use $r=n_H+1$
in~(\ref{defB}) (with only the antisymmetric part), and thus
\begin{equation}
  \widehat B_{XY}=\frac{1}{4(n_H+2)}R_{[XY]}\ \Rightarrow\
   \widehat{R}^{\rm Ric}_{\rm antis}{}_{XYZ}{}^W= \frac{n_H+1}{n_H+2}
  R^{\rm Ric}_{\rm antis}{}_{XYZ}{}^W.
 \label{BhatB}
\end{equation}

The Weyl parts of the curvatures are determined by the traceless
$\widehat{{\cal W}}$ and ${\cal W}$ tensors, but, as mentioned above
in~(\ref{hatcalW}), in the hypercomplex space this is dependent on the
trace $\widehat{W}_{ABC}{}^C$. We can extract the Weyl part of the
curvature of the quaternionic space from~(\ref{hatRRRic}) and find
\begin{eqnarray}\label{hatWiscW}
\widehat{W}_{ABC}{}^D=\mathcal{W}_{ABC}{}^D+\frac3{2(n_H+1)}\delta^D_{(A}\widehat{W}_{BC)E}{}^E.
\end{eqnarray}
Hence, $\mathcal{W}_{ABC}{}^D$ is the traceless part of
$\widehat{W}_{ABC}{}^D$.

Therefore, the main results are:
\begin{enumerate}
  \item The antisymmetric Ricci tensor of the hypercomplex manifold is the same
   as the antisymmetric Ricci tensor of the quaternionic manifold.
  \item The symmetric Ricci part of the quaternionic manifold
   is a universal expression in terms
  of the candidate metric $h$ (the same as for $\mathbb{H}P^{n_H}$).
  \item The traceless $\mathcal{W}$ tensor of the quaternionic space is the
  traceless part of the $\widehat{W}$ tensor of the hypercomplex space.
\end{enumerate}
As mentioned, we derived everything for a special choice of $\xi $.
However, we have seen in Sect.~\ref{ss:xi} that the Weyl tensor is not
changed by a $\xi $-transformation. Hence, the last conclusion is valid
for any $\xi $. Symbolically, we can represent the dependence of parts of
the curvature tensors on basic tensors as follows:
\begin{equation}
  \begin{array}{ccccccc}
    \widehat{R} & = &   &   & \widehat{R}^{\rm Ric}_{\rm antis} & + & \widehat{R}^{(\rm W)} \\
            &   &   &   & \uparrow & \nearrow & \uparrow \\
     &   & h_{XY} &   & \widehat{W}_{ABC}{}^C &   & \mathcal{W}_{ABC}{}^D \\
      &   & \downarrow &   & \downarrow &   & \downarrow \\
    R & = & R^{\rm Ric}_{\rm symm}  & + & R^{\rm Ric}_{\rm antis}  & + & R^{(\rm W)} \\
  \end{array}
 \label{schemCurvDecomp}
\end{equation}

\bigskip

For the mapping between a hyper-K{\"a}hler manifold and a quaternionic-K{\"a}hler
manifold, there is no antisymmetric part of the Ricci tensor, and hence
also $\widehat{W}_{ABC}{}^D$ is traceless. Hence, the relation
(\ref{hatWiscW}) reduces to
\begin{equation}
  \widehat{W}_{ABC}{}^D={\cal W}_{ABC}{}^D.
 \label{hatWisW}
\end{equation}
It implies that the hyper-K{\"a}hler curvature components along the
quaternionic directions are the Weyl part of the quaternionic-K{\"a}hler
curvature
\begin{equation}
  \widehat{R}_{XYZ}{}^W= R^{(\rm W)}{}_{XYZ}{}^W.
 \label{hatRisRW}
\end{equation}
The Ricci part of this quaternionic-K{\"a}hler curvature is the same
expression as for $\mathbb{H}P^{n_H}$. In fact, we find
\begin{equation}
  B_{XY}=-\ft14h_{XY}.
 \label{Bsymm}
\end{equation}
The metric of the small space inherited from the large space is
$g_{XY}=z^0h_{XY}$, see~(\ref{QK}). Comparing~(\ref{Bsymm}) with the
relation~(\ref{BinqK}), using this metric implies
\begin{equation}
  \nu =-\frac{1}{\Red{z^0}} .
 \label{valuenu}
\end{equation}
In the context of supergravity, the value of $\Red{z^0}$ determines the
normalization of the Einstein term and is fixed to $z^0=\kappa ^{-2}$,
where $\kappa $ is the gravitational coupling constant.

Finally, note that for the 1-dimensional case, $n_H=1$, we had restricted
the definition of quaternionic manifolds with special requirements in
Appendix~B.4 of~\cite{Bergshoeff:2002qk}, as was also done in the
mathematical literature~\cite{Alekseevsky1975}.  Here we find that these
relations are automatically fulfilled in the embedded quaternionic
manifolds. Hence, they are unavoidable in a supergravity context.
%%%%%%%%%%%%%%%%%%%%%%%%%%%%%%%%%%%%%%%%%%%%%%%
\section{Reduction of the Symmetries}\label{ss:symmetries}
%%%%%%%%%%%%%%%%%%%%%%%%%%%%%%%%%%%%%%%%%%%%%%%

\subsection{Symmetries and moment maps}

The main part of this subsection is a summary of the results on
symmetries given in~\cite{Bergshoeff:2002qk}. In the case of manifolds
where there is no (good) metric, the question of defining symmetries
needs some careful consideration. In general, we consider transformations
$\delta q^X=\Blue{k^X_I(q)}\Lambda ^I$, where the index $I$ runs over the
set of possible symmetries. We will define when these are called
`symmetries', and then we define `quaternionic symmetries'.

A set of vector fields $k_I$ are symmetry generators if the following
condition on the connection and curvature are met:
\begin{equation}\label{RkI}
\mathfrak{D}_X\mathfrak{D}_Yk_I^Z=R_{XWY}{}^Zk_I^W .
\end{equation}
For Riemannian manifolds with a metric $g_{XY}$, this is just the
integrability condition that follows from the Killing equation
$\mathfrak{D}_{(X}k_{Y)I}=0$, where $k_{XI}=g_{XY}k^Y_I$. On the other
hand, (\ref{RkI}) does not imply a Killing equation as it is independent
of a choice of metric. E.g. the conformal Killing vectors, which do not
satisfy the Killing equation, satisfy (\ref{RkI}). However, it is a
sufficient condition to define `symmetries' if there is no metric
available.

The `physical' origin for this condition is the following. For
simplicity, consider the equations of motion for a rigid non-linear sigma
model:
\begin{equation}
\Box q^X=\partial_\mu \partial^\mu q^X+\partial_\mu q^Y \partial^\mu q^Z
\Gamma_{YZ}{}^X=0\,.
\end{equation}
Consider the transformation $\delta q^X=k_I^X\Lambda ^I$ on this field
equation:
\begin{equation}
\delta \Box q^X=\partial_Yk_I^X\Lambda ^I \Box q^Y+\partial_\mu q^Y
\partial^\mu q^Z \mathfrak{D}_Z\mathfrak{D}_Y k_I^X\Lambda ^I-\partial_\mu
q^Y\partial^\mu q^Zk_I^V\Lambda ^IR_{ZVY}{}^X\,.
\end{equation}
Hence, the set of equations of motion is left invariant iff the defining
condition (\ref{RkI}) for a `symmetry' is satisfied .

Symmetry generators are quaternionic if the vector fields normalize the
complex structures, which means
\begin{equation}\label{kInormalJ}
\mathcal{L}_{\Blue{k_I}} \vec J_X{}^Y=\vec{r}_I\times \vec J_X{}^Y,
\end{equation}
for some 3-vectors $\vec{r}_I$. Using the $\SU(2)$ connections, this
defines moment maps $ \nu \Maroon{\vec P_I}(q)$ by\footnote{We define
here $\nu \vec{P}_I$, where $\nu $ is a number. For quaternionic-K{\"a}hler
manifolds, this number is defined by (\ref{BinqK}), while it is not
specified in quaternionic manifolds. This normalization is convenient for
comparison with other papers, and, as will be shown below, because the
formulae are then applicable to hyper-K{\"a}hler manifolds upon setting $\nu
=0$.}
\begin{equation}
\nu \Maroon{\vec{P}_I}\equiv -\ft12   \vec{r}_I-
\Blue{k_I^X}\OliveGreen{\vec{\omega }_X}.
 \label{defP}
\end{equation}
See \cite{Bergshoeff:2002qk} for more information, where it was also
shown that this leads to a decomposition of the derivatives of the
$k_I^X$ as
\begin{equation}
 \mathfrak{D}_X\Blue{k^Y_I}=\nu \vec J_X{}^Y\cdot \Maroon{\vec P_I} +
L_X{}^Y{}_A{}^B \Red{t_{IB}{}^A}.
 \label{DXkIYdecomp}
\end{equation}
The so-called \textit{moment maps} $\vec P_I(q)$ describe the $\SU(2)$
content of the symmetry and the $t_{IB}{}^A(q)$ describe the
$\Gl(r,\mathbb{H})$ content. The $L_X{}^Y{}_A{}^B$ symbols are defined as
in (\ref{defL}). We can extract $\nu \vec P_I$ from~(\ref{DXkIYdecomp})
as
\begin{equation}\label{PeqJDkI}
4r\nu \vec P_I=-\vec J_X{}^Y \mathfrak{D}_Yk_I^X.
\end{equation}
If we project the curvature tensors along the symmetry vectors, we get
the following relations:\footnote{The first of these equations holds for
any choice of $\SU(2)$-connection $\vec{\omega} _X$, if a corresponding
$\vec{P}_I$ is defined by (\ref{defP}).\label{fn:arbitromega}}
\begin{equation}
 \OliveGreen{\vec {\cal R}_{XY}} \Blue{k^Y_I}= -\nu \mathfrak{D}_X
\Maroon{\vec P_I}, \qquad
  \Red{{\cal R}_{XYB}{}^A}\Blue{k^Y_I}= \mathfrak{D}_X \Red{t_{IB}{}^A}.
 \label{Rkin2} %[B.84]
\end{equation}
In supersymmetric models, the condition of quaternionic symmetries is
necessary for the invariance of the full field equations including the
fermions.

The vector fields generate a Lie-algebra:
\begin{equation}\label{G-alg} %[2.95]
2k_{[I}^Y\partial_Yk_{J]}^X=-f_{IJ}{}^Kk_K^X.
\end{equation}
The vector at the left-hand side of this equation for any $[IJ]$
satisfies the two conditions mentioned above, (\ref{RkI}) and
(\ref{kInormalJ}), and this equation is thus a statement of completeness
of the set of quaternionic symmetries. It leads to an important property
of the moment maps, which is the `equivariance relation'
\begin{equation}
  -2\nu ^2 \vec P_I \times \vec P_J + \vec {\cal
  R} _{YW} k_I^Y k_J^W-\nu f_{IJ}{}^K \vec P_K=0.
 \label{equivariance} %[B.87]
\end{equation}

The absence of the $\SU(2)$ curvature parts for hypercomplex (and
hyper-K{\"a}hler) manifolds implies that we can take $\nu =0$ for these
manifolds. As also $\vec \omega _X=0$ in that case, we see
from~(\ref{kInormalJ}) and (\ref{defP}) that the Lie derivative along the
symmetry generators of the complex structures vanishes. The symmetries
are then called triholomorphic.

Though $\nu =0$ implies that the moment maps do not appear for
hyper-K{\"a}hler manifolds in the above relations, moment maps still appear
in the Lagrangian for sigma models on these manifolds. They are defined
by a relation consistent with the equations for quaternionic manifold
(\ref{Rkin2}) and (\ref{RlambdaJ}) for any $\nu $:
\begin{equation}
\vec J_{XY} \Blue{k^Y_I}= -2 \partial _X \Maroon{\vec P_I}.
 \label{defPHK}
\end{equation}
Notice that we have used the existence of a metric here, so this relation
only defines moment maps for hyper-K{\"a}hler manifolds. They should still
satisfy the $\nu =0$ limit of the equivariance
relation~(\ref{equivariance}):
\begin{equation}
  \Red{k_I^X} \vec J_{XY} \Red{k_J^Y}= 2f_{IJ}{}^K \Maroon{\vec P_K}.
 \label{equivarianceHK}
\end{equation}

We can summarize the various cases as follows:
\begin{itemize}

\item Hypercomplex : Triholomorphic symmetries must satisfy (\ref{RkI}) and
\begin{equation}
\mathcal{L}_{\Blue{k_I}} \vec J_X{}^Y=0.
\end{equation}
There exist no moment maps.

\item Hyper-K{\"a}hler : conditions for the triholomorphic symmetries as for
hypercomplex. The condition for a triholomorphic isometry can be
translated to the existence of a triplet of moment maps (\ref{defPHK}).

\item Quaternionic-K{\"a}hler : in this case, {\it any}
isometry normalizes the quaternionic structure
\cite{Wit:2001bk,Galicki:1987ja}. Indeed, if we define the moment maps
for any Killing vector $k_I$ as in (\ref{PeqJDkI}), and take a covariant
derivative, the first equation of (\ref{Rkin2}) follows from (\ref{RkI}),
the covariant constancy of $\vec J$ and the decomposition
(\ref{RdecompJ}). The proportionality of the $\SU(2)$ curvature and
complex structure, (\ref{RlambdaJ}), implies that this can be written as
a covariant version of (\ref{defPHK}). Then the Killing equation implies
\begin{equation}
 \vec{J}_X{}^Z\mathfrak{D}_Zk_I^Y  =  -  \vec{J}_X{}^Z\mathfrak{D}^Yk_{IZ}=
2\mathfrak{D}^Y\mathfrak{D}_X\vec{P}_I, \qquad
 -\mathfrak{D}_Xk_I^Z\vec{J}_Z{}^Y =  -2 \mathfrak{D}_X\mathfrak{D}_Y\vec{P}_I.
\end{equation}
It is straightforward to show that (\ref{kInormalJ}) is satisfied, as
this reduces to the sum of these two expressions, see
\cite[(B.80)]{Bergshoeff:2002qk}. Thus, all isometries on a
quaternionic-K{\"a}hler manifold are quaternionic.

\item Quaternionic manifolds: Not all symmetries are quaternionic, i.e.
satisfying (\ref{kInormalJ}). However, when they are, moment maps exist
and are given by (\ref{PeqJDkI}).
%\begin{equation}
%4r\nu \vec P_I=-\vec J_X{}^Y \mathfrak{D}_Yk_I^X.
%\end{equation}
Not all $\xi $-transformations preserve the symmetries. Indeed, the
condition (\ref{RkI}) is not invariant under general $\xi
$-transformations. Writing (\ref{RkI}) with $\xi $-transformed
connections modifies it proportional to
\begin{equation}
  S_{XY}^{ZU}\left[ \xi _W\mathfrak{D}_U k_I ^W +k_I^W\mathfrak{D}_W\xi _U\right].
 \label{xitransfsymm}
\end{equation}
We thus conclude that a symmetry is a symmetry after a
$\xi$-transformation iff
\begin{equation}
  {\cal L}_{k_I} \xi_X=k_I^Y\partial _Y\xi _X +\xi _Y\partial _X k_I ^Y =0.
 \label{LKxi0}
\end{equation}
The condition for a symmetry to be quaternionic, (\ref{kInormalJ}), is
invariant if $\vec{r}_I$ is invariant. We can obtain the transformation
of the moment map from  (\ref{PeqJDkI}). This leads, using
(\ref{SJcontr}), to
\begin{equation}
  \nu \widetilde {\vec{P}}_I = \nu \vec{P}_I -\vec{J}_X{}^Y k_I^X\xi _Y.
 \label{xitransfP}
\end{equation}
This implies indeed that $\vec{r}_I$ is an invariant for
$\xi$-transformations.

\end{itemize}

\subsection{Conformal hypercomplex}

As explained in Sect.~\ref{CHC}, hypercomplex and hyper-K{\"a}hler manifolds
with a homothetic Killing vector (\ref{homothetic}) have additional
properties. First of all, as follows from (\ref{konRis0}), both
dilatations and $\SU(2)$ transformations define symmetries in the sense
of (\ref{RkI}). Notice that, when there is a metric, these symmetries are
not always isometries of the metric. Precisely, the dilatations are of
this type, as they satisfy the conformal Killing equation instead of the
Killing equation. The dilatation symmetry is triholomorphic. The $\SU(2)$
transformations are `quaternionic symmetries', with $\vec{r}_I=-\unity_3
$ as matrix in the 3 components of vectors and three values of $I$.

For the remainder of this section, we concentrate on possible additional
symmetries, other than the dilatations and $\SU(2)$ transformations. We
require such symmetries to commute with the dilatational symmetry. This
condition is expressed as
\begin{equation}\label{kIcommk}
k^{\hat{Y}}\widehat{\mathfrak{D}}_{\hat{Y}}\widehat{k}_I^{\hat{X}}=\ft32\widehat{k}_I^{\hat{X}}.
\end{equation}
For triholomorphic symmetries, this implies~\cite{Bergshoeff:2002qk}
\begin{equation}
  {\vec k}^{\hat{Y}}\widehat{\mathfrak{D}}_{\hat{Y}}\widehat{k}_I^{\hat{X}}=
  \ft12\widehat{k}_I^{\hat{Y}} \widehat{\vec{J}}_{\hat{Y}}{}^{\hat{X}},
 \label{veckkI}
\end{equation}
which expresses the fact that they also commute with the $\SU(2)$
transformations.

For conformal hyper-K{\"a}hler manifolds, one can deduce more identities. In
fact, contracting (\ref{kIcommk}) with $k_{\hat{X}}$, one finds
\begin{equation}
k^{\hat{X}}\widehat{g}_{\hat{X}\hat{Y}}\widehat{k}^{\hat{Y}}_I=0.
\end{equation}
Moreover, the consistency with the conformal symmetry allows us to
integrate (\ref{defPHK}) in terms of the moment maps \cite{deWit:1999fp},
such that
\begin{equation}
  -6\widehat{\vec P}_I=k^{\hat{X}}\widehat{\vec J}_{\hat{X}\hat{Y}}  \widehat{k}_I^{\hat{Y}}
  = -\ft 23k^{\hat{X}}k^{\hat{Z}}\widehat{\vec J}_{\hat{Z}}{}^{\hat{Y}}
  \widehat{\covder}_{\hat{Y}} \widehat{k}_{I\hat{X}}.
\label{momentmapconfhK}
\end{equation}
A possible integration constant in $\widehat{\vec P}_I$ (a
Fayet-Iliopoulos term) is not allowed in conformally invariant actions,
see \cite{Wit:2001bk}.

\subsection{The map of the symmetries}
In this section, we will discuss the reduction of triholomorphic
symmetries of the large space to quaternionic symmetries of the small
space. We also illustrate how triholomorphic isometries and moment maps
on hyper-K{\"a}hler spaces descend to quaternionic isometries on
quaternionic-K{\"a}hler manifolds.

\subsubsection{Conformal hypercomplex and quaternionic symmetries}

We now show that the components of the higher-dimensional symmetry
generators lying along the quaternionic directions are bona-fide symmetry
generators of the quaternionic space.

First of all, for any triholomorphic symmetry on a conformal hypercomplex
manifold, the reduction of the Eqs.~(\ref{kIcommk}) and (\ref{veckkI})
leads to the $z^0$ and $z^\alpha$ dependence of the higher-dimensional
symmetry generators. The form of the symmetry vectors is as follows:
\begin{equation}
  \widehat{k}^0_I =z^0 V_I(q),\qquad \widehat{k}^\alpha _I=\vec{k}^\alpha \cdot \vec{Q}_I,
  \qquad
  \widehat{k}^X_I=k_I^X(q),
 \label{hatkvolledig}
\end{equation}
where all $z^0$-dependence is explicitly indicated, and
\begin{equation}
   \left( \partial _\alpha
  +\vm\alpha \times \right)\vec Q_I=0\,,
 \label{dalQI}
\end{equation}
while $V_I$ and $k_I^X$ are independent of $z^\alpha $.

This can now be used to study the normalization of the complex structures
(\ref{kInormalJ}) on the quaternionic space.  Using the triholomorphicity
of ${\widehat{k}}_I^{\hat{X}}$, together with (\ref{allhatJ}),
(\ref{dXJ}) and (\ref{hatkvolledig}), it is easy to show that the vector
field with components $k_I^X=\widehat{k}_I^X$ normalizes the quaternionic
complex structures,
\begin{equation}\label{defnormcomplstruct}
  {\mathcal L}_{{k}_I} \vec J_X{}^Y =\vec{Q}_I\times
   \vec J_X{}^Y .
\end{equation}
This identifies $\vec{Q}_I$ with the vector $\vec{r}_I$ in
(\ref{kInormalJ}). The  ${}_X{}^0$ component of the normalization
condition gives
\begin{equation}
  -\vec{J}_X{}^Y\partial _YV_I+  \left(\partial
  _X-\vec{A}_X\times \right)\left(\vec{A}_Y k_I^Y-\vec{Q}_I\right)
  -2k_I^Y\big[ \vec{R}(-\ft12\vec{A})\big]_{YX} =0.
 \label{X0triholcond}
\end{equation}
The relations (\ref{Rkin2}), using footnote \ref{fn:arbitromega} with
$\vec{\omega} _X=-\ft12\vec{A}_X$,  implies that the last two terms
cancel, and we find
\begin{equation}
  \partial _X V_I=0.
 \label{dXV0}
\end{equation}
The constant contribution in $V_I$ can be set to zero. Indeed, this just
reflects that the dilatation vector (\ref{dilat-coord}) is a
triholomorphic symmetry vector. We can indicate this as the symmetry with
label $I=0$:
\begin{equation}
  \widehat{k}_0^{\hat{X}}=3z^0\delta_0^{\hat{X}}.
 \label{0symm}
\end{equation}
We can thus subtract this from all the other symmetries, redefining them
as
\begin{equation}
  \widehat{k}_I^{\prime
  \hat{X}}=\widehat{k}_I^{\hat{X}}-3z^0\delta_0^{\hat{X}}V_I.
 \label{redefkI}
\end{equation}
This redefined symmetry vector is of the form (\ref{hatkvolledig}) with
$V_I=0$.

One may now verify explicitly that the condition for a symmetry in the
large space reduces to the condition that $k_I^X$ is a symmetry in the
small space. Excluding the dilatation symmetry, the result is thus that
any triholomorphic symmetry that commutes with the dilatations is of the
form
\begin{equation}
  \widehat{k}^0_I =0,\qquad \widehat{k}^\alpha _I=\vec{k}^\alpha \cdot \vec{r}_I,
  \qquad
  \widehat{k}^X_I=k_I^X(q),
 \label{hatkconclusion}
\end{equation}
where $k_I^X$ is a quaternionic symmetry of the small space, satisfying
(\ref{kInormalJ}). With this formula, we can thus also uplift any
quaternionic symmetry of the small space to a triholomorphic symmetry of
the large space preserving dilatations.

Further, we consider the algebra in the large space:
\begin{equation}
  2 \hat k_{[I}^{\widehat Y} \partial_{\widehat Y} \hat k_{J]}^{\widehat X} = - f_{IJ}{}^K \hat k_K^{\widehat X}.
\end{equation}
Reducing this equation for the different values of $\hat{X}$ leads to
\begin{eqnarray}
 X      &:& 2 k_{[I}^Y \partial_Y k_{J]}^X = - f_{IJ}{}^K k_K^X,\nonumber\\
 \alpha &:& - 2 \nu^2 \vec P_I \times \vec P_J + \vec {\mathcal R}_{YW} k_I^Y k_J^W
 - \nu f_{IJ}{}^K \vec P_K = 0, \nonumber\\
 0   &:&  f_{IJ}{}^K \widehat{k}_K^0=0 .
\end{eqnarray}
The first equation says that the algebra in the small space is the same
as the algebra in the large space, excluding dilatations. The second
equation is the equivariance condition (\ref{equivariance}). The third
one says that the dilatations do not appear in the right-hand side of
commutator relations.

Remark that all equations obtained in this section are invariant under
$\xi $-transformations satisfying (\ref{LKxi0}).

\subsubsection{Isometries on conformal hyper-K{\"a}hler and quaternionic-K{\"a}hler spaces}

When there is a metric, we find that the only non-trivial part of the
Killing equation in the large space is
\begin{equation}
  \widehat{\mathfrak D}_{(X} \hat k_{Y)I} =z^0 {\mathfrak D}_{(X}
  k_{Y)I}.
 \label{isometrymap}
\end{equation}
Hence, a triholomorphic symmetry preserving the dilatations is an
isometry if and only if it is an isometry of the quaternionic-K{\"a}hler
space.

For conformal hyper-K{\"a}hler manifolds, the moment map is defined in
(\ref{momentmapconfhK}). Using the decomposition of the symmetry vector
(\ref{hatkconclusion}) and (\ref{valuenu}), we find
\begin{equation}
  \widehat{\vec P}_I=-\ft16 k^{\hat{X}}\widehat{\vec J}_{\hat{X}}{}^{\hat{Y}}
  \widehat{k}_{I\hat{Y}}
  =-\ft12\vec{k}^{\hat{X}}\widehat{g}_{\hat{X}\hat{Y}}\widehat{k}_I^{\hat{Y}}
  =\ft12z^0 \left( \vec{r}_I - \vec{A}_Xk^X_I\right)= \vec{P}_I-z^0\left(
  2\vec{\omega }_X+\vec{A}_X\right)k^X_I.
 \label{Pmap}
\end{equation}
As in the quaternionic-K{\"a}hler spaces we have (\ref{vecomegaisA}), the
last term vanishes and the moment map in the hyper-K{\"a}hler space is equal
to the moment map in the quaternionic-K{\"a}hler space.

\subsubsection{The conformal hyper-K{\"a}hler manifold of quaternionic dimension~1.}

In quaternionic manifolds, the moment map is completely determined, see
(\ref{PeqJDkI}). Fayet-Iliopoulos (FI) terms are in general undetermined
constants in the moment maps (see the review~\cite{VanProeyen:2004xt}).
Hence, they are not present in quaternionic-K{\"a}hler manifolds, except for
the `trivial situation' $n_H=0$. The latter corresponds to a large space
of quaternionic dimension~1, which is hyper-K{\"a}hler as the metric is given
in the first line of (\ref{metricinverse}):
\begin{equation}
  \widehat g_{00}=-\frac1{z^0},\qquad \widehat{g}_{0\alpha
}=0,\qquad  \widehat g_{\alpha \beta}=-z^0\vm\alpha \cdot \vm\beta.
 \label{metricd1}
\end{equation}

This metric has $\SU(2)\times \SU(2)$ isometries. The first factor is
generated by $\vec{k}^\alpha $, but these are not triholomorphic.
However, there is a commuting set of $\SU(2)$ Killing vectors, $k^\alpha
_{-I}$, with $I=1,2,3$, and the minus sign indicating the second $\SU(2)$
factor in the holonomy group, different from $\vec{k}_\alpha $. The
generators are thus
\begin{equation}
  \widehat{k}^0_I=0,\qquad \widehat{k}^\alpha _I=k^\alpha_{-I}.
 \label{generatorsextraSU2}
\end{equation}
These are triholomorphic with respect to the complex structures defined
by the first Killing vectors:
\begin{equation}
   \begin{array}{ll}
    \widehat{\vec J}_0{}^0=0 ,\qquad \qquad & \widehat{\vec J}_\alpha{}^0=
    -z^0\vm\alpha  ,
    \\
    \widehat{\vec J}_0{}^\beta=\frac{1}{z^0}\vec k^\beta   ,& \widehat{\vec J}_\alpha{}^\beta=
     \vec k^\beta \times\vm\alpha .
  \end{array}
 \label{hatJd1}
\end{equation}
This leads to the moment maps
\begin{equation}
  -2\nu \vec{P}_I=\vec{r}_I= \vm\alpha k^\alpha_{-I}.
 \label{PSU2n1}
\end{equation}
When a gauge fixing is taken, i.e. the $z^\alpha $ are fixed to a value,
then this leads to some constants. These are the $\SU(2)$ FI terms.

\section{Summary and Discussion}\label{ss:discussion}

\begin{figure}[p]
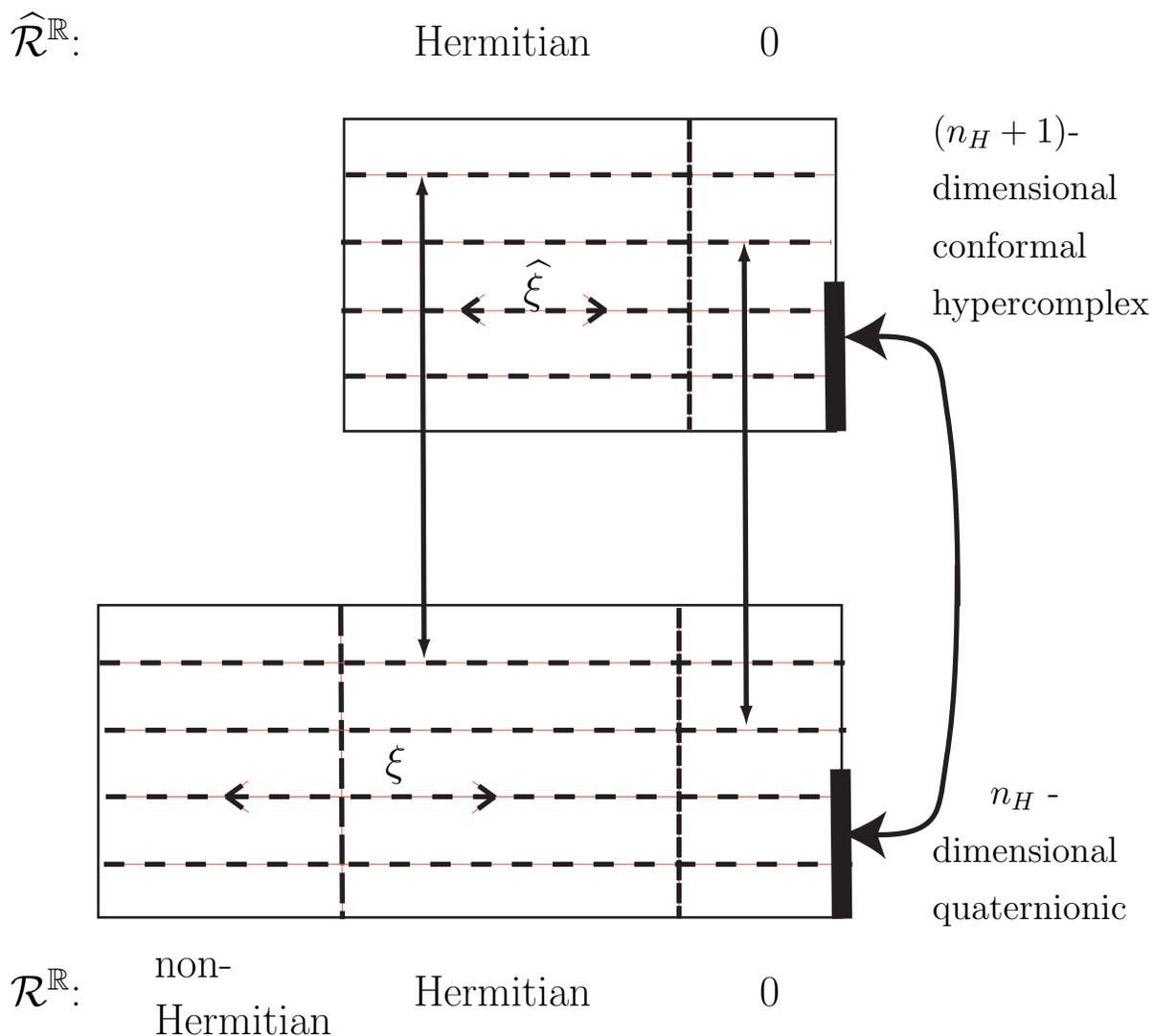

\unitlength=0.8mm \caption{\it The map schematically. The two blocks
represent the families of large (upper block) and small spaces (lower
block), where the horizontal lines indicate how they are related by
$\widehat{\xi }$, resp. $\xi $, transformations. They connect
parametrizations of the same manifold with different complex structures
for hypercomplex manifolds, and different affine and $\SU(2)$ connections
for the quaternionic manifolds. At the right, the spaces have no
$\mathbb{R}$ curvature, and part of these are hyper-K{\"a}hler, resp.
quaternionic-K{\"a}hler. The latter two classes are indicated by the thick
lines. The vertical arrows represent the map described in this paper,
connecting the manifolds with similar parametrizations. They are a
representation of the map between the horizontal lines, which is the map
between hypercomplex and quaternionic manifolds. The thick arrow
indicates the map between hyper-K{\"a}hler and quaternionic-K{\"a}hler spaces.
 \label{fig:mapoverview}}
\begin{center}
\begin{picture}(150,190)(0,0)
 \ifpdf
 \put(-53,-46){
  \includegraphics{maphkctoqk.pdf}}
 \else
  \put(-10,20){\leavevmode
   \epsfxsize=120mm
 \epsfbox{maphkctoqk.eps}
  } \fi
\put(-25,5){{\Large ${\cal R}^{\mathbb{R}}$:}}
 \put(0,10){\Large non-}
 \put(0,0){\Large Hermitian}
 \put(45,5){\Large Hermitian}
 \put(105,5){\Large 0}
\put(-25,170){{\Large $\widehat{{\cal R}}{}^{\mathbb{R}}$:}}
 \put(45,170){\Large Hermitian}
 \put(105,170){\Large 0}
 \put(135,155){\large $(n_H+1)$-}
 \put(135,145){\large dimensional}
 \put(135,135){\large conformal}
 \put(135,125){\large hypercomplex}
 \put(145,40){\large $n_H$ -}
 \put(135,30){\large dimensional}
 \put(135,20){\large quaternionic}
\put(64,127){\Large $\widehat{\xi }$ }
 \put(40,45){\Large $\xi $}
\end{picture}
\end{center}
\end{figure}

For the convenience of the reader we summarize the main results of this
paper. The map is schematically represented in
Fig.~\ref{fig:mapoverview}, which will be further explained in this
section.

\subsection{The map from hypercomplex manifolds}
The setup starts from a $4(n_H+1)$-dimensional space that is hypercomplex
and has a conformal symmetry. The latter is mathematically expressed as
the presence of a closed homothetic Killing vector
\begin{equation}
  { \widehat{\mathfrak{D}}}_{ \hat{Y}} k^{\hat X} \equiv \partial_{\hat Y}
k^{\hat X} + {\widehat{\Gamma}_{{\hat Y}{\hat Z}}}{}^{\hat X} k^{\hat Z}
= \ft32 \delta_{\hat Y}{}^{\hat X}.
 \label{s:homothetic}
\end{equation}
Observe that the coordinates and covariant derivatives in this large
space are indicated by hatted quantities. The vector $k$, which is the
generator of dilatations, yields also generators of $\SU(2)$
transformations:
\begin{equation}
  \vec k^{\widehat X} \equiv \ft13 {\widehat {\vec J}}_{\widehat Y}{}^{\widehat X}
k^{\widehat Y}.
 \label{s:defveck}
\end{equation}
We use coordinates in this large space that are adapted to these
quantities:
\begin{eqnarray}
&&q^{\widehat X}=\left\{z^0 ,\,y^p\right\} = \left\{ \Red{z^0,\,
z^\alpha},\, \Blue{q^X}\right\}, \qquad \alpha =1,2,3,\quad
X=1,\ldots ,4n_H,\nonumber\\
&&  k^{\widehat X}=3z^0 \delta _0^{\widehat{X}}, \qquad \vec k^0=\vec
k^X=0.
 \label{s:splitcoord}
\end{eqnarray}
This leads to the following form of the complex structures:
\begin{equation}
   \begin{array}{lll}
    \widehat{\vec J}_0{}^0=0 ,\qquad \qquad & \widehat{\vec J}_\alpha{}^0=
    -z^0\vm\alpha  ,
     & \widehat{\vec J}_X{}^0=z^0 \vec A_X,\\
    \widehat{\vec J}_0{}^\beta=\frac{1}{z^0}\vec k^\beta   ,& \widehat{\vec J}_\alpha{}^\beta=
     \vec k^\beta \times\vm\alpha ,\qquad&
    \widehat{\vec J}_X{}^\beta=\vec A_X \times \vec k^\beta
                               +\vec J_X{}^Z(\vec A_Z\cdot k^\beta)  ,    \\
    \widehat{\vec J}_0{}^Y=0 ,& \widehat{\vec J}_\alpha{}^Y=0 ,& \widehat{\vec J}_X{}^Y=\vec
    J_X{}^Y.
  \end{array}
 \label{s:allhatJ}
\end{equation}
Here, $\vec{k}^\alpha$ are $\SU(2)$ Killing vectors, and $\vm\alpha $ is
the inverse as a $3\times 3$ matrix. The result thus depends on $z^0$,
these $\SU(2)$ Killing vectors, complex structures on the small space
$\vec{J}_X{}^Y$ and a vector $\vec{A}_X$. The unique torsionless affine
connection is the Obata connection
\begin{equation}
 \Purple{\widehat{\Gamma}{}_{\hat{X}\hat{Y}}{}^{\hat{Z}}}=-\ft16
 \left( 2 \partial_{(\hat{X}}\widehat{\vec
J}_ {\hat{Y})}{}^{\hat{W}}+\widehat{\vec J}_{(\hat{X}}{}^{\hat{U}}\times
\partial_{|\hat{U}|} \widehat{\vec J}_{\hat{Y})}{}^{\hat{W}}\right) \cdot
\widehat{\vec J}_{\hat{W}}{}^{\hat{Z}}. \label{s:Obata}
\end{equation}
The condition that this connection preserves the complex structures
implies that
\begin{eqnarray}
&&\partial _0\vec{A} _X=0, \qquad \left( \partial _\alpha +\vm\alpha
\times \right) \vec A_X+\partial
_X\vm\alpha =0,\nonumber\\
&&\partial_0\vec J_X{}^Y=0,\qquad \left( \partial _\alpha +\vm\alpha
\times \right) \vec
J_X{}^Y=0, \nonumber\\
&&  \big[\vec R(-\ft12\vec A )\big]_{XY}\equiv -\partial _{[X}\vec
  A_{Y]}+ \ft12\vec A_X\times \vec A_Y=\ft12 h_{V[X} \vec J_{Y]}{}^V,
  \label{s:RhJA}
\end{eqnarray}
for some symmetric tensor $h_{XY}$. Furthermore it implies that the
$\vec{J}_X{}^Y$ are a quaternionic structure on the small space.

Hypercomplex manifolds have no $\SU(2)$ curvature, but can have a
non-vanishing  $\mathbb{R}$ curvature, which is Hermitian, i.e. it
commutes with the hypercomplex structure, see Appendix
\ref{app:projmatrices}. This $\mathbb{R}$ curvature is equal to the Ricci
tensor of these manifolds, which is antisymmetric,
\begin{equation}
  {\cal R}^{\mathbb{R}}_{\hat{X}\hat{Y}}=-R_{\hat{X}\hat{Y}}.
 \label{s:RU1isRicci}
\end{equation}

We found that in these conformal hypercomplex manifolds there is a
transformation $\widehat{\xi }_{\hat X}$ that preserves the hypercomplex
structure,
\begin{equation}
  (\widehat{\vec J}_{\widehat \xi})_{ \widehat X}{}^{\widehat Y}=\widehat{\vec{J}}_{\widehat X}{}^{\widehat Y}
   +\ft23\left[ \widehat {\vec{J}}_{\widehat X}{}^{\widehat Z}(\widehat \xi
  _{\widehat Z}k^{\widehat Y})-
  (\widehat \xi _{\widehat X}k^{\widehat Z})\widehat {\vec{J}}_{\widehat Z}{}^{\widehat
  Y}\right].
 \label{s:transfoJxi}
\end{equation}
The one-form with components $\widehat \xi _{\widehat X}$ satisfies
conditions implying, in the parametrization of (\ref{s:allhatJ}), that
the vector has only coordinates $\widehat{\xi}_X$. Furthermore this
vector can only depend on $q^X$ and is such that $\partial
_{[X}\widehat{\xi}_{Y]}$ is Hermitian. The transformation is in this
basis generated by
\begin{equation}
  \delta(\widehat{\xi }) \vec{A}_X = 2\vec{J}_X{}^Z\widehat{\xi }_Z,
 \label{s:hatxiA}
\end{equation}
while $\vec{k}^\alpha $ and $\vec{J}_X{}^Y$ are invariant. These
transformations can always be used to obtain $\widehat{\cal
R}^{\mathbb{R}}=0$. They are represented by the horizontal lines in the
upper part of Fig.~\ref{fig:mapoverview}.

Another invariance that is present in the conformal hypercomplex
manifolds is a \textit{local} $\SU(2)$, which acts as
\begin{equation}
  \delta_{\rm SU(2)} \vec A_X = \partial_X \vec{\ell}   + \vec{\ell} \times \vec
  A_X, \qquad\delta_{\rm SU(2)} \vec{J}_X{}^Y=\vec{\ell }\times\vec{J}_X{}^Y.
 \label{s:SO3quat}
\end{equation}

If the manifold admits a metric, and hence would be hyper-K{\"a}hler, it must
be equal to
\begin{equation}
  \widehat{g}_{\widehat{X}\widehat{Y}}=
2\widehat{\Gamma}_{\widehat{X}\widehat{Y}}{}^0.
 \label{s:candmetric}
\end{equation}
In the parametrization that we use, the full form is
\begin{eqnarray}\label{s:QK}
\rmd \widehat s^2  & = & -\frac{(\rmd z^0)^2}{z^0}
 +\Big\{ z^0h_{XY} (q)\rmd q^X \rmd q^Y
 \nonumber\\
&&\qquad \qquad + \widehat{g}_{\alpha\beta} [\rmd z^\alpha - \vec
A_X(z,q)\cdot \vec k^\alpha \rmd q^X][\rmd z^\beta - \vec A_Y(z,q)\cdot
\vec k^\beta \rmd q^Y]\Big\},
\end{eqnarray}
where
\begin{equation}
\widehat{g}_{\alpha \beta}=-\frac1{z^0}\vec{k}_{\alpha}\cdot
\vec{k}_\beta=-z^0\vm\alpha\cdot  \vm\beta, \qquad h_{XY}\equiv
\frac{1}{z^0} \widehat{g}_{XY}+ \vec A_X\cdot\vec A_Y.
\end{equation}
$h_{XY}$ is the quantity that appears in (\ref{s:RhJA}). These formulae
are very reminiscent of a Kaluza-Klein reduction on an $\SU(2)$ group
manifold.

\subsection{The quaternionic manifold}

We have proven that the $4n_H$-dimensional subspace described by the
$q^X$ is a quaternionic space. This means that we restrict the $4(n_H+1)$
dimensional space by 4 gauge choices for the dilatations and $\SU(2)$
transformations spanned by $\vec{k}^\alpha $. In the context of
superconformal tensor calculus, these are the transformations that are
present in the superconformal group and should be gauge-fixed by the
compensating hypermultiplet~\cite{deWit:1980gt,deWit:1985px}. The gauge
fixing of dilatations is done by fixing a value of $z^0$, which will set
the scale of the manifold, see below. On the other hand, $\SU(2)$ is
gauge-fixed by choosing a value for the $z^\alpha $ coordinates. For any
value of $z^\alpha $ we thus find a quaternionic manifold. These are
related by $\SU(2)$ transformations. This means that objects that depend
on $z^\alpha $ are gauge-dependent. Some intrinsic quantities, like the
affine connection and the metric on the small space, turn out to be
$z^\alpha $-independent.

The geometric quantities in the small space are not uniquely defined. In
particular, we can take different choices for the affine connection and
for the $\SU(2)$ connection $\vec{\omega }_X$, which is the gauge field
of the transformations (\ref{s:SO3quat}), restricted to the small space
\begin{equation}
  \delta_{\rm SU(2)} \vec{\omega }_X=-\ft12\partial _X\vec{\ell }
    +\vec{\ell }\times\vec{\omega}_X.
 \label{s:SO3omega}
\end{equation}
One particular choice is the \emph{Oproiu connection} \cite{Oproiu1977},
\begin{eqnarray}
 \vec { \omega}^{\rm Op} _X &=& -\ft16\left( 2\vec A_X + \vec A_Y\times \vec
  J_X{}^Y\right),\nonumber\\
\Purple{\Gamma ^{\rm Op}{}_{XY}{}^Z}&\equiv&  \Purple{\Gamma^{\rm
Ob}{}_{XY}{}^Z}-\vec J_ {(X}{}^Z\cdot \Green{{\vec\omega}^{\rm Op}
_{Y)}},\label{s:Oproiu}
\end{eqnarray}
where $\Purple{\Gamma^{\rm Ob}{}_{XY}{}^Z}$ is the \emph{Obata
connection} on  the small space. These connections can be changed by $\xi
$-transformations~\cite{AM1996,Fujimura:1976,Oproiu1984},
\begin{equation}
\Purple{\Gamma _{XY}{}^Z}\rightarrow \Purple{\Gamma _{XY}{}^Z}+2\delta
^Z_{(X}\Blue{\xi_{Y)}}-2\vec J_{(X}{}^Z\cdot \vec J_{Y)}{}^W\Blue{\xi
_W},\qquad  \OliveGreen{\vec \omega_X} \rightarrow \OliveGreen{\vec
\omega_X } +\vec J_X{}^W\Blue{\xi _W}.
 \label{s:changeGammaomega}
\end{equation}
These $\xi $-transformations are represented by the horizontal lines in
the lower part of Fig.~\ref{fig:mapoverview}. As shown in that picture,
they can e.g. be used to remove the $\mathbb{R}$ curvature in
quaternionic manifolds. They can also be used to obtain a simple form for
the connections and curvatures:
\begin{equation}
  \vec \omega _X=-\ft12\vec A_X, \qquad \Gamma_{XY}{}^Z=\widehat{\Gamma}_{XY}{}^Z + \vec A_{(X} \cdot \vec
J{}_{Y)}{}^Z.
 \label{s:specialchoiceOmega}
\end{equation}
This choice of a quaternionic connection will turn out to be special for
two different reasons. With this choice
\begin{equation}
 \widehat{R}_{XY}=- \widehat {\cal R}^{\mathbb{R}}_{XY}=-{\cal R}^{\mathbb{R}}_{XY}=R_{[XY]}.
 \label{s:RU1updown}
\end{equation}
This equality implies that the map can be represented by vertical arrows
in Fig.~\ref{fig:mapoverview}. Since in the hypercomplex space this
curvature is proportional to the Ricci tensor, which is Hermitian, this
implies that the $\mathbb{R}$ curvature on the small space is Hermitian.
For uplifting a quaternionic manifold to a hypercomplex manifold using
(\ref{s:allhatJ}) one thus first has to apply $\xi $-transformations such
that this condition is satisfied, as one can see in
Fig.~\ref{fig:mapoverview}. The $\SU(2)$ curvature is
\begin{equation}
  \vec {\mathcal{R}}_{XY} = 2\vec J_{[X}{}^Z B_{Y]Z}, \qquad \mbox{where}\qquad
  B_{XY}=\frac1{4(n_H+1)}\widehat{R}_{XY}-\frac14h_{XY}.
 \label{s:RSU2B}
\end{equation}

A quaternionic space is quaternionic-K{\"a}hler if and only if the Ricci
tensor is Hermitian, and its symmetric part is invertible. This implies
that the antisymmetric part vanishes, and the symmetric part is
proportional to the metric. In the basis that we presented, the induced
metric in the small space is
\begin{equation}
  g_{XY}= z^0 h_{XY}. %= -\frac{1}{\nu }h_{XY}= \frac{1}{\kappa ^2}h_{XY},
 \label{s:gquat}
\end{equation}
This metric does not depend on $z^\alpha $. In supergravity, $z^0$ is
fixed to $\kappa ^{-2}$, where $\kappa $ is the gravitational coupling
constant. The value of $z^0$ fixes the scale of the manifold in the sense
that
\begin{equation}
  \vec{{\cal R}}_{XY}=\ft12\nu \vec{J}_{XY}, \qquad R=4n_H(n_H+2)\nu, \qquad
  z^0=-\frac{1}{\nu },
 \label{s:scale}
\end{equation}
where $R$ is the Ricci scalar. The Levi-Civita connection of the metric
is the one in (\ref{s:specialchoiceOmega}), and this fixes the
$\xi$-transformations. Therefore, these manifolds can be represented by
the vertical thick lines in Fig.~\ref{fig:mapoverview}.

\subsection{Curvatures}

The curvatures of conformal hypercomplex manifolds have as only
non-vanishing components
\begin{equation}
  \widehat{R}_{XYZ}{}^W,\qquad\widehat{R}_{XYZ}{}^0,\qquad
  \widehat{R}_{XYZ}{}^\alpha.
 \label{s:Rhc}
\end{equation}
The latter two do not contribute in the map to the quaternionic space.
The first one is expressed using vielbeins in a $W$-tensor, which plays
an important role in supersymmetry:
\begin{equation}
\widehat{W}_{CDB}{}^A \equiv \ft12 \varepsilon ^{ij}
\widehat{f}{}^{\hat{X}}_{jC} \widehat{f}{}^{\hat{Y}}_{iD}
\widehat{f}_{kB}^{\hat{Z}} \widehat{f}_{\hat{W}}^{kA}
R_{\hat{X}\hat{Y}\hat{Z}}{}^{\hat{W}}.\label{s:WfromR}
\end{equation}
It is symmetric in its lower indices. It is however not traceless in
general. Its trace determines the Ricci tensor, and is thus zero for
hyper-K{\"a}hler.

The curvatures can be split into a Ricci part and a Weyl part. The former
further splits into a `symmetric' and an `antisymmetric' contribution,
\begin{eqnarray}
\widehat{R}_{XYZ}{}^{W}&=&
 \left( \widehat{R}^{\rm Ric}_{\rm antis}\right)
  _{XYZ}{}^W+\widehat{R}^{(\rm W)}{}_{XYZ}{}^W, \nonumber\\
   R_{XYZ}{}^W&=& \left( R^{\rm Ric}_{\rm symm}+R^{\rm Ric}_{\rm antis}\right)
  _{XYZ}{}^W+R^{(\rm W)}{}_{XYZ}{}^W.\label{s:splitR}
\end{eqnarray}
There is no Ricci-symmetric part for the hypercomplex space.

With the special $\xi$-choice as in (\ref{s:specialchoiceOmega}), we can
relate the different parts as follows:
\begin{eqnarray}
   R^{\rm Ric}_{{\rm symm}\,XYZ}{}^W&=&-\ft12\delta _{[X}^Wh_{Y]Z}
  +\ft12\vec J_Z{}^{W}\cdot \vec J_{[X}{}^{V}h_{Y]V}
  +\ft12\vec J_Z{}^{V}\cdot \vec J_{[X}{}^{W}h_{Y]V},\nonumber\\
 {R}^{\rm Ric}_{\rm antis}{}_{XYZ}{}^W&=& \frac{n_H+2}{n_H+1}
   \widehat{R}^{\rm Ric}_{\rm antis}{}_{XYZ}{}^W, \nonumber\\
 R^{(\rm W)}{}_{XYZ}{}^W&=&- \ft 12 f^{iA}_X
\varepsilon_{ij}f^{jB}_Yf_Z^{kC}f^W_{kD}
  \Magenta{\mathcal{W}_{ABC}{}^D},
 \label{s:decompR}
\end{eqnarray}
where
\begin{equation}
 \mathcal{W}_{ABC}{}^D=
\widehat{W}_{ABC}{}^D-\frac3{2(n_H+1)}\delta^D_{(A}\widehat{W}_{BC)E}{}^E.
 \label{calWinhatW}
\end{equation}

Observe that the symmetric Ricci part is a universal expression in terms
of $h_{XY}$. The Weyl part of quaternionic manifolds is determined by a
tensor ${\cal W}_{ABC}{}^D$, which is the traceless part of the one
mentioned in (\ref{s:WfromR}). The Weyl tensor of quaternionic manifolds
is invariant under the $\xi $-transformations. For the case of
hyper-K{\"a}hler and quaternionic-K{\"a}hler manifolds, the antisymmetric parts
are absent, and the Weyl parts are identical.

\subsection{Symmetries}

Triholomorphic symmetries in the hypercomplex space are either the
dilatations or they can be decomposed as
\begin{equation}
  \widehat{k}^0_I =0,\qquad \widehat{k}^\alpha _I=\vec{k}^\alpha \cdot \vec{r}_I,
  \qquad
  \widehat{k}^X_I=k_I^X(q),
 \label{s:hatkconclusion}
\end{equation}
where $k_I^X$ is a quaternionic symmetry of the small space, such that
\begin{equation}\label{s:kInormalJ}
\mathcal{L}_{\Blue{k_I}} \vec J_X{}^Y=\vec{r}_I\times \vec J_X{}^Y.
\end{equation}
Using the $\SU(2)$ connections, this defines moment maps $ \nu
\Maroon{\vec P_I}$ by
\begin{equation}
\nu \Maroon{\vec{P}_I}\equiv -\ft12   \vec{r}_I-
\Blue{k_I^X}\OliveGreen{\vec{\omega }_X}.
 \label{s:defP}
\end{equation}
For hyper-K{\"a}hler manifolds there is also a moment map, which is equal to
the one in the corresponding quaternionic-K{\"a}hler space.

Only $\xi $-transformations such that
\begin{equation}
  {\cal L}_{k_I} \xi_X=k_I^Y\partial _Y\xi _X +\xi _Y\partial _X k_I ^Y =0
 \label{s:LKxi0}
\end{equation}
preserve symmetries. The vectors in (\ref{s:hatkconclusion}) do not
change under these $\xi $-transformations. The moment map transforms as
\begin{equation}
  \nu \widetilde {\vec{P}}_I = \nu \vec{P}_I -\vec{J}_X{}^Y k_I^X\xi _Y.
 \label{s:xitransfP}
\end{equation}

A conformal 1-dimensional hypercomplex manifold is always hyper-K{\"a}hler.
In this case there is no quaternionic manifold after the map, but there
are possible constant moment maps after a point $z^\alpha $ is fixed by
the $\SU(2)$ gauge. This is the origin of the Fayet-Iliopoulos terms in
supergravity when there are no physical hypermultiplets.

\subsection{Remarks}

In this paper, we have treated the case of negative scalar curvature for
the quaternionic-K{\"a}hler space. This corresponds to $\nu <0$, see
(\ref{s:scale}). If there are continuous isometries, this implies that
the space is non-compact. Supergravity restricts us to those manifolds,
as is clear from the relation with $\kappa ^2$. This is obtained by an
indefinite signature in the hyper-K{\"a}hler space, as can be seen in
(\ref{s:QK}). The terms for $z^0$ and $z^\alpha $ have negative
signature, and we choose $h_{XY}$ to be positive definite. However, we
can as well generalize the equations to $z^0<0$, which then implies a
positive definite metric for the hyper-K{\"a}hler manifold. The scalar
curvature of the quaternionic-K{\"a}hler manifold is then positive, and this
allows compact isometry groups. The only place where this is non-trivial
is for vielbeins in Sect.~\ref{mapviel}, where $\sqrt{z^0}$ appears.
However, this can be cured by inserting appropriate $\rmi$ factors.
Furthermore, we never used properties of positive definiteness of
$h_{XY}$, so that this choice can easily be relaxed, and the analysis
applies to any signature. The signature of $h$ and the sign of $\nu $
determine the signature of the large space.

We have developed this paper in relation to matter couplings to
supergravity in 5 spacetime dimensions. The procedure is, however,
independent whether it is applied to 3, 4, 5 or 6 dimensional
supersymmetry with 8 supercharges~\cite{Rosseel:2004fa}. Therefore the
analysis of this paper can be applied to superconformal tensor calculus
in general.

The generalization of supersymmetric theories with hypercomplex or
quaternionic target spaces is related to the idea that physical systems
can be defined by field equations without necessity of an action. Indeed,
the metric is only necessary for the construction of an action, while
supersymmetry transformations and field equations are independent of a
metric.

Quaternionic-K{\"a}hler manifolds appear as moduli spaces for type II
superstring compactifications on Calabi-Yau 3-folds. It would be very
interesting to understand the role of general quaternionic manifolds in
the context of such compactifications.

In mathematics, the map that is described in this paper was investigated
in~\cite{Swann,PedersenPS1998}. We have pointed out that the
corresponding manifolds in the hypercomplex/hyper-K{\"a}hler picture are
conformal, and discovered some new properties. E.g. the $\widehat{\xi }$
and $\SU(2)$ transformations in these manifolds played an important role.
We hope that this stimulates further investigations in such conformal
hypercomplex manifolds. Furthermore, we gave new contributions to the
understanding of symmetries, generalizing isometries, in quaternionic
manifolds.

%%%%%%%%%%%%%%%%%%%%%%%%%%%%%%%%
\medskip
\section*{Acknowledgments.}

\noindent We are grateful to D. Alekseevsky, V. Cort{\'e}s, J.
Figueroa-O'Farrill, G.~Gibbons, M. Lled{\'o}, G.~Papadopoulos and Ph. Spindel
for interesting and stimulating discussions. We thank the University of
Torino for hospitality in the final stage of the preparation of this
paper.  Work supported in part by the European Community's Human
Potential Programme under contract MRTN-CT-2004-005104 `Constituents,
fundamental forces and symmetries of the universe'. The work of J.G. is
supported by the FWO-Vlaanderen to which he is affiliated as postdoctoral
researcher. The work of E.B. and T.d.W. is part of the research program
of the Stichting voor Fundamenteel Onderzoek der materie (FOM). The work
of S.C., J.G. and A.V.P. is supported in part by the Federal Office for
Scientific, Technical and Cultural Affairs through the `Interuniversity
Attraction Poles Programme -- Belgian Science Policy' P5/27. The work is
supported by the Italian M.I.U.R. under the contract P.R.I.N. 2003023852,
`Physics of fundamental interactions: gauge theories, gravity and
strings'.

\newpage
%%%%%%%%%%%%%%%%%%%%%%%%%%%
\appendix
\section{Projections Depending on the Complex Structure}
\label{app:projmatrices}

The complex structures are defined from the vielbeins as
\begin{equation}
\vec J_X{}^Y \equiv -\rmi f_X^{iA}\vec \sigma _i{}^j
f_{jA}^Y,\label{defJfapp}
\end{equation}
where $\vec \sigma$ are the three Pauli matrices. Similar matrices are
\begin{equation}
  L_W{}^Z{}_A{}^B\equiv f^Z_{iA}f_W^{iB}.
 \label{defL}
\end{equation}
They project e.g. the curvature to the $\Gl(r,\mathbb{H})$ factor as in
(\ref{RdecompJ}). The matrices $L_A{}^B$ and $\vec J$ commute and their
mutual trace vanishes %[B.39]
\begin{equation}
\vec J_X{}^Y L_Y{}^Z{}_A{}^B=
  L_X{}^Y{}_A{}^B\vec J_Y{}^Z ,\qquad \vec J_Z{}^Y L_Y{}^Z{}_A{}^B=0.
 \label{JLcommute}
\end{equation}
Other useful properties are
\begin{eqnarray}
 && L_X{}^Y{}_A{}^BL_Y{}^Z{}_C{}^D=L_X{}^Z{}_C{}^B \delta
  _A{}^D,\nonumber\\
  &&L_X{}^X{}_A{}^B=2\delta_A{}^B,\qquad
  L_X{}^Y{}_A{}^B L_Y{}^X{}_C{}^D=2\delta _C{}^B\delta_A{}^D,\nonumber\\
&&L_Z{}^W{}_A{}^B L_X{}^Y{}_B{}^A =\ft12\left( \delta _X{}^W\delta
_Z{}^Y-\vec J_X{}^W\vec J_Z{}^Y\right)  .\label{orthogJL}
\end{eqnarray}

Bilinear forms are projected to Hermitian ones using
\begin{equation}
  \Pi _{XY}{}^{ZW}\equiv \ft14\left( \delta _X{}^Z\delta _Y{}^W+\vec J_X{}^Z
  \cdot \vec J_Y{}^W   \right).
\end{equation}
As a projection operator, it squares to itself:
\begin{equation}
  \Pi _{XY}{}^{ZW}\Pi _{ZW}{}^{UV}=\Pi _{XY}{}^{UV}.
\label{Pi2Pi}
\end{equation}
Useful relations are
\begin{eqnarray}
 4 \vec J_Z{}^X \Pi _{XY}{}^{UV}&=&\vec J_Z{}^U\delta _Y{}^V-\delta_Z{}^U\vec
   J_Y{}^V- \vec J_Z{}^U\times \vec J_Y{}^V\nonumber\\
   &=& \vec J_{[Z}{}^X \Pi
   _{|X|Y]}{}^{(UV)}+ \vec J_{(Z}{}^X \Pi
   _{|X|Y)}{}^{[UV]}, \label{JPi}\\
  4 \Pi _{XY}{}^{ZV}\vec J_V{}^U &=&\delta _X{}^Z\vec J_Y{}^U-\delta_Y{}^U\vec
   J_X{}^Z- \vec J_X{}^Z\times \vec J_Y{}^U\nonumber\\
   &=& \Pi _{[XY]}{}^{(Z|V|}\vec J_V{}^{U)} +  \Pi
    _{(XY)}{}^{[Z|V|}\vec J_V{}^{U]}. \label{PiJ}
\end{eqnarray}

A \textit{Hermitian bilinear form} is a tensor $F_{XY}$ such that
$J^\alpha{} _X{}^ZJ^\alpha {}_Y{}^WF_{ZW}=F_{XY}$ (no sum over $\alpha
$). Any bilinear form can be projected to the space of Hermitian bilinear
forms by the projection $\Pi $. It is easy to prove that if
\begin{equation}
 \Pi_{XY}{}^{ZW}F_{ZW}=F_{XY},
 \label{PiFF}
\end{equation}
this implies the hermiticity of $F$. Thus, Hermitian bilinear forms are
those that satisfy~(\ref{PiFF}). Moreover, one can prove that they
satisfy
\begin{equation}
  F_{XU}\vec J_Y{}^U=\ft12\vec J_X{}^U\times \vec J_Y{}^VF_{UV},
 \label{propHermBil}
\end{equation}
and thus also
\begin{equation}
  F_{XU}\vec J_Y{}^U+\vec J_X{}^UF_{UY}=0.
 \label{FJTis-JF}
\end{equation}
Inversely, this equation [or~(\ref{propHermBil})] is also sufficient for
a form to be Hermitian.

Finally, another useful matrix between bilinear forms is
\begin{equation}
   S_{XY}{}^{ZW}\equiv 2\delta ^Z_{(X}\delta ^W_{Y)}-2 \vec J_X{}^{(Z}\cdot
   \vec J_Y{}^{W)}= 4\delta_{(XY)}{}^{ZW}-8 \Pi _{(XY)}{}^{ZW}.
\label{defSapp}
\end{equation}
It satisfies the following identities:
\begin{eqnarray}
   &&  S_{WX}{}^{WV}=4(r+1)\delta _X^V, \label{Scontr}\\
   && S_{XY}{}^{ZV}\vec J_V{}^U= 4\delta _{(X}{}^{(Z}\vec J_{Y)}{}^{U)}+2
   \vec J_{(X}{}^Z\times \vec J_{Y)}{}^U, \label{SJ}\\
   &&  S_{ZW}{}^{XV}\vec J_V{}^Y-\vec J_W{}^V S_{ZV}{}^{XY}=2
\vec J_Z{}^X\times \vec J_W{}^Y,\label{SJ-JS}\\
&& S_{XZ}{}^{YW}\vec{J}_W{}^Z= 4n_H \vec{J}_X{}^Y,\label{SJcontr}\\
 && S_{T[X}{}^{(WU}S_{Y]Z}{}^{V)T}=0 \label{SS},
 \end{eqnarray}
which are used at various places in the main text.
%%%%%%%%%%%%%%%%%%%%%%%%%%%%
\providecommand{\href}[2]{#2}\begingroup\raggedright\endgroup

%%%%%%%%%%%%%%%%%%%%%%%%%%%%%%%%%%%%%%%%%%%%%%%%%%%%%%%%
%\bibliography{refd5conf}
%\bibliographystyle{toine}
\end{document}